%% file: main.tex
\documentclass[sn-basic,iicol]{sn-jnl}

\jyear{2021}%

\theoremstyle{thmstyleone}%

\usepackage[english]{babel}
\usepackage{natbib}
\usepackage{program}
\catcode`\|=12\relax
\usepackage{bbm}
\usepackage{afterpage}
\usepackage{float}
\usepackage[utf8]{inputenc}
\usepackage{graphicx}
\usepackage{amsmath}
\usepackage{mathtools}
\usepackage{amsthm}
\usepackage{amssymb}

\usepackage{csvsimple}
\usepackage{adjustbox}
\usepackage{longtable}
\usepackage{booktabs}

\usepackage{blkarray}

\usepackage{float}

\usepackage{placeins}

\usepackage[shortcuts]{extdash}

\usepackage{caption}
\usepackage{subcaption}
\usepackage{csquotes}
\usepackage{algorithm}
\usepackage{listings}

\usepackage{color}

\usepackage{optidef}
\usepackage{bm}
\usepackage{caption}
\usepackage{subcaption}

\usepackage{microtype}

\usepackage{mathtools}
\usepackage[capitalize,noabbrev]{cleveref}

\theoremstyle{plain}

\theoremstyle{definition}

\theoremstyle{remark}

\usepackage{bbm}

\usepackage{tikz}
\usepackage{tikzsymbols}
\usepackage{pgfplots}
\pgfplotsset{height=6cm, 
	width=12cm,
	compat=1.16,
	legend style = {fill = white, draw = black},
	contour/every contour label/.style={
			every node/.style={mapped color!50!white,fill=white}}
}
\usetikzlibrary{shapes}
\usetikzlibrary{arrows}
\usetikzlibrary{positioning}
\usetikzlibrary{calc}
\usetikzlibrary{automata} 
\usetikzlibrary{intersections}
\usetikzlibrary{backgrounds}
\usetikzlibrary{external}
\tikzset{
	declare function={
			discreteuniform(\a,\b)=1/(\b-\a+1);%
			binomial(\n,\p)=\n!/(x!*(\n-x)!)*\p^x*(1-\p)^(\n-x);%
			poisson(\l)=(\l^x)*exp(-\l)/(x!);%
			negativebinomial(\r,\p)=((x+\r-1)!/((\r-1)!*x!))*((1-\p)^x*\p^\r);%
			continuousuniform(\a,\b)=1/(\b-\a);%
			gaussian(\m,\s)=1/(\s*sqrt(2*pi))*exp(-((x-\m)^2)/(2*\s^2));%
			lognormal(\m,\s)=1/(x*\s*sqrt(2*pi))*exp(-((ln(x)-\m)^2)/(2*\s^2));%
			exponential(\l)=\l*exp(-\l*x);%
			gamma(\z)=2.506628274631*sqrt(1/\z)+ 0.20888568*(1/\z)^(1.5)+ 0.00870357*(1/\z)^(2.5)- (174.2106599*(1/\z)^(3.5))/25920- (715.6423511*(1/\z)^(4.5))/1244160)*exp((-ln(1/\z)-1)*\z;%
			gammadist(\a,\t)=(x^(\a-1))*exp(-x/\t)*gamma(\a)*\t^\a;%
			student(\n)=gamma((\n+1)/2.)/(sqrt(\n*pi)*gamma(\n/2.))*((1+(x*x)/\n)^(-(\n+1)/2.));%
			cauchy(\m,\s)=(1/(pi*\s*(1+((x-\m)/\s)^2)));%
			beta(\a,\b)=(x^(\a-1)*(1-x)^(\b-1)/((gamma(\a)*gamma(\b))/gamma(\a+\b));%
			binormal(\ma,\sa,\mb,\sb,\ro)=exp(-(((x-\ma)/\sa)^2+((y-\mb)/\sb)^2-(2*\ro)*((x-\ma)/\sa)*((y-\mb)/\sb))/(2*(1-\ro^2)))/(2*pi*\sa*\sb*(1-\ro^2)^0.5);%
			conditionalbinormal(\yc,\ma,\sa,\mb,\sb,\ro)=exp(-(((x-\ma)/\sa)^2+((\yc-\mb)/\sb)^2-(2*\ro)*((x-\ma)/\sa)*((\yc-\mb)/\sb))/(2*(1-\ro^2)))/(2*pi*\sa*\sb*(1-\ro^2)^0.5);%
			sumtwonormals(\ma,\sa,\wa,\mb,\sb,\wb)=(\wa*gaussian(\ma,\sa))+(\wb*gaussian(\mb,\sb));%
			normcdf(\m,\s)=1/(1 + exp(-0.07056*((x-\m)/\s)^3 - 1.5976*(x-\m)/\s));%
		}
}

\lstdefinelanguage{Julia}%
{morekeywords={abstract,begin,break,case,catch,const,continue,do,else,elseif,end,%
			end,export,false,for,function,immutable,import,importall,if,in,%
			macro,module,otherwise,quote,return,switch,true,try,type,typealias,%
			using,while},%
	sensitive=true,%
	alsoother={$},%
	morecomment=[l]\#,%
	morecomment=[n]{\#=}{=\#},%
	morestring=[s]{"}{"},%
	morestring=[m]{'}{'},%
}[keywords,comments,strings]%

\definecolor{codegreen}{rgb}{0,0.6,0}
\definecolor{codegray}{rgb}{0.5,0.5,0.5}
\definecolor{codepurple}{rgb}{0.58,0,0.82}
\definecolor{backcolour}{rgb}{0.95,0.95,0.92}

\lstdefinestyle{mystyle}{
	language         = Julia,
	backgroundcolor=\color{backcolour},
	commentstyle=\color{codegreen},
	keywordstyle=\color{blue},
	numberstyle=\tiny\color{codegray},
	stringstyle=\color{codepurple},
	basicstyle=\ttfamily\footnotesize,
	breakatwhitespace=false,
	breaklines=true,
	captionpos=b,
	keepspaces=true,
	numbers=left,
	numbersep=5pt,
	showspaces=false,
	showstringspaces=false,
	showtabs=false,
	tabsize=2,
        escapeinside=\{\},
}
\usepackage{hyphenat}
\usepackage{graphicx}
\usepackage{tabularx}
\usepackage{booktabs}
\usepackage{bookmark}
\usepackage{multicol} 

\usepackage{amsmath}
\usepackage{amssymb}
\usepackage{wasysym}

\usepackage{lstbayes} 

\usepackage[normalem]{ulem}

\begin{document}

\input{sections/0-title_and_abstract}
\input{sections/1-introduction}
\input{sections/2-bayesian-workflow}
\input{sections/3-background-and-intuition}
\input{sections/4-example_models}
\input{sections/5-conclusion_and_acknowledgements}

\clearpage

\bibliography{refs} 




\end{document}

%% file: sections/0-title_and_abstract.tex

\title[Pumas Bayesian]{A Practitioner's Guide to Bayesian Inference in Pharmacometrics using Pumas}


\author*[1,2]{\fnm{Mohamed} \sur{Tarek}}\email{mohamed@pumas.ai}

\author[1,3]{\fnm{Jose} \sur{Storopoli}}\email{jose@pumas.ai}

\author[4]{\fnm{Casey} \sur{Davis}}\email{cbdavis33@gmail.com}

\author[1,5]{\fnm{Chris} \sur{Elrod}}\email{chrise@pumas.ai}

\author[1]{\fnm{Julius} \sur{Krumbiegel}}\email{juliusk@pumas.ai}

\author[1,5,6]{\fnm{Chris} \sur{Rackauckas}}\email{crackauc@mit.edu}

\author[1,7]{\fnm{Vijay} \sur{Ivaturi}}\email{vijay@pumas.ai}


\affil*[1]{\orgname{Pumas-AI Inc.}, \orgaddress{\country{USA}}}

\affil[2]{\orgdiv{Business School}, \orgname{University of Sydney}, \orgaddress{\country{Australia}}}

\affil[3]{\orgdiv{Department of Computer Science}, \orgname{Universidade Nove de Julho - UNINOVE}, \orgaddress{\country{Brazil}}}

\affil[4]{\orgname{Gilead Sciences Inc.}, \orgaddress{\country{USA}}}

\affil[5]{\orgname{JuliaHub (formerly JuliaComputing) Inc.}, \orgaddress{\country{USA}}}

\affil[6]{\orgdiv{Computer Science and Artificial Intelligence Laboratory}, \orgname{Massachusetts Institute of
Technology}, \orgaddress{\country{USA}}}

\affil[7]{\orgdiv{School of Pharmacy}, \orgname{University of Maryland Baltimore}, \orgaddress{\country{USA}}}



\abstract{This paper provides a comprehensive tutorial for Bayesian practitioners in pharmacometrics using Pumas workflows. We start by giving a brief motivation of Bayesian inference for pharmacometrics highlighting limitations in existing software that Pumas addresses. We then follow by a description of all the steps of a standard Bayesian workflow for pharmacometrics using code snippets and examples. This includes: model definition, prior selection, sampling from the posterior, prior and posterior simulations and predictions, counter-factual simulations and predictions, convergence diagnostics, visual predictive checks, and finally model comparison with cross-validation.
Finally, the background and intuition behind many advanced concepts in Bayesian statistics are explained in simple language. This includes many important ideas and precautions that users need to keep in mind when performing Bayesian analysis. Many of the algorithms, codes, and ideas presented in this paper are highly applicable to clinical research and statistical learning at large but we chose to focus our discussions on pharmacometrics in this paper to have a narrower scope in mind and given the nature of Pumas as a software primarily for pharmacometricians.}

\keywords{Bayesian inference, statistical software, pharmacometrics, workflow}



\maketitle
\thispagestyle{empty}

\begin{figure*}
\setcounter{tocdepth}{2}
\tableofcontents
\end{figure*}

\clearpage
\pagenumbering{arabic}

%% file: sections/1-introduction.tex

\section{Motivation}

Fully Bayesian approaches have become important tools in a pharmacometrician's toolbox \citep{lee2011bayesian,lecanemab} because they enable the rigorous and flexible quantification of uncertainty in all of the model's parameters as well as the use of knowledge from previous similar studies which have applications in rare disease drug approvals, pediatric studies, precision dosing and adaptive trial design. The Bayesian workflow implemented in Pumas \citep{rackauckasAcceleratedPredictiveHealthcare2020} was designed to be flexible and easily accessible using an intuitive clean syntax. We hope that this paper can be a good learning resource for any pharmacometrician interested in learning about and using fully Bayesian methods.

\section{Introduction} \label{sec:introduction}

In this section, we discuss the need for a fully Bayesian approach in pharmacometrics and describe where it fits in the spectrum of methods typically used in the field.
The standard models, data, algorithms, workflows, and software used in pharmacometrics will be briefly presented to set the context for future discussions.
Finally, the main contributions of Pumas and the layout of this paper will be summarized.

\subsection{Pharmacometrics Workflow}

Pharmacometrics is a field of study that includes various quantitative analysis techniques used to understand interactions between drugs and patients.
A typical pharmacometrics workflow involves:
\begin{enumerate}
  \item Prepare analysis-ready datasets from clinical trial source datasets,
  \item Exploratory data analysis and scientific reasoning of observed data via summaries and plots,
  \item Developing parametric models for disease-drug-patient interaction,
  \item Fitting the models' parameters to the data and estimating uncertainty in the parameters,
  \item Diagnosing the models' fits and evaluating the quality of their fits,
  \item Comparing models and selecting the best model, and
  \item Using the best model to predict/simulate alternative scenarios for existing or new patients or to answer key drug development questions.
\end{enumerate}

\subsection{Data} 

The data used in pharmacometrics typically includes:
\begin{enumerate}
	\item Patient covariates, e.g. age or sex, sometimes including time-varying covariates,
	\item Drug dosing regimen and administration schedule for each patient, and
	\item Observed response for each patient, e.g. the measured drug concentration in the blood and/or some clinical endpoints, typically observed at multiple time points.
\end{enumerate}

\subsection{Models}

The kinds of models used in pharmacometrics are typically:
\begin{enumerate}
	\item Structural models to capture the observed response, e.g, dynamic-based models where the pharmacokinetics are modeled using ordinary differential equations (ODEs)
        \item Covariate models predicting the observed response conditional on the covariates, and
	\item Hierarchical models with population-level parameters and patient-specific parameters,\footnote{
		      In this paper, we use the terminology ``population-level'' parameters and ``patient-specific'' (or subject-specific) parameters instead of ``fixed effects'' and ``random effects'' which are more common in pharmacometrics.
		      This is because the definition of fixed and random effects in the statistical literature is ambiguous (see page 21 in \cite{Gelman_2005}), and in Bayesian modeling, every parameter is modeled using a random variable, which further increases the ambiguity.
	      }
\end{enumerate}

These are not mutually exclusive.
For example, a covariate-based model can also be a hierarchical model with a dynamic-based ODE component.



The following showcases a classic model of Theophylline dynamics via a 1-compartment pharmacokinetic (PK) oral absorption model.
The model describes the dynamics of drug absorption into the bloodstream and its clearance.
Initially, the drug is assumed to be administered as a bolus into a single depot compartment, e.g.\ when it's first ingested into the gut.
The drug then gradually enters the bloodstream with an absorption rate $\text{Ka} \times \text{Depot}$ where $\text{Depot}$ is the amount of drug remaining in the gut.
The central compartment represents the drug amount absorbed in the bloodstream over time.
The drug is also assumed to be cleared from the central compartment by a rate $\frac{\text{CL}}{V} \times \text{Central}$, where $\text{V}$ is the volume of distribution of this compartment.

The model has population parameters (\boldsymbol{$\theta$}, \boldsymbol{$\Omega$}, $\sigma$) and patient-specific parameters $\boldsymbol{\eta}_i$.
\boldsymbol{$\theta$} is a vector of $4$ numbers, \boldsymbol{$\Omega$} is a $3 \times 3$ positive definite covariance matrix and $\sigma$ is a standard deviation parameter for the error model.
In this model, each patient $i$ has weight and sex covariates: 

\begin{equation}
	Z_i =
	\begin{bmatrix}
		\text{wt}_i, \\ \text{sex}_i, \\
	\end{bmatrix},
\end{equation}
where $\text{wt}_i$ is a positive real number and $\text{sex}_i$ is an indicator variable which is 0 for males and 1 for females, has individual coefficients:

\begin{equation}
	\begin{bmatrix}
		\text{Ka} \\ \text{CL} \\ V \\
	\end{bmatrix}
	=
	\begin{bmatrix}
		\theta_1 e^{\eta_{i,1}} \\ \theta_2 (\frac{\text{wt}_i}{70})^{0.75} \theta_{4}^{\text{sex}_i} e^{\eta_{i,2}} \\ \theta_3 e^{\eta_{i,3}} \\
	\end{bmatrix},
\end{equation}
has internal dynamics: 

\begin{align*}
	\frac{d[\text{Depot}]}{dt} & = -\text{Ka} [\text{Depot}] \\ \frac{d[\text{Central}]}{dt} & = \text{Ka} [\text{Depot}] - \frac{\text{CL}}{V} [\text{Central}],
\end{align*}
where $\text{Depot}$ and $\text{Central}$ are the depot and central compartments, and has normally distributed residual error in the observed drug concentration $\text{conc}$, also known as the error model: 

\begin{align*}
	\text{conc} & \sim \text{Normal} \left( \frac{\text{Central}}{V}, \sigma \right).
\end{align*}

\subsection{Algorithms} 

In this section, we briefly present an overview of the various algorithms commonly used to fit models to data in pharmacometrics.

\subsubsection{Marginal Maximum Likelihood Estimation (MLE)} 

When fitting parametric models to data in pharmacometrics, marginal MLE algorithms are the most popular.
The patient-specific parameters are marginalized and the marginal likelihood is maximized.
There are two families of algorithms to do MLE with the marginal likelihood:
\begin{enumerate}
  \item Approximate integration of the conditional likelihood \citep{wang_derivation_2007} which includes:
  \begin{enumerate}
      \item Laplace method, and
      \item First order conditional estimation (FOCE).
  \end{enumerate}
  \item Stochastic approximation expectation maximization (SAEM) \citep{delyon_convergence_1999,KUHN20051020}.
\end{enumerate}

\subsubsection{Marginal Maximum-a-Posteriori (MAP) Estimation}

Marginal MAP is an alternative to marginal MLE that also gives a point estimate for the population parameters but instead of maximizing the marginal likelihood, it maximizes the product of the prior probability of the population parameters and the marginal likelihood.

\subsubsection{Fully Bayesian Analysis}

Marginal likelihood maximization can give us, as a by-product, the conditional posterior of the patient-specific parameters $\eta_i$ given the optimal population parameters, or an approximation of it.
However, a fully Bayesian approach can give us samples from the joint posterior of all the population parameters and patient-specific parameters simultaneously.
The sampling from the joint posterior is typically done using an MCMC algorithm such as the No-U-Turn sampler (NUTS) \citep{hoffman2014no,multlinomial} algorithm which is a variant of the Hamiltonian Monte Carlo (HMC) \citep{neal2011mcmc} MCMC algorithm.
We will cover Bayesian notation and algorithms in more detail in later sections of the paper.

Besides the ability to sample from the joint posterior and easily simulate from the posterior uncertainty, a fully Bayesian approach allows modellers to: 

\begin{enumerate}
  \item Incorporate domain knowledge and insights from previous studies using prior distributions.
  \item Quantify the so-called epistemic uncertainty, which is the uncertainty in the model parameters' values in cases where the model is non-identifiable\footnote{if a parameter is unidentifiable, that means their values cannot be uniquely determined from the available data.} and/or there are not many data points available.
\end{enumerate}

The above are advantages of Bayesian analysis which the non-Bayesian workflow typically doesn't have a satisfactory answer for. Bayesian inference as a paradigm uses the established theory of probability to rigorously quantify the uncertainty in the parameter values with fewer assumptions about the model and data.

Bayesian workflows empower the analyst to use, when available,  domain knowledge to quantify the epistemic uncertainty of model parameters, thus providing immense flexibility in analysis pipelines.

Since Bayesian analysis is a conceptual replacement for bootstrapping, the performance of Bayesian inference should be compared to that of bootstrapping rather than that of a single model fit using Laplace/FOCE/SAEM. This is important to set the users' expectations because Bayesian inference will typically be one or more orders of magnitude slower than running a single model fit with Laplace/FOCE/SAEM.

\subsection{Software}

A number of software implement all the classic MLE-based analysis workflows, e.g. NONMEM \citep{Beal2009NONMEMUG}, Monolix, and Pumas \citep{rackauckasAcceleratedPredictiveHealthcare2020}.
For fully Bayesian analysis, a few options exist including: Pumas, Stan \citep{carpenter_stan_2015}, Torsten \citep{margossianFlexibleEfficientBayesian2022}, BUGS \citep{goudie2020multibugs}, Turing.jl \citep{pmlr-v84-ge18b}, PyMC \citep{pymc}, Pyro \citep{pyro} and NONMEM's relatively new Bayesian module.

In Bayesian statistics, Stan has grown to be the de-facto standard software for Bayesian analysis when only continuous parameters exist, which is typically the case in pharmacometric models.
Stan has been tested and used by many statisticians working in different fields and with different kinds of models over many years.
Of all the above software for Bayesian analysis, Stan has the largest community and most mature ecosystem of R packages for a full Bayesian analysis workflow.

\subsection{Limitations of Other Software}

The main limitation of generic probabilistic programming languages (PPLs), such as Stan or Turing.jl, in pharmacometrics (PMx) is that they are not tailored to PMx users and the kinds of models and data found in PMx.
For instance, as of the time of this writing, parallelism over subjects is difficult to write in Stan\footnote{In order to parallelize over subjects, you need to use the \texttt{reduce\_sum} function in the likelihood which is not a trivial task. For examples on using the \texttt{reduce\_sum} function, you can refer to \href{stanpmx.github.io}{https://stanpmx.github.io}.}.

Torsten tries to bridge the gap between PMx users and the Stan software, e.g. by making it easier to define dosing regimens and simplifying parallelism over subjects using its group solver. For more on the Torsten group solver, you can refer to \href{metrumresearchgroup.github.io/Torsten/function/ode-pop/}{https://metrumresearchgroup.github.io/Torsten/function/ode-pop/}.
However, the APIs of both Stan and Torsten do not provide the best user experience for PMx users because they are based on C++, an arguably difficult-to-use, non-interactive programming language.
Additionally, as of the time of this writing, it is unnecessarily difficult to define dynamics initial conditions in Torsten that depend on the model's parameters, while it is easier to do that in the parent PPL Stan.

More broadly, once you assume a particular model structure, there are numerous computational optimizations, e.g. automatic parallelism and automatic ODE linearity detection, that can be employed which we do in Pumas, but which are difficult to do in other more generic software.

The second limitation of other software is that they do not support other non-Bayesian workflows that are common in PMx which means that users with a PMx background need to translate their models from other software to the PPL to compare the Bayesian and non-Bayesian results.
This is often a time consuming and error-prone process.

\subsection{Related Works and Contributions}

This paper is not the first attempt to describe a Bayesian workflow in the context of pharmacometrics. Works like \cite{wakefield1999} and \cite{Lunn2002} were early attempts using the BUGS software \citep{goudie2020multibugs}. More recently, the Torsten paper \citep{margossianFlexibleEfficientBayesian2022} and Bayesian workflow paper \citep{bayesian_workflow} are also excellent resources with a focus on using Stan \citep{carpenter_stan_2015}. \cite{bayesian_workflow} is particularly focused on the workflow aspect of developing Bayesian models and performing analysis and diagnostics, and while it is not customized for pharmacometrics, the Torsten paper \citep{margossianFlexibleEfficientBayesian2022} attempts to fill this gap. There is also \cite{Elmokadem2023} which provides a guide to using Stan/Torsten or Turing.jl \citep{pmlr-v84-ge18b} for Physiologically-based pharmacokinetic (PBPK) models. Given the above excellent resources, we identify the two main contributions of this paper as:
\begin{enumerate}
    \item Offering a complete pharmacometrics-oriented Bayesian workflow including scripts and examples of common models using the Pumas software.
    \item Offering an intuitive explanation of various Bayesian statistical and computational concepts that enable the fine tuning of algorithm's options and making sense of software outputs without being too technical.
\end{enumerate}

We believe the Pumas Bayesian workflow implementation addresses many of the issues with other software by:

\begin{enumerate}
  \item Providing a complete, user-friendly Bayesian analysis workflow that includes:
    \begin{enumerate}
        \item model and dosing regimen definition,
	\item MCMC sampling,
        \item diagnostics,
        \item simulation,
        \item counter-factual predictions, and
        \item customizable cross-validation,
    \end{enumerate}
  using a few lines of code;
  \item Using the same user-friendly, compact model syntax for hierarchical, dynamics-based models used in both Bayesian and non-Bayesian analyses so there is no need to translate models;
  \item Using Julia as the core programming language which is a fast, interactive, and easy-to-use language, providing an excellent user experience;
  \item Automating the definition and computational optimization of PMx models, e.g. using automatic parallelism over subjects and automatic ODE linearity and stiffness detection, delivering high performance with a neat syntax; and
  \item Using the open-source, modular implementation, AdvancedHMC.jl \citep{xu2020advancedhmc}, of the same MCMC algorithm that Stan uses, which is also used in Turing.jl \citep{pmlr-v84-ge18b}).
\end{enumerate}


\subsection{Paper Layout and Reading Plan}

The rest of this paper is organized as follows. In Section \ref{sec:workflow}, we describe the Bayesian workflow in Pumas. This should help users who are already familiar with the Bayesian workflow and theory to get started quickly using Pumas. In Section \ref{sec:background}, we then a take a step back and give a reasonably comprehensive, intuition-oriented introduction to all of the main concepts and ideas invoked in the Bayesian workflow including the intuition behind the algorithm used and general points of caution when selecting priors. This section can help users make sense of highly technical concepts using mostly English and some light mathematics which can be useful to make informed decisions when using Pumas. Finally, Section \ref{sec:example_models} contains a number of common example models and Section \ref{sec:conclusion} includes some additional reading material and resources for current and future Pumas users.

This paper is meant to serve both as a tutorial paper for users of Pumas and as an accessible resource for pharmacometricians learning Bayesian inference. If you are an experienced Bayesian and would like to learn how to use Pumas, then you can read Section \ref{sec:workflow} only and skip Section \ref{sec:background}. If you are still learning Bayesian theory, we recommend reading at least some parts of Section \ref{sec:background} first. You can then use Table \ref{tab:reading_plan} to switch to reading the relevant parts in Section \ref{sec:workflow} if you prefer interweaving theory and code. For more focused tutorials and scripts, you can also check the Pumas tutorials website (\href{tutorials.pumas.ai}{https://tutorials.pumas.ai}).

\begin{table*}
    \centering
    \begin{tabular}{|c|c|c|}
        \hline
        \textbf{Theme} & \textbf{Section \ref{sec:workflow}} & \textbf{Section \ref{sec:background}} \\
        \hline
        Model definition and prior choice & \ref{sec:defmodel} & \ref{sec:notation}, \ref{sec:bayes_stats}, \ref{sec:priors} \\
        \hline
        Prior and posterior predictions and simulations & \ref{sec:prior_sims_and_preds}, \ref{sec:posterior_queries}-\ref{sec:nca} & \ref{sec:mcmc_intuition} \\
        \hline
        Inference algorithm and options & \ref{sec:fit}-\ref{sec:basic_summary} & \ref{sec:nuts}-\ref{sec:summary_stats3} \\
        \hline
        Convergence and diagnostics & \ref{sec:nsamples}-\ref{sec:notconverging} & \ref{sec:convergence} \\
        \hline
        Crossvalidation and model Selection & \ref{sec:crossvalidation}-\ref{sec:information} & \ref{sec:model_selection} \\
        \hline
    \end{tabular}
    \caption{This table shows the relevant groups of subsections from Section \ref{sec:workflow} and Section \ref{sec:background}.}
    \label{tab:reading_plan}
\end{table*}

%% file: sections/2-bayesian-workflow.tex

\section{Bayesian Workflow in Pumas} \label{sec:workflow}

A general Bayesian analysis workflow was presented in \cite{bayesian_workflow} which summarizes many of the current best practices in Bayesian statistics.
There are many similarities between the general Bayesian workflow and the standard pharmacometrics workflow based on marginal MLE.
There are a number of differences between them worth highlighting though:
\begin{enumerate}
	\item The Bayesian workflow focuses on what a modeller can do when MCMC sampling fails for a model.
	      This could be failure because the sampler is taking too long to complete its sampling, the sampler is often triggering numerical errors,or the samples are failing to pass the diagnostic tests after sampling.
	      Things like simplifying the model or using less or more data can help pinpoint the issue and improve sampling.
	      Much of this is also applicable if MLE/MAP optimization fails but optimization failing is less common than MCMC failing with one of its many failure modes.
	\item The Bayesian workflow includes prior selection diagnostics, e.g. prior predictive checks.
	      Marginal MLE optimization does not use priors on population parameters, and there are well-established standard priors often used for patient-specific parameters so this is also less of an issue outside the Bayesian context.
	\item The general Bayesian workflow mentions using the model and MCMC samples to make predictions, but it naturally does not specify what kinds of predictions which is very much context-specific.
	      In pharmacometrics, we are typically interested in identifying the effect of using different dose amounts for new or existing patients or in predicting/simulating future responses given past data.
\end{enumerate}
In this section, we will borrow components from the general Bayesian workflow, customize it for pharmacometrics, and present the Pumas syntax that can help a pharmacometrician follow the current best practices when performing Bayesian analysis. The syntax presented here is using Pumas 2.4 (to be released in June 2023), but Pumas 2.3 has most of the workflow implemented already. For updated syntax, options, and features, please refer to the Pumas documentation (\href{docs.pumas.ai}{https://docs.pumas.ai}).

\subsection{Defining a Model}
\label{sec:defmodel}

\subsubsection{Overview}

In Pumas you can define models using the \texttt{@model} macro, which is composed of model blocks.
We will cover six model blocks in this tutorial:
\begin{enumerate}
    \item \texttt{@param}: where we define the population parameters of our model, along with their priors.
    \item \texttt{@random}: where we define the subject-specific parameters of our model, along with their priors. This is an optional block which can be dropped for single subject models.
    \item \texttt{@covariates}: where we declare the subject's covariates. This is an optional block which can be dropped for covariate-free models.
    \item \texttt{@pre}: here we do all sorts of pre-computations and other statistical transformations, e.g. calculating the individual PK parameters.
    \item \texttt{@dcp}: here you can define dose control parameters (DCPs) in the model. This is an optional block which can be dropped if no DCPs exist in the model.
    \item \texttt{@dynamics}: where we define all of our dynamics, e.g. the system of ODEs that governs the relationship between the PK/PD compartments. This is an optional block which can be dropped for models without ODEs.
    \item \texttt{@derived}: this is where we defined our observed response's distribution used to calculate the likelihood.
    \item \texttt{@observed}: this is where we define additional variables to be computed during simulation but not fitting, e.g a non-compartmental analysis (NCA) based on the concentration curve can be performed here. This is an optional block.
\end{enumerate}

With these blocks, you can code almost all PK/PKPD/PBPK models in Pumas.
We'll cover the functionality of all of these model blocks.

First, the \texttt{@param} block is where we include all of the population parameters of our model.
We begin by specifying a \texttt{begin ... end} where we insert one parameter per line.
For each parameter, we give a prior with the \texttt{$\sim$} operator followed by a distribution.
Listing \ref{lst:param} shows an example of an \texttt{@param} block with three parameters and their priors.
\texttt{tvcl} has a \texttt{LogNormal} prior with log-scale mean $\log(2.5)$ and unit log-scale standard deviation $1$,
\texttt{tvvc} has a positive-constrained \texttt{Normal} prior with mean $70$ and standard deviation $10$,
and \texttt{$\sigma$} has an \texttt{Exponential} prior with rate $3$.\footnote{To write LaTeX symbols like $\sigma$ and $\theta$ in Pumas, you can write the LaTeX form, e.g. $\backslash sigma$, followed by the Tab key.}
\begin{lstlisting}[caption=\texttt{@param} block example,label=lst:param]
@param begin
    tvcl {$\sim$} LogNormal(log(2.5), 1)
    tvvc {$\sim$} Constrained(Normal(70, 10); lower = 0)
    {$\sigma$} {$\sim$} Exponential(3)
end
\end{lstlisting}

Next, the \texttt{@random} block holds all of the subject-specific parameters and their priors.
Similar to the \texttt{@param} block, we also begin by specifying a \texttt{begin ... end} where we insert one parameter per line; and each parameter is assigned a prior with the \texttt{$\sim$} operator followed by a distribution.
In Listing \ref{lst:random}, we have an example of an \texttt{@random} block with a single parameter $\eta$ which has an \texttt{MvNormal} (multivariate normal) prior with mean $0$ and identity covariance matrix.
\begin{lstlisting}[caption=\texttt{@random} block example,label=lst:random]
@random begin
    {$\eta$} {$\sim$} MvNormal([1 0; 0 1])
end
\end{lstlisting}

The \texttt{@covariates} block is used to specify the subject's covariates.
This is only a declaration block that follows the same approach by declaring one covariate per line inside the \texttt{begin ... end} statements.
You can find an example in Listing \ref{lst:covariates} where we are declaring two subject covariates: \texttt{WT} for the subject's weight, and \texttt{SEX} for the subject's sex.
\begin{lstlisting}[caption=\texttt{@covariates} block example,label=lst:covariates]
@covariates begin
    WT
    SEX
end
\end{lstlisting}

We continue with the \texttt{@pre} block, where any pre-computations or statistical transformations necessary for our model are done.
In this block, we can use all of the parameters and covariates declared in the previous blocks.
The approach is similar to the other blocks so far: one pre-computation or transformation per line inside the \texttt{begin ... end} statements.
Listing \ref{lst:pre} provides an example of the \texttt{@pre} block,
in which we compute the individual PK parameters for each subject.
\begin{lstlisting}[caption=\texttt{@pre} block example,label=lst:pre]
@pre begin
    CL = tvcl * exp({$\eta$}[1]) * (WT / 70)^0.75
    Vc = tvvc * exp({$\eta$}[2]) * (WT / 70)
end
\end{lstlisting}

The fifth block, \texttt{@dynamics}, is where we specify the ODE system that governs the dynamics of our model.
In this block, we declare one ODE per line inside \texttt{begin ... end} statements.
On the left-hand side (LHS) of the equation is the derivative of the compartment, i.e. the rate of change. We are free to give each compartment any name. On the LHS the compartment name is followed with a \texttt{'} (prime) operator to denote the rate of change in that compartment.
The prime operator is an intuitive way to declare the derivative of the compartment.
On the right-hand side (RHS) of the equation is some combination of the compartments and the individual parameters specified in the \texttt{@pre} block.
In the example in Listing \ref{lst:dynamics}, we specify the dynamics of the Theophylline model presented in section \ref{sec:introduction} with parameters \texttt{Ka}, \texttt{CL} and \texttt{Vc}..
\begin{lstlisting}[caption=\texttt{@dynamics} block example,label=lst:dynamics]
@dynamics begin
    Depot' = -Ka * Depot
    Central' = Ka * Depot - CL/Vc * Central
end
\end{lstlisting}
Pumas supports automatic ODE linearity and stiffness detection. So even linear ODEs can be written in the above readable syntax with no performance loss. Additionally if ODE stiffness is detected, Pumas will switch to a stiff ODE solver as appropriate.

The \texttt{@derived} block is where we define our likelihood term.
We can use two types of assignments in this block: the deterministic assignment \texttt{=} and the probabilistic assignment \texttt{$\sim$}.
For the deterministic assignments, the RHS is a deterministic quantity, whereas for the probabilistic assignment the RHS is a probabilistic quantity represented as a distribution.
In Listing \ref{lst:derived}, we have two variables being defined\footnote{We are using the macro \texttt{@.} to vectorize all of the function calls to the right of it. Since \texttt{Central} and \texttt{cp} are vectors of values at the observed time points.}.
The first is \texttt{cp} defined using the deterministic assignment, and the second is the \texttt{conc} defined using the probabilistic assignment while also using \texttt{cp} as one of the parameters of the log-normal distribution.
In this example, \texttt{conc} is observed and is called the dependent variable.
We define the distribution of \texttt{conc} to be a log-normal distribution with log-scale mean $\log \texttt{cp}$ and log-scale standard deviation $\sigma$ (from our \texttt{@param} block).

\begin{lstlisting}[caption=\texttt{@derived} block example,label=lst:derived]
@derived begin
    cp = @. Central / Vc
    conc {$\sim$} @. LogNormal(log(cp), {$\sigma$})
end
\end{lstlisting}

The \texttt{@observed} block can be used to compute additional quantities such as the NCA parameters from the concentration curve.

\begin{lstlisting}[caption=\texttt{@observed} block example,label=lst:observed]
@observed begin
    nca := @nca cp
    auc = NCA.auc(nca)
    cmax = NCA.cmax(nca)
end
\end{lstlisting}

In addition to the above blocks, there are also more blocks available for:
\begin{itemize}
    \item \texttt{@init} for initializing the dynamics manually, or
    \item \texttt{@vars} for defining short-hand notation variables for use in the \texttt{@dynamics} and \texttt{@derived} blocks.
\end{itemize}  
For more on the various components of a Pumas model, please refer to the Pumas documentation (\href{docs.pumas.ai}{https://docs.pumas.ai}).

\subsubsection{Example: PK Model}

For illustration purposes, consider the following 1-compartment model with first-order absorption:

\begin{align*}
    \text{Depot}^{\prime}   & = -\text{Ka} \cdot \text{Depot} \\
    \text{Central}^{\prime} & = \text{Ka} \cdot \text{Depot} -\frac{\text{CL}}{V_C} \cdot \text{Central} \\
\end{align*}
 where $\text{CL}$ is the elimination clearance from the $\text{Central}$ compartment; $V_C$ is the volume of the $\text{Central}$ compartment; and $\text{Ka}$ is absorption rate constant.

If we had one subject only, this model can be coded in Pumas using all the blocks we've seen except for the \texttt{@random} which is not necessary for single-subject models.
Listing \ref{lst:pkmodel} shows the code for this model.

\begin{figure*}[t]
\begin{lstlisting}[caption=PK 1-compartment single-subject model example,label=lst:pkmodel]
@model begin
    @param begin
        tvcl {$\sim$} LogNormal(log(2), 0.25)
        tvvc {$\sim$} LogNormal(log(70), 0.25)
        tvka {$\sim$} LogNormal(log(1.5), 1)
        {$\sigma$}    {$\sim$} Constrained(Cauchy(), lower = 0)
    end
    @pre begin
        CL = tvcl
        Vc = tvvc
        Ka = tvka
    end
    @dynamics begin
        Depot'   = -Ka * Depot
        Central' =  Ka * Depot - (CL/Vc) * Central 
    end
    @derived begin
        cp  := @. Central / Vc
        conc {$\sim$} @. LogNormal(log(cp), {$\sigma$})
    end
end
\end{lstlisting}
\end{figure*}

This is a complete Bayesian Pumas model.
We are specifying the parameters along with their priors in the \texttt{@param} block.
We only have a single subject, so there is no need for the inclusion of a \texttt{@random} block, which in turn makes the PK individual parameters defined in the \texttt{@pre} block being the same as the population parameters.
In the \texttt{@dynamics} block, we are declaring two ODEs that govern our model dynamics. The dynamics have 2 compartments named: \texttt{Depot} and \texttt{Central}.
Finally, in the \texttt{@derived} block, we calculate \texttt{cp} as a function of the \texttt{Central} compartment divided by the PK parameter \texttt{Vc} with a deterministic assignment, and we define our observed response \texttt{conc} as following a log-normal distribution with log-scale mean $\log \texttt{cp}$ and log-scale standard deviation $\sigma$.

The model in Listing \ref{lst:pkmodel} is a single-subject model, but most of the time we have multiple subjects in the data and need to define a population model.
This can be accomplished with the addition of a \texttt{@random} block to define the subject-specific parameters $\eta$. We can then use the population and subject-specific parameters together to define the individual PK parameters in the \texttt{@pre} block.

We chose to assign a Gaussian prior distribution to $\eta$s with a covariance matrix parameterized using correlations and standard deviations as explained in \ref{sec:corvscov}.
The model in Listing \ref{lst:poppkmodel} is an example of such parameterization.
It builds upon the previous single-subject PK model by adding 2 more population parameters: a correlation matrix $C$ and a vector of standard deviations $\omega$ in the \texttt{@param} block.
The $C$ correlation matrix has a Cholesky-parameterized LKJ prior (the recommended prior) and the $\omega$ vector of standard deviations has a positive-constrained multivariate-normal distribution with a diagonal covariance matrix with pre-truncation mean equal to 0 and variances equal to $0.4^2$.
We build the $\eta$s in the \texttt{@random} block by using a multivariate Gaussian prior with a covariance matrix built from the correlation and standard deviations using the Pumas' function \texttt{cor2cov}.
Finally, in the \texttt{@pre} block we defined the individual PK parameters as a transformation of the population and the subject-specific parameters combined.

\begin{figure*}[t]
\begin{lstlisting}[caption=Population PK 1-compartment first-order absorption model example,label=lst:poppkmodel]
@model begin
    @param begin
        tvcl {$\sim$} LogNormal(log(2), 0.25)
        tvvc {$\sim$} LogNormal(log(70), 0.25)
        tvka {$\sim$} LogNormal(log(1.5), 1)
        {$\sigma$}    {$\sim$} Constrained(Cauchy(); lower=0)
        C    {$\sim$} LKJCholesky(3, 1.0)
        {$\omega$}    {$\sim$} Constrained(
            MvNormal(
                zeros(3),
                Diagonal(0.4^2 * ones(3)),
            ),
            lower = zeros(3),
            upper = fill(Inf, 3),
            init=ones(3),
        )
    end
    @random begin
        {$\eta$} {$\sim$} MvLogNormal(Pumas.cor2cov(C, {$\omega$}))
    end
    @pre begin
        CL = tvcl * {$\eta$}[1]
        Vc = tvvc * {$\eta$}[2]
        Ka = tvka * {$\eta$}[3]
    end
    @dynamics begin
        Depot'   = -Ka * Depot
        Central' =  Ka * Depot - (CL/Vc) * Central 
    end
    @derived begin
        cp  := @. Central / Vc
        conc {$\sim$} @. LogNormal(log(cp), {$\sigma$})
    end
end
\end{lstlisting}
\end{figure*}

\subsubsection{Selecting Prior Distributions}

\begin{table*}
	\centering
	\begin{tabular}
		{|c|p{0.6\linewidth}|}
		\hline
		\textbf{Support}           & \textbf{Distributions}                                                                                                                                                                                                                                                                                                             \\
		\hline
		$(0, 1)$                   & \texttt{Beta}, \texttt{KSOneSided}, \texttt{NoncentralBeta}, \texttt{LogitNormal}                                                                                                                                                                                                                                                  \\
		\hline
		$(0, \infty)$              & \texttt{BetaPrime}, \texttt{Chi}, \texttt{Chisq}, \texttt{Erlang}, \texttt{Exponential}, \texttt{FDist}, \texttt{Frechet}, \texttt{Gamma}, \texttt{InverseGamma}, \texttt{InverseGaussian}, \texttt{Kolmogorov}, \texttt{LogNormal}, \texttt{NoncentralChisq}, \texttt{NoncentralF}, \texttt{Rayleigh}, \texttt{Weibull}           \\
		\hline
		$(-\infty, \infty)$        & \texttt{Cauchy}, \texttt{Gumbel}, \texttt{Laplace}, \texttt{Logistic}, \texttt{Normal}, \texttt{NormalCanon}, \texttt{NormalInverseGaussian}, \texttt{PGeneralizedGaussian}, \texttt{TDist}                                                                                                                                        \\
		\hline
		Real vectors               & \texttt{MvNormal}                                                                                                                                                                                                                                                                                                                  \\
		\hline
		Positive vectors           & \texttt{MvLogNormal}                                                                                                                                                                                                                                                                                                               \\
		\hline
		Positive definite matrices & \texttt{Wishart}, \texttt{InverseWishart}                                                                                                                                                                                                                                                                                          \\
		\hline
		Correlation matrices       & \texttt{LKJ}, \texttt{LKJCholesky}                                                                                                                                                                                                                                                                                                                       \\
		\hline
		Other                      & \texttt{Constrained}, \texttt{truncated}, \texttt{LocationScale}, \texttt{Uniform}, \texttt{Arcsine}, \texttt{Biweight}, \texttt{Cosine}, \texttt{Epanechnikov}, \texttt{Semicircle}, \texttt{SymTriangularDist}, \texttt{Triweight}, \texttt{Pareto}, \texttt{GeneralizedPareto}, \texttt{GeneralizedExtremeValue}, \texttt{Levy} \\
		\hline
	\end{tabular}
	\caption{The table shows some of the most popular prior distributions available in Pumas and their corresponding support domains.
		You can learn more about each distribution using ?
		followed by the distribution name in the Pumas command line prompt.
	} \label{tab:priors2}
\end{table*}

Choosing a new prior for a parameter can be a daunting task.
In general, there is no one prior that fits all cases and it might be a good practice to follow a previous similar study's priors where a good reference can be found.
However, if you are faced with the task of choosing good prior distributions for an all-new model, it will generally be a multi-step process consisting of: 

\begin{enumerate}
	\item \textbf{Deciding the support of the prior}.
	      The support of the prior distribution must match the domain of the parameter.
	      For example, different priors can be used for positive parameters than those for parameters between 0 and 1.
	      Table \ref{tab:priors2} can help narrow down the list of options available based on the domain of the parameter.
	\item \textbf{Deciding the center of the prior}, e.g. mean, median or mode.
	\item \textbf{Deciding the strength of the prior}.
	      This is often controlled by a standard deviation or scale parameter in the corresponding distribution function.
	      A small standard deviation or scale parameter implies low uncertainty in the parameter value which leads to a stronger prior.
	      A large standard deviation or scale parameter implies high uncertainty in the parameter value which leads to a weaker prior.
	      It is recommended that each prior distribution that is considered to be used should be assessed carefully before using it. This will ensure that the strength of the prior reflects your confidence level in the parameter values.
	      Refer to the discussion in Section \ref{sec:priors} on prior selection for more details.
	\item \textbf{Deciding the shape of the prior}.
	      Some distributions are left-skewed, others are right skewed and some are symmetric.
	      Some have heavier tails than others, e.g. the student's T-distribution is known for its heavier tail compared to a normal distribution.
	      The shape of the probability density function (PDF) should reflect knowledge about the parameter value prior to observing the data.
\end{enumerate}

When selecting new priors, besides the discussion in Section \ref{sec:priors}, you may also find the tips and recommendations in \href{github.com/stan-dev/stan/wiki/Prior-Choice-Recommendations}{https://github.com/stan-dev/stan/wiki/Prior-Choice-Recommendations} to be useful. For a more advanced discussion of various prior choices, also see \cite{advanced_priors}.

For univariate distributions, you can plot the distribution's probability density curve using the \texttt{PumasPlots.lines}\footnote{which is a part of the \texttt{PumasUtilities} package} function, e.g:
\begin{lstlisting}
	using PumasUtilities
	dist = Normal(0.0, 1.0)
	PumasPlots.lines(dist)
\end{lstlisting}
For multivariate and matrix-variate distributions, you can use the \texttt{rand}, \texttt{mean} and \texttt{var} functions to sample from the distribution and make sense of the values' distributions. For example:
\begin{lstlisting}
	dist = LKJ(3, 1.0)
	x = rand(dist, 100) # 100 samples
	mean(x) # mean
	var(x) # element-wise variance
\end{lstlisting}

The following is a description of some of the most popular prior distributions available in Pumas:
\begin{enumerate}
	\item \texttt{Normal($\mu$, $\sigma$)}: univariate normal distributions with support $(-\infty, \infty)$, mean $\mu$ and standard deviation $\sigma$.
	
	\item \texttt{LogNormal($\mu$, $\sigma$)}: univariate log normal distribution with support $(0, \infty)$ and a log-scale mean $\mu$ and log-scale standard deviation $\sigma$.
	
	\item \texttt{MvNormal($\mu$, $\Sigma$)}: multivariate normal distribution with mean vector $\mu$ and covariance matrix $\Sigma$. The matrix $\Sigma$ can also be a diagonal matrix, e.g. \texttt{Diagonal([1.0, 1.0])}. You can also pass $\Sigma$ alone as a matrix, e.g. \texttt{MvNormal($\Sigma$)}, and the means will be assumed to be 0.
	
	\item \texttt{MvLogNormal($\mu$, $\Sigma$)}: a multivariate log-normal distribution over positive vectors with log-scale mean vector $\mu$ and log-scale covariance matrix $\Sigma$ as defined in the \texttt{MvNormal} case above.

        \item \texttt{Cauchy($\mu$, $\sigma$)}: a univariate Cauchy distribution with support $(-\infty, \infty)$, location $\mu$, and scale $\sigma$.
	
        \item \texttt{Constrained(dist, lower = l, upper = u)}: a constrained prior distribution with a fixed support \texttt{(l, u)} and a fixed base distribution \texttt{dist} that could be any univariate or multivariate distribution.
	\texttt{lower} and \texttt{upper} set the lower and upper bounds on the random variables' support, respectively, defaulting to \texttt{-Inf} ($-\infty$) and \texttt{Inf} ($\infty$), respectively.
	When \texttt{dist} is a univariate distribution, \texttt{lower} and \texttt{upper} should be scalars.
    When constraining multivariate distributions, \texttt{lower} and \texttt{upper} can be vectors or scalars.
    If set to a scalar, the same bound will be used for all random variables.
    There is also a \texttt{truncated} distribution which is different from \texttt{Constrained} in that it allows the base distribution to be a function of the model's parameters but \texttt{truncated} only supports univariate base distributions.
    In general, it's recommended to use \texttt{Constrained} in the \texttt{@param} block and \texttt{truncated} in the \texttt{@random} and \texttt{@derived} blocks.
	Examples:
	  \begin{itemize}
	    \item \texttt{Constrained(Normal(0.0, 1.0), lower = 0.0)} is a half normal distribution.\\
		\item \texttt{Constrained(Cauchy(0.0, 1.0), lower = 0.0}) is a half Cauchy distribution.
		\item \texttt{Constrained(MvNormal([0.0, 0.0], [1.0 0.0; 0.0 1.0]), lower = 0.0)} is a constrained multivariate normal distribution.
	  \end{itemize}
	The \texttt{init} keyword argument can also be set to specify the initial value of the parameter, e.g. \texttt{Constrained(Normal(), lower = 0.0, init = 1.0)}
	
        \item \texttt{truncated(dist, lower, upper)}: similar to \texttt{Constrained} with fixed lower and upper bounds \texttt{lower} and \texttt{upper}, respectively, and a base distribution \texttt{dist}.
	Setting \texttt{upper} is optional and it defaults to \texttt{Inf} ($\infty$) when not set.
	In \texttt{truncated}, the base distribution \texttt{dist} is allowed to depend on the model's parameters and the normalization constant is computed in every log probability evaluation.
	However, the lower and upper bounds must be fixed constants and \texttt{truncated} only supports univariate base distribution.
	Examples: \texttt{truncated(Normal(0, 1), 0.0, Inf)} is a half normal distribution.
	\texttt{truncated(Cauchy(), 0.0, Inf)} is a half Cauchy distribution.
	\texttt{truncated(Normal(), -Inf, 0.0)} is a negative half normal distribution.
	
	\item \texttt{Uniform($l$, $u$)}: a univariate uniform distribution with lower and upper bounds $l$ and $u$ respectively.
	
        \item \texttt{LKJ($d$, $\eta$)}: a matrix-variate LKJ prior over correlation matrices of size $d \times d$. $\eta$ is the positive shape parameter of the LKJ prior. Decreasing $\eta$ results in samples with correlations closer to $\pm1$. There is also \texttt{LKJCholesky} which is semantically identical to \texttt{LKJ} but has some advantages. See below. 
        
        \item \texttt{LKJCholesky($d$, $\eta$)}: a Cholesky-factorized version of the LKJ distribution where the matrix sampled is in factorized form. This is recommended over \texttt{LKJ} for use inside the model for performance reasons.
	
        \item \texttt{Wishart($\nu$, $S$)}: a matrix-variate Wishart distribution over $d \times d$ positive definite matrices with $\nu$ degrees of freedom and a positive definite $S$ scale matrix.
	
        \item \texttt{InverseWishart($\nu$, $\Psi$)}: a matrix-variate inverse Wishart distribution over $d \times d$ positive definite matrices with $\nu$ degrees of freedom and a positive definite scale matrix $\Psi$.
	
        \item \texttt{Beta($\alpha$, $\beta$)}: a univariate Beta distribution with support from 0 to 1 and shape parameters $\alpha$ and $\beta$.
	
        \item \texttt{Gamma($\alpha$, $\theta$)}: a univariate Gamma distribution over positive numbers with shape parameter $\alpha$ and scale $\theta$.
	
        \item \texttt{Logistic($\mu$, $\theta$)}: a univariate logistic distribution with support $(-\infty, \infty)$, location $\mu$ and scale $\theta$.
	
        \item \texttt{LogitNormal($\mu$, $\sigma$)}: a univariate logit normal distribution with support $(0, 1)$ and a base normal distribution with mean $\mu$ and standard deviation $\sigma$.
	
        \item \texttt{TDist($\nu$)}: a univariate Student's T distribution with support $(-\infty, \infty)$, $\nu$ degrees of freedom and mean 0. To change the mean of the T distribution, you can use a \texttt{LocationScale} distribution (shown below).
	
        \item \texttt{LocationScale($\mu$, $\sigma$, $d$)}: a scaled and translated univariate distribution with a base distribution $d$. The base distribution's random variable is first scaled by $\sigma$ and then translated by $\mu$. Example: \texttt{LocationScale(1.0, 2.0, TDist(2))} is a scaled and translated Student's $t$ distribution. The mean of the \texttt{LocationScale} distribution is $\mu + \sigma \times \text{mean(d)}$ and the standard deviation is $\sigma \times \text{std(d)}$.
	
        \item \texttt{Laplace($\mu$, $\sigma$)}: a univariate Laplace distribution with support $(-\infty, \infty)$, location $\mu$ and scale $\sigma$.
	
        \item \texttt{Exponential($\theta$)}: a univariate exponential distribution with support $(0, \infty)$ and scale $\theta$.
	
        \item (Improper) flat priors: instead of using a distribution, one can specify a domain instead such as a \texttt{VectorDomain} for vector parameters, \texttt{PSDDomain} for positive definite parameters or \texttt{CorrDomain} for correlation matrix parameters. Those domains are treated in Pumas as flat priors. If the domain is open, this would be an improper prior. For more about domains, see the Pumas documentation (\href{docs.pumas.ai}{https://docs.pumas.ai}).
\end{enumerate}

\subsection{Prior Simulations and Predictions} \label{sec:prior_sims_and_preds}

After defining a model with the priors, a prior predictive check may be run to check how close the prior predictions or simulations are to the real data. Prior simulations can be run using the \texttt{simobs} function passing in the model, \texttt{model}, and subject/population, \texttt{data}, as arguments.
\begin{lstlisting}[caption=Prior simulation,label=lst:prior_sims]
sims = simobs(model, data; samples = 100)
\end{lstlisting}
The \texttt{samples} keyword argument specifies the number of samples to simulate from the prior distributions for each subject in \texttt{data}.

In the \texttt{simobs} function, there's also a keyword argument: \texttt{simulate\_error}. If it is set to \texttt{true} (the default value), Pumas will sample from the response's error model in the \texttt{@derived} block (aka simulation), otherwise it will return the expected value of the error distribution (aka prediction).
The latter is equivalent to using the \texttt{predict} function.

There are various statistics and queries that can be run given the simulation results. These are explained in Section \ref{sec:sim_queries}. The simulation results can also be plotted using a visual predictive check (VPC) as explained in Section \ref{sec:vpc}. When the simulations are from the prior model, the VPC is usually called a prior predictive check.

\subsection{Fitting a Model}
\label{sec:fit}

Now that we have a model, we need to fit it using Pumas.
This is the role of the \texttt{fit}  function which takes four positional arguments:
\begin{enumerate}
    \item A Pumas model
    \item A Pumas population
    \item The initial parameter estimates
    \item A fitting algorithm
\end{enumerate}

The fitting algorithm in an MLE setting can be an instance of \texttt{FOCE} or \texttt{Laplace} for example. However to run Bayesian inference using MCMC instead, it can be set to an instance of either:
\begin{itemize}
    \item \texttt{BayesMCMC}
    \item \texttt{MarginalMCMC}
\end{itemize}

\texttt{MarginalMCMC} samples from the marginal posterior by integrating out the subject-specific parameter first, whereas \texttt{BayesMCMC} samples from the joint posterior.
\texttt{MarginalMCMC} can be much faster than \texttt{BayesMCMC} in some cases but it is still experimental and will be improved in the future.
The options in \texttt{BayesMCMC} and \texttt{MarginalMCMC} are passed when constructing an instance of the algorithm using keyword arguments, e.g \texttt{BayesMCMC(nsamples = 2000)}.
The main options that can be set in both \texttt{BayesMCMC} and \texttt{MarginalMCMC} are:
\begin{itemize}
	\item \texttt{target\_accept}: target acceptance ratio for the NUTS algorithm, defaults to 0.8
	
        \item \texttt{nsamples}: number of Markov Chain Monte Carlo (MCMC) samples to generate, defaults to 2000
	
        \item \texttt{nadapts}: number of adaptation steps in the NUTS algorithm, defaults to 1000
	
        \item \texttt{nchains}: number of MCMC chains to sample, defaults to 4
	
        \item \texttt{ess\_per\_chain}: target effective sample size (ESS) per chain, sampling terminates if the target is reached, defaults to \texttt{nsamples}
	
        \item \texttt{check\_every}: the number of samples after which the ESS per chain is checked
	
        \item \texttt{time\_limit}: a time limit for sampling in seconds, sampling terminates if the time limit is reached, defaults to \texttt{Inf} (which is $\infty$ in Julia)
	
        \item \texttt{ensemblealg}: can be set to \texttt{EnsembleSerial()} for serial sampling, \texttt{EnsembleThreads()} for multi-threaded sampling or \texttt{EnsembleDistributed()} for multi-processing (aka distributed parallelism) sampling. By default parallelism over both chains and subjects will be turned on if enough threads/processes are available.
	
        \item \texttt{parallel\_chains}: can be set to \texttt{true} or \texttt{false}. If set to \texttt{false}, the chains will not be sampled in parallel. If set to \texttt{true}, the chains will be sampled in parallel using either multi-threading or multi-processing depending on the value of \texttt{ensemblealg}. The default value is \texttt{true}.
	
        \item \texttt{parallel\_subjects}: can be set to \texttt{true} or \texttt{false}. If set to \texttt{false}, the log probability computation will not be paralellized. This is preferred when the number of subjects is small. If set to \texttt{true}, the log probability computation will be parallelized over the subjects using either multi-threading or multi-processing depending on the value of \texttt{ensemblealg}. The default value is \texttt{true} if enough threads/processes are available to do both parallelism over chains and subjects.
	
        \item \texttt{rng}: the random number generator used
	
        \item \texttt{diffeq\_options}: a NamedTuple of all the differential equations solver's options, e.g \texttt{diffeq\_options = (alg = Rodas5(),)} can be used to force Pumas to use the stiff ODE solver \texttt{Rodas5} instead of relying on the automatic stiffness detection and auto-switching behaviour of Pumas.
	
        \item \texttt{constantcoef}: a \texttt{NamedTuple} of the parameters to be fixed during sampling. This can be used to sample from conditional posteriors fixing some parameters to specific values, e.g. \texttt{constantcoef = ($\sigma$ = 0.1,)} fixes the \texttt{$\sigma$} parameter to \texttt{0.1} and samples from the posterior of the remaining parameters conditional on \texttt{$\sigma$}.
\end{itemize}

The \texttt{MarginalMCMC} algorithm also has a keyword argument \texttt{marginal\_alg} which defaults to \texttt{LaplaceI()} but can also be \texttt{FOCE()} or \texttt{FO()}. By default both \texttt{BayesMCMC} and \texttt{MarginalMCMC} will run 4 Markov chains in parallel, using the remaining computing resources to parallelize the computations across subjects. By default, 2,000 MCMC iterations will be run using the first 1,000 samples of each chain as burn-in.
Pumas does not automatically discard the burn-in samples.
Hence, the user needs to use the function \texttt{discard} to discard the burn-in samples. If you are using Pumas < 2.4, instead of \texttt{discard}, you can use \texttt{Pumas.truncate}.
Listing \ref{lst:fitbayesMCMC} shows a Bayesian Pumas model fitting example.
We save the result to \texttt{res} and we call \texttt{discard} on it with the keyword argument \texttt{burnin} set to 1,000 samples.
This will output a truncated fit by discarding the first 1,000 samples per chain.
Note that in Julia, 1\_000 and 1000 are equivalent.

\begin{lstlisting}[caption=Fitting a Bayesian model in Pumas,label=lst:fitbayesMCMC]
res = fit(model, pop, iparams, BayesMCMC(nsamples = 2_000, nadapts = 1_000))
tres = discard(res; burnin=1_000)
\end{lstlisting}

You can also pass a \texttt{ratio} keyword argument to the \texttt{discard} function to drop \texttt{(1 - ratio)} $\times 100\%$ of the samples. This is known as thinning and it works by selecting 1 sample from every $1 / \text{ratio}$ samples in each chain.
Generally speaking, thinning is usually discouraged in the final analysis because it leads to some loss of information. However, in the initial exploratory phase when many exploratory simulations/predictions are run, it may be desired to do thinning to do faster iterations. Another example is in Listing \ref{lst:fitmarginalMCMC} where we are using \texttt{MarginalMCMC}.

\begin{lstlisting}[caption=Fitting a Bayesian model in Pumas with custom arguments,label=lst:fitmarginalMCMC]
res = fit(model, pop, iparams, MarginalMCMC(nsamples=100, nadapts=10))
tres = discard(res; burnin=10)
\end{lstlisting}

\begin{figure*}[h]
    \centering
    \includegraphics[width=0.8\textwidth]{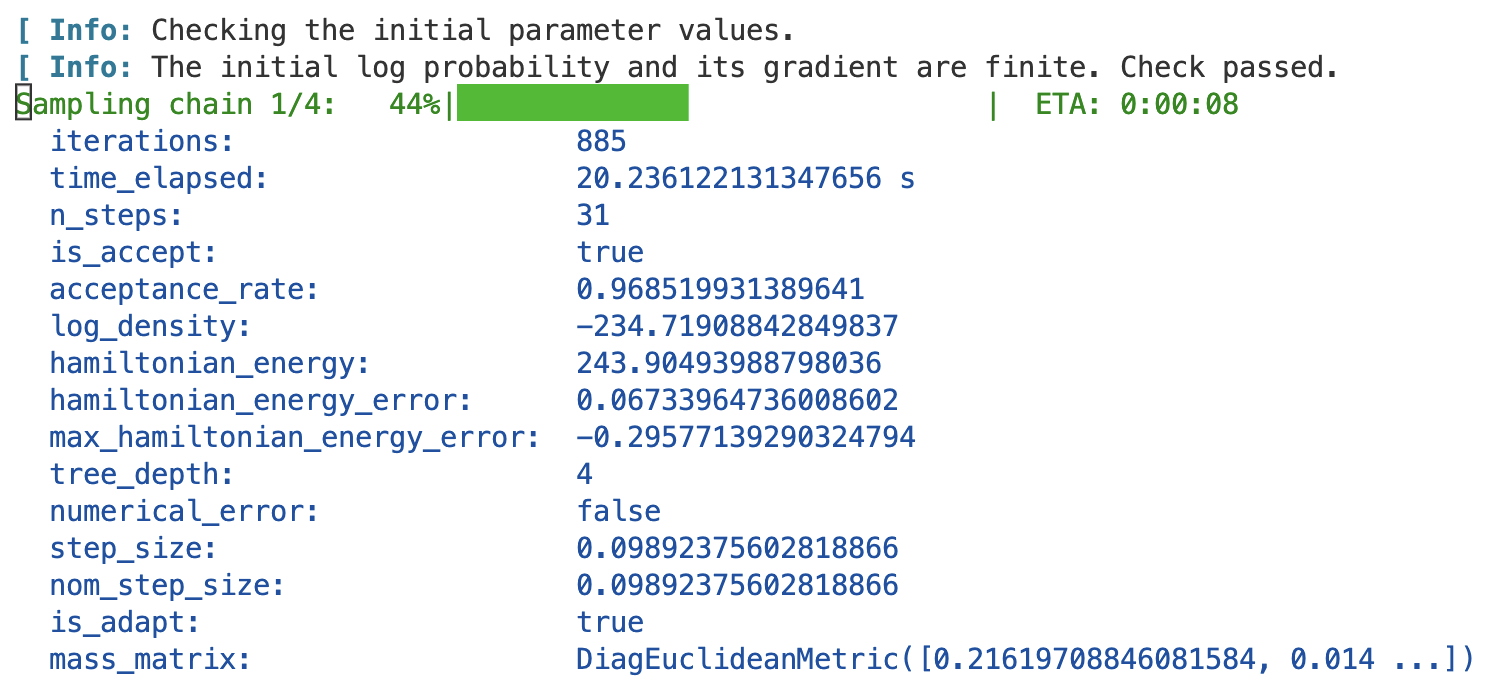}
    \vspace{5pt}
    \caption{Live progress information displayed during sampling using Pumas.}
    \label{fig:live_progress}
\end{figure*}

\begin{figure*}[h]
    \centering
    \includegraphics[width=0.8\textwidth]{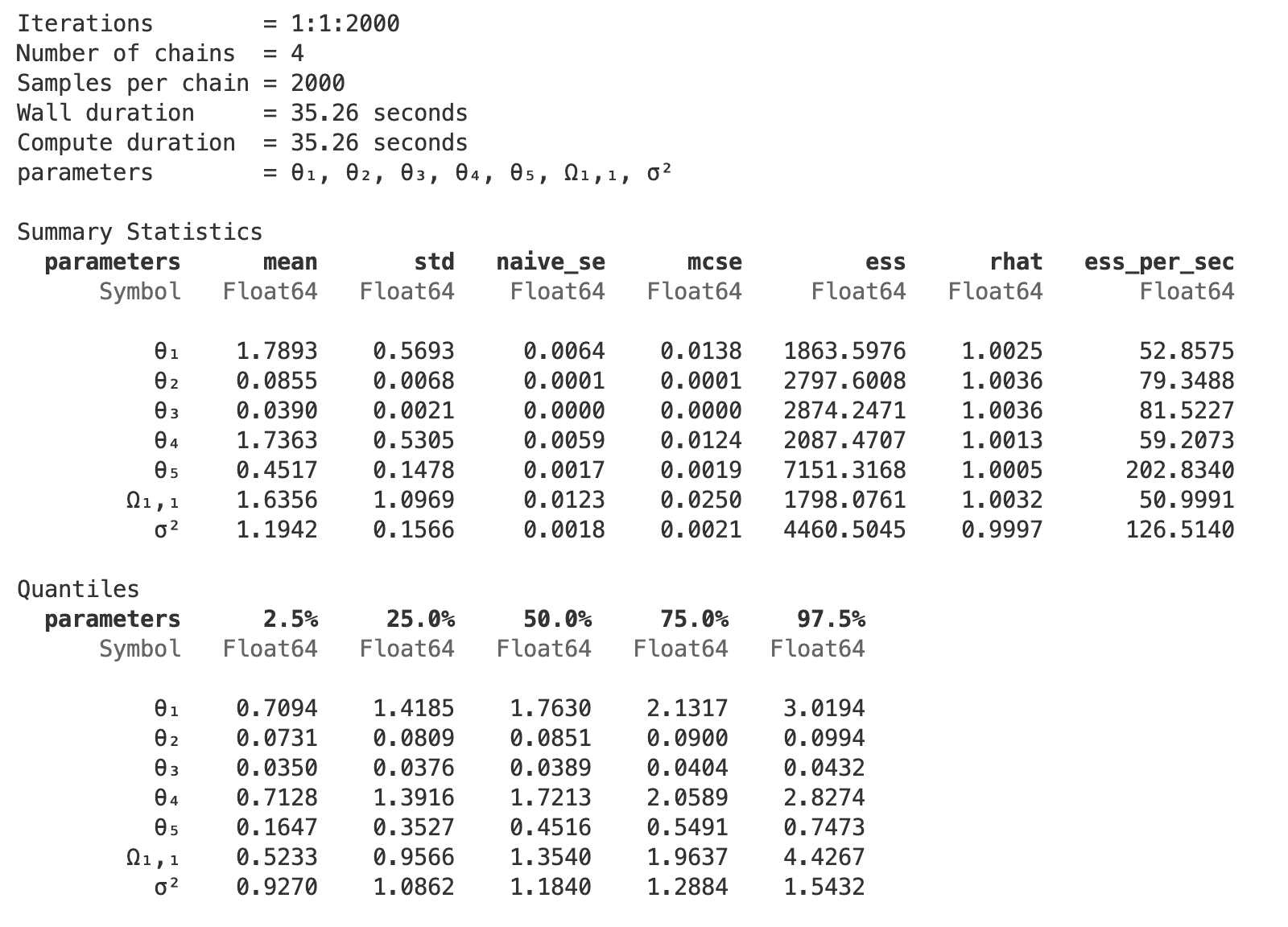}
    \vspace{5pt}
    \caption{An example of the display of the Bayesian fit result, output from the \texttt{fit} function in Pumas.}
    \label{fig:fit_result}
\end{figure*}

When fitting the model using \texttt{BayesMCMC} or \texttt{MarginalMCMC}, you will be able to view the progress of the sampler using live progress information displayed as shown in Figure \ref{fig:live_progress}. When using multi-threading or distributed parallelism, an interval or ratio is displayed for each field instead of a value. The following is a description of the most important fields displayed in Figure \ref{fig:live_progress}:
\begin{itemize}
    \item \texttt{iterations} refers to how many MCMC iterations are completed including both the adaptation and sampling phases.
    \item \texttt{n\_steps} is the number of time steps taken in the last proposal. If this is too large, the NUTS is sampler will be very slow and inefficient. For more on the number of steps, see Section \ref{sec:uturns}.
    \item \texttt{is\_accept} is true if the last proposal was accepted and false otherwise. For more on proposals and the NUTS algorithm, see Section \ref{sec:nuts}.
    \item \texttt{acceptance\_rate} refers to the average acceptance rate of all the past proposals. This should converge to a similar value as the \texttt{target\_accept} option set after the adaptation phase. For more on the acceptance ratio, see Section \ref{sec:exploration}.
    \item \texttt{log\_density} refers to the log joint probability of the parameters and observations. If this is monotonically increasing late during the sampling phase of the fit, this is a sign that the sampler likely didn't converge to the area(s) of high posterior probability mass during adaptation and the chains likely would not converge. For more on signs of lack of convergence, see Section \ref{sec:convergence} and for more on monotonically increasing log densities (aka optimization behaviour), see Section \ref{sec:exploration}.
    \item \texttt{tree\_depth} is the maximum tree depth reached when generating the last proposal. For more on this, see Section \ref{sec:uturns}.
    \item \texttt{step\_size} is the time step size in the NUTS algorithm which is adapted during the adaptation phase and fixed during the sampling phase. For more on the step size and its connection to the target acceptance ratio, see Section \ref{sec:exploration}.
    \item \texttt{is\_adapt} is true during the adaptation phase and false during the sampling phase.
\end{itemize}

An example of the result of the \texttt{fit} function is shown in Figure \ref{fig:fit_result}. A number of summary statistics for the population parameters are displayed automatically. You can also use other Pumas functions to query specific summary statistics programmatically, rather than only in display. For more on summary statistics functions, see Sections \ref{sec:basic_summary} and \ref{sec:posterior_queries}.

\subsection{Numerical Errors and Debugging}
\label{sec:simulation}

\subsubsection{Numerical Instability in the Model}

Each evaluation of the log-likelihood at specific parameters values $(\eta, \theta)$ involves a full evaluation of the model, including the structural model (\verb+@pre+ block), numerically solving the differential equations (\verb+@dynamics+ block), and the computing of the likelihood (\verb+@derived+ block). In order to perform effective Bayesian inference, one needs to ensure that all the model blocks are numerically stable and do not lead to \texttt{Inf}, \texttt{-Inf} or \texttt{NaN} values. Numerical instability can result from many causes but some usual suspects include:
\begin{enumerate}
    \item Dividing by a very small number or 0, e.g. if a parameter is in the denominator and is allowed to be 0 during the fitting.
    \item Calling the exponential function with a large exponent, e.g. due to bad initial parameter values.
    \item Some observations may have 0 or approximately 0 probability according to your model at a specific $(\eta, \theta)$. For example, if a Bernoulli response distribution was used and the probability parameter of the Bernoulli distribution was exactly 0 (or 1), when a 1 (or 0) observation exists in the data.
    \item The ODE solver is failing to converge to a solution  because the dynamics are not stable for a particular choice of extreme $(\eta, \theta)$.
    \item The response's distribution has 0 standard deviation, e.g. when using a proportional error model and the concentration drops to 0.
    \item Taking log or square root of a negative parameter.
\end{enumerate}

Initial parameter values that cause numerical errors are often more important to watch out for than intermediate bad values during the fitting. This is because bad intermediate models will be rejected automatically when they lead to numerical errors. When this happens, one may see the following warning occur:

\begin{lstlisting}
Warning: The current proposal will be rejected due to numerical error(s).
   isfinite.(({$\theta$}, r, l{$\pi$}, l{$\kappa$})) = (true, false, false, false)
\end{lstlisting}

This warning is not necessarily a bad thing or a failure of the estimation process. What it represents is that the numerical estimation of $p(\eta, \theta | D)$ has failed for a particular run. In many cases, this warning is benign and expected from the Bayesian estimation process, e.g. it could be that extreme parameters of an ODE model led to unstable dynamics causing the simulator to diverge. The MCMC stepping process will recover from this by rejecting the step and proposing new $(\eta, \theta)$ values to try. Thus if one only sees a few of these warnings during the estimation process, there may be nothing to worry about.

However, if excessive warnings are displayed, then this could mean that many steps are being rejected causing the MCMC process to not effectively explore the posterior, potentially leading to a lack of convergence. One further indication of this is if the stepping process also displays the additional warning:

\begin{lstlisting}
NaN dt detected. Likely a NaN value in the state, parameters, or derivative value caused this outcome.
\end{lstlisting}

This warning implies that the ODE solver failed. This warning is shown because the calculation of the initial ODE time step $dt$, which is the first part of the ODE solver process, resulted in NaN. This is usually caused by NaN or Inf values in the ODE parameters. If this is the case, it may be a good idea to investigate whether the individual parameters in the \verb|@pre| block have reasonable values or not. One quick way to do this is to instrument the \verb|@model| definition to print out the current values that are being used in the simulation process. For example, the line

\begin{lstlisting}
VC = {$\theta$}[3] * exp({$\eta$}[2]
\end{lstlisting}

can be changed to

\begin{lstlisting}
VC = @pumasdebug {$\theta$}[3] * exp({$\eta$}[2]
\end{lstlisting}

which will print out the value that is calculated at every step. Using these printouts, one can directly see the values of the ODE parameters being used in the \verb|@dynamics| block. \texttt{@pumasdebug} is a Pumas 2.4 feature which is not available prior to that.

Some of the most common issues found through this form of debugging are due to incorrectly setting parameter bounds. For example, a parameter defined by dividing by a fixed or random effect may cause a parameter value to be (nearly) infinite if the denominator is close to zero. Thus a fix is to ensure that a lower bound is appropriately defined for the effect causing the (near) 0 denominator.

\subsubsection{Numerical Instability in the ODE Solver}

If the ODE parameters seem to be realistic candidates that are being rejected, then one may need to ensure that the ODE solver process is appropriate for the equation at hand. Any of the following warnings is usually a sign that the ODE solver is failing for algorithmic reasons rather than NaN or Inf values in the parameters:

\begin{lstlisting}
Warning: Interrupted. Larger maxiters is needed.
\end{lstlisting}

\begin{lstlisting}
Warning: dt(x) <= dtmin(y) at t=z. Aborting.
\end{lstlisting}

\begin{lstlisting}
Warning: Instability detected. Aborting
\end{lstlisting}

For debugging such situations, it can be helpful to recreate the ODE solve with the exact parameters generated by \verb|@pre| and directly call the ODE solver functions from DifferentialEquations.jl (\cite{DifferentialEquations.jl-2017}). However, some common guidance is:

\begin{itemize}
    \item Reduce the tolerances. Generally a very robust set of tolerances (at the cost of some performance) is \verb|abstol=1e-12, reltol=1e-12|. This can be set as part of the \texttt{diffeq\_options} keyword argument in the sampling algorithm, e.g. \texttt{BayesMCMC(diffeq\_options = (abstol=1e-12, reltol=1e-12,))}.
    \item Change the ODE solver to a method specifically designed for stiff differential equations. A common choice is \verb|Rodas5P()|. Once again, this can be set using the \texttt{diffeq\_options} keyword argument, e.g. \texttt{BayesMCMC(diffeq\_options = (alg = Rodas5P(),))}
\end{itemize}

One common reason for numerical failures of ODE solvers is due to a property known as stiffness in the ODE. Stiffness is difficult to rigorously define but can loosely be defined as large time-scale differences in the rates of change in an ODE. For example, if one value has a derivative of $1$ while the other has a derivative of $10^9$. This can lead to difficulties in the ODE solver and by consequence in the simulation and Bayesian inference process. Pumas, by default, uses an ODE solver which switches between a less robust but faster method for non-stiff ODEs, and a more robust but slower method for stiff ODEs. However, this default behaviour at times can be less stable than requiring all steps to use a stiff ODE solver, hence the second recommendation to manually switch the ODE solver.

\subsection{Updating the Posterior with New Data}

There are algorithms that can efficiently update the posterior samples given new observations per subject or new subjects, such as sequential Monte Carlo. As of the time of this writing, Pumas does not implement these algorithms. To update the posterior given new data for existing subjects or new subjects, you would currently have to refit the model to the entire dataset. Alternatively, you can approximate the posterior samples using a tractable distribution family, e.g. a multivariate Gaussian distribution, and refit the model to the new data only using the posterior approximation as the prior distribution. In future releases of Pumas, we intend to implement such efficient methods for updating the posterior samples. Please refer to the Pumas documentation (\href{docs.pumas.ai}{https://docs.pumas.ai}) for a list of the latest features.

\subsection{Basic Summary Statistics}
\label{sec:basic_summary}

To query a number of basic summary statistics for the population parameters, you can use:
\begin{lstlisting}
summarystats(tres)
\end{lstlisting}
where \texttt{tres} is the result from \texttt{fit} or \texttt{discard}. This will output the sample mean, sample standard deviation, Monte Carlo standard error (MCSE), effective sample size (ESS), $\hat{R}$ and ESS per second.

To get the same summary statistics for the subject-specific parameters of the $i^{th}$ subject, you can use the \texttt{subject} keyword argument:
\begin{lstlisting}
summarystats(tres, subject = i)
\end{lstlisting}

\subsection{How Many Samples are Needed?} \label{sec:nsamples}

The number of samples needed to accurately estimate various quantities can be different.
For instance, to be able to estimate the probability of rare events or some extreme quantiles, you will need many more samples than the number of samples needed to estimate the mean of the posterior.
By default, Pumas will generate 4 chains with 2000 samples per chain, 1000 of which will be used for adaptation. Depending on what you are trying to estimate, you may need to run the sampler for longer and check that the result does not significantly change as you increase the number of samples in your chains.

More concretely, an ESS of 400 was recommended as a good target in \cite{ess_mcse}. With the default 4 chains Pumas runs, this is an ESS of 100 per chain. The same ESS recommendations were also reported in old and recent editions of \cite{gelman2013bayesian}. While this is just a general guideline and it doesn't apply to extreme quantile estimation or estimating probabilities of rare events, it can be a good initial target to aim for. The \texttt{ess\_per\_chain} option discussed in Section \ref{sec:fit} can be used to set a target ESS per chain.

Beside a target ESS, one should also ensure that the $\hat{R}$ diagnostic is less than 1.1. It is even better if it were less than 1.01 as recommended in \cite{ess_mcse}. Estimating the ESS, $\hat{R}$ and MCSE for the purpose of estimating different quantities other than the posterior mean, e.g. parameter quantiles, is currently not implemented in Pumas but it is an upcoming feature. This can give users more insights into the estimation accuracy of their MCMC samples.

\subsection{Diagnostic Plots}

There are several diagnostic plots that help you identify lack of convergence including: trace plot, cumulative mean plot, and auto-correlation plot. All of the diagnostic plots require the loading of the \texttt{PumasUtilities} package first:
\begin{lstlisting}
using PumasUtilities
\end{lstlisting}
Assume \texttt{tres} is the output from \texttt{fit} or \texttt{discard}.

\subsubsection{Trace Plot}

The trace plot of a parameter shows the value of the parameter in each iteration of the MCMC algorithm.
A good trace plot is one that:
\begin{itemize}
    \item is noisy, not an increasing or decreasing line.
    \item has a fixed mean.
    \item has a fixed variance.
    \item shows all chains overlapping with each other, also known as chain mixing\footnote{Chain mixing refers to the case when different chains include samples from the same regions in the posterior as opposed to each chain including samples from a separate region of the posterior.}.
\end{itemize}

\begin{figure}[h]
    \centering
    \includegraphics[width=0.4\textwidth]{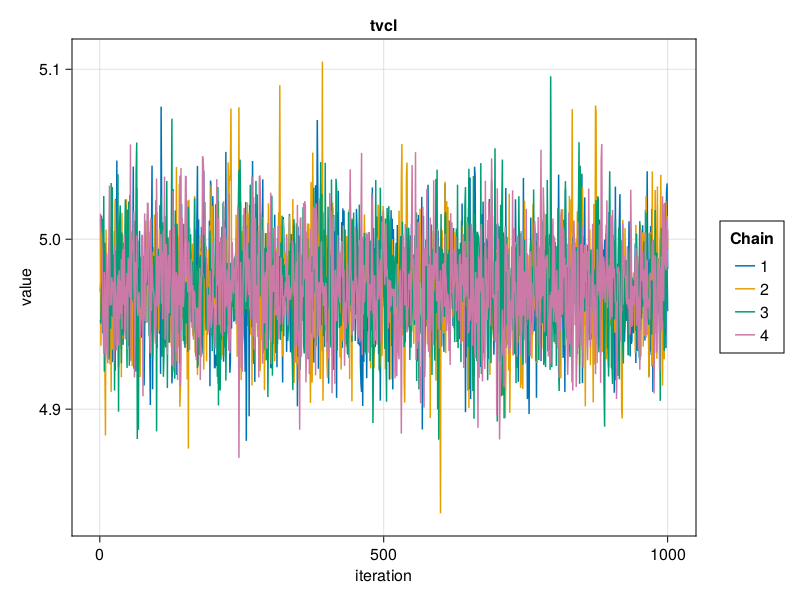}
    \caption{Example of a trace plot.}
    \label{fig:trace}
\end{figure}

You can plot trace plots with the function \texttt{trace\_plot}, e.g:

\begin{lstlisting}
trace_plot(tres; parameters = [:tvcl])
\end{lstlisting}

Figure \ref{fig:trace} shows the resulting trace plot for the parameter \texttt{tvcl}. You can add more parameter names to the \texttt{parameters} keyword argument, e.g. \texttt{parameters = [:tvcl, :tvvc]} to plot more parameters.
As you can see the trace plot shown has many of the desired properties of a good trace plot.

When the \texttt{parameters} keyword argument is not specified, all the population parameters' trace plots will be displayed. To plot the trace plot of the subject-specific parameters of a group of subjects, you can set the \texttt{subjects} keyword argument instead of setting the \texttt{parameters} keyword argument, e.g:
\begin{lstlisting}
trace_plot(tres; subjects = [1, 2])
\end{lstlisting}

See the Pumas documentation (\href{docs.pumas.ai}{https://docs.pumas.ai}) for more details and examples.

\subsubsection{Cumulative Mean Plot}

The cumulative mean plot of a parameter shows the mean of the parameter value in each MCMC chain up to a certain iteration.
An MCMC chain converging to a stationary posterior distribution should have the cumulative mean of each parameter converge to a fixed value.
Furthermore, all the chains should be converging to the same mean for a given parameter, the posterior mean.
If the cumulative mean curve is not converging or the chains are converging to different means, this is a sign of non-convergence.

\begin{figure}[h]
    \centering
    \includegraphics[width=0.4\textwidth]{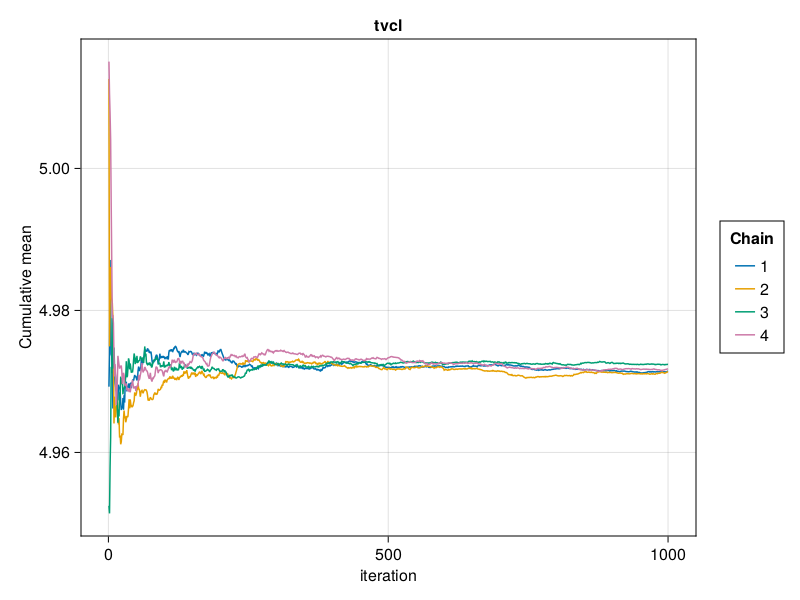}
    \caption{Example of a cumulative mean plot.}
    \label{fig:cummean}
\end{figure}

You can plot a cumulative mean plot for the population parameter \texttt{tvcl} using:
\begin{lstlisting}
cummean_plot(tres; parameters = [:tvcl])
\end{lstlisting}

Figure \ref{fig:cummean} shows the resulting trace plot for the parameter \texttt{tvcl}. Much like in the trace plot, you can add more parameter names to the \texttt{parameters} keyword argument or leave it out completely which will plot all the population-level parameters. Similarly, the same plot can be plotted for the subject-specific parameters using the \texttt{subjects} keyword argument instead of the \texttt{parameters} keyword argument.

\subsubsection{Auto-correlation Plot}

MCMC chains are prone to auto-correlation between the samples because each sample in the chain is a noisy function of the previous sample. The auto-correlation plot shows the correlation between every sample with index $s$ and the corresponding sample with index \texttt{s + lag} for all \texttt{s $\in$ 1:N-lag} where \texttt{N} is the total number of samples. For each value of \texttt{lag}, we can compute a correlation measure between the samples and their lag-steps-ahead counterparts. The correlation is usually a value between 0 and 1 but can sometimes be between -1 and 0 as well. The auto-correlation plot shows the \texttt{lag} on the x-axis and the correlation value on the y-axis. For well behaving MCMC chains when \texttt{lag} increases, the corresponding correlation gets closer to 0. This means that there is less and less correlation between any 2 samples further away from each other. The value of \texttt{lag} where the correlation becomes close to 0 can be used to guide the thinning of the MCMC samples to extract mostly independent samples from the auto-correlated samples. The \texttt{discard} function can be used to perform thinning with the ratio keyword set to \texttt{1 / lag} for an appropriate value of \texttt{lag}.
\begin{lstlisting}
discard(tres; ratio = 1/lag)
\end{lstlisting}
That said, generally speaking, thinning is usually discouraged in the final analysis because it leads to some loss of information. However, in the initial exploratory phase when many exploratory simulations/predictions are run, it may be desired to do thinning to do faster iterations.

\begin{figure}[h]
    \centering
    \includegraphics[width=0.4\textwidth]{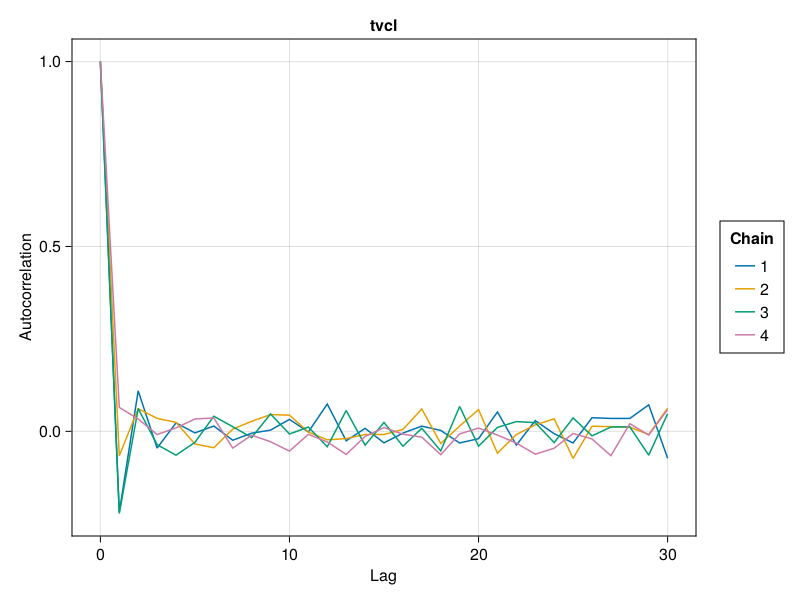}
    \caption{Example of an auto-correlation plot.}
    \label{fig:autocor}
\end{figure}

You can plot an auto-correlation plot for the population parameter \texttt{tvcl} using:
\begin{lstlisting}
autocor_plot(pk_1cmp_tfit; parameters = [:tvcl])
\end{lstlisting}

Figure \ref{fig:autocor} shows the resulting auto-correlation plot for the parameter \texttt{tvcl}. Much like in the trace plot, you can add more parameter names to the \texttt{parameters} keyword argument or leave it out completely which will plot all the population-level parameters. Similarly, the same plot can be plotted for the subject-specific parameters using the \texttt{subjects} keyword argument instead of the \texttt{parameters} keyword argument.

\subsection{More Diagnostics} \label{sec:more_diagnostics}

A number of other diagnostics exist to help you identify:
\begin{itemize}
    \item When the MCMC algorithm hasn't converged, or
    \item How many samples to throw away as burn-in.
\end{itemize}
In general, we recommend running these diagnostics after removing the adaptation steps using the \texttt{discard} function. Some of the diagnostics we present here can then tell you how many more samples to remove as burn-in after removing the adaptation steps. The \texttt{discard} function can be used again on its own output to remove the additional samples as burn-in.

\subsubsection{Geweke Diagnostic}

\begin{lstlisting}
gewekediag(tres; subject = nothing, first = 0.1, last = 0.5)
\end{lstlisting}

The above function computes the Geweke diagnostic \citep{Geweke1991} for each chain outputting a p-value per parameter. \texttt{tres} is the output from \texttt{fit} or \texttt{discard} and the remaining keyword arguments have the default values shown above. If the \texttt{subject} keyword argument is set to \texttt{nothing} (the default value) or left out, the chains diagnosed are those of the population parameters. If \texttt{subject} is set to an integer index, the chains diagnosed are those of the subject-specific parameters corresponding to the subject with the input index.

The Geweke diagnostic compares the sample means of two disjoint sub-chains $X_1$ and $X_2$ of the entire chain using a normal difference of means hypothesis test where the null and alternative hypotheses are defined as:
\begin{align*}
H_0: \mu_1 = \mu_2 \\
H_1: \mu_1 \neq \mu_2
\end{align*}
where $\mu_1$ and $\mu_2$ are the population means. The first sub-chain $X_1$ is taken as the first \texttt{(first * 100)\%} of the samples in the chain, where \texttt{first} is a keyword argument defaulting to 0.1. The second sub-chain $X_2$ is taken as the last \texttt{(last * 100)\%} of the samples in the chain, where \texttt{last} is a keyword argument defaulting to 0.5.

The test statistic used is: 
\begin{align*}
z_0 = (\bar{x}_1 - \bar{x}_2) \Big/ \sqrt{s_1^2 + s_2^2}
\end{align*}
where $\bar{x}_1$ and $\bar{x}_2$ are the sample means of $X_1$ and $X_2$ respectively, and $s_1$ and $s_2$ are the Markov Chain standard error (MCSE) estimates of $X_1$ and $X_2$ respectively. Auto-correlation is assumed within the samples of each individual sub-chain, but the samples in $X_1$ are assumed to be independent of the samples in $X_2$. The p-value output is an estimate of $P(|z| > |z_0|)$, where $z$ is a standard normally distributed random variable.

Low p-values indicate one of the following:
\begin{itemize}
    \item The first and last parts of the chain are sampled from distributions with different means, i.e. non-convergence,
    \item The need to discard some initial samples as burn-in, or
    \item The need to run the sampling for longer due to lack of samples or high auto-correlation.
\end{itemize}

High p-values (desirable) indicate the inability to conclude that the means of the first and last parts of the chain are different with statistical significance. However, this alone does not guarantee convergence to a fixed posterior distribution because:
\begin{itemize}
    \item Either the standard deviations or higher moments of $X_1$ and $X_2$ may be different, or
    \item The independence assumption between $X_1$ and $X_2$ may not be satisfied when high auto-correlation exists.
\end{itemize}

\subsubsection{Heidelberger and Welch diagnostic}

\begin{lstlisting}
heideldiag(tres; subject = nothing, alpha = 0.05, eps = 0.1, start = 1)
\end{lstlisting}

The above function computes the Heidelberger and Welch diagnostic \citep{Heidelberger1983} for each chain. If the \texttt{subject} keyword argument is set to \texttt{nothing} (default value) or left out, the chains diagnosed will be those of the population parameters. If \texttt{subject} is set to an integer index, the chains diagnosed will be those of the subject-specific parameters corresponding to the subject with the input index. The output of this function is a dataframe whose columns are explained below. Intuitively, the Heidelberger diagnostic attempts to:
\begin{itemize}
    \item Identify a cutoff point for the initial transient phase for each parameter, after which the samples can be assumed to come from a steady-state distribution. The initial transient phase can be removed as a burn-in. The cutoff point for each parameter is given in the \texttt{burnin} column of the output dataframe.
    \item Estimate the relative confidence interval for the mean of the steady-state posterior distribution of each parameter, assuming such steady-state distribution exists in the samples. The relative confidence interval is computed by dividing the lower and upper bounds of the confidence interval by the mean value of the parameter. A large confidence interval implies either the lack of convergence to a stationary distribution or the lack of samples. Half the relative confidence interval is given in the \texttt{halfwidth} column of the output dataframe. The \texttt{test} column will be \texttt{true} (1) if the halfwidth is less than the input target \texttt{eps} (default is 0.1) and \texttt{false} (0) otherwise. Note that parameters with a mean value close to 0 can have erroneously large relative confidence intervals because of the division by the mean. The test value can therefore be expected to be \texttt{false} (0) for those parameters without concluding a lack of convergence.
    \item Quantify the extent to which the distribution of the samples is stationary using statistical testing. The returned p-value, shown in the \texttt{pvalue} column of the output dataframe, can be considered a measure of mean stationarity. A p-value lower than the input threshold \texttt{alpha} (default is 0.05) implies a lack of stationarity of the mean, i.e. the posterior samples did not converge to a steady-state distribution with a fixed mean.
\end{itemize}

The Heidelberger diagnostic only tests for the mean of the distribution. Therefore, much like other diagnostics, it can only be used to detect the lack of convergence and not to prove convergence. In other words, even if all the numbers seem normal, one cannot conclude that the chain converged to a stationary distribution or that it converged to the true posterior.





\subsection{What if the Chains Are Not Converging?} \label{sec:notconverging}

If the chains seem to not be converging, there are things you can try to help your Markov chains converge:
\begin{itemize}
    \item Lower the target acceptance ratio from the default $0.8$.
    \item Re-parameterize your model to have less parameter dependence.
    \item Fix some parameter values to known good values,
  e.g. values obtained by \textit{maximum-a-posteriori} (MAP) optimization.
    \item Initialize the sampling from good parameter values.
    \item Use a stronger prior around suspected good parameter values.
    \item Simplify your model, e.g. using simpler dynamics.
    \item Try the marginal MCMC algorithm \texttt{MarginalMCMC} instead of the full joint MCMC algorithm \texttt{BayesMCMC}.
\end{itemize}

\subsection{Advanced Posterior Queries}
\label{sec:posterior_queries}

\subsubsection{Summary Statistics}

After you fit your Bayesian Pumas model, there are a number of functions and plots you can call on the output of the \texttt{fit} or \texttt{discard} functions.
Often you want to execute posterior queries. Beside the basic summary statistics that one can get using the \texttt{summarystats} function as discussed in Section \ref{sec:basic_summary}, one can also compute more advanced statistics based on the posterior.

A common advanced posterior query is the probability that a certain parameter $\theta$ is higher than $0$ which can be written as an expectation problem:
\begin{align*}
\operatorname{E}[\theta > 0 \mid \text{data}]
\end{align*}

The way you can do this is using the \texttt{mean} function with a convenient \texttt{do} operator.
Listing \ref{lst:postquery} shows an example of a posterior query using the \texttt{do} operator where we are testing if the parameter \texttt{tvcl} is higher than $0$.
It outputs a valid probability estimate, i.e. $\in [0,1]$.
\begin{lstlisting}[caption=Example of a posterior query with the \texttt{do} operator,label=lst:postquery]
mean(tres) do p
    p.tvcl > 0
end
\end{lstlisting}
Instead of \texttt{mean}, one can also use \texttt{var} to compute the variance, or use \texttt{cov} and \texttt{cor} to compute the covariance and correlation matrices, respectively, if multiple outputs are returned from the \texttt{do} block. Listing \ref{lst:multipleout} shows an example where the correlation matrix between the \texttt{tvcl} and \texttt{tvvc} parameters is estimated using the posterior samples.

\begin{lstlisting}[caption=Posterior queries from multiple outputs,label=lst:multipleout]
cor(tres) do p
    [p.tvcl, p.tvvc]
end
\end{lstlisting}

Note that any transformation of the parameters can be done in the \texttt{do} block. For example, we can get the mean value of the lower triangular Cholesky factor of a correlation matrix parameter \texttt{C} using the code in Listing \ref{lst:meanchol}.

\begin{lstlisting}[caption=Mean Cholesky factor of correlation matrix,label=lst:meanchol]
mean(tres) do p
    getchol(p.C).L
end
\end{lstlisting}

This is sometimes needed to compare the results of Pumas and Stan because Stan's equivalent to \texttt{LKJCholesky} reports the lower triangular factors in the MCMC samples instead of the actual correlation matrices which Pumas reports.

To compute summary statistics of the subject-specific parameters of subject $i$ instead of the population parameters, you can use the \texttt{subject} keyword argument as shown in Listing \ref{lst:subject_summary}.

\begin{lstlisting}[caption=Mean subject-specific parameters,label=lst:subject_summary]
mean(tres, subject = i) do p
    p.{$\eta$}std
end
\end{lstlisting}

\subsubsection{Quantiles} \label{sec:quantiles}

You can query the estimate of the posterior quantiles for population-level or subject-specific parameters using the \texttt{quantile} function:
\begin{lstlisting}
quantile(tres)
\end{lstlisting}
This will display the 2.5\%, 25\%, 50\%, 75\%, and 97.5\% quantiles of all the population-level parameters by default. To display the quantiles of the subject-specific parameters of subject $i$ instead, you can use the \texttt{subject} keyword argument as such:
\begin{lstlisting}
quantile(tres, subject = i)
\end{lstlisting}
To change the quantiles computed, you can also manually input the desired quantiles using the \texttt{q} keyword argument. For example, the following returns the 10\% and 90\% quantiles of the subject-specific parameters of subject $i$:
\begin{lstlisting}
quantile(tres, subject = i, q = [0.1, 0.9])
\end{lstlisting}

\subsubsection{Credible Intervals}

A credible interval is an interval containing a pre-specified probability mass in the posterior distribution. For instance, an estimate of the 95\% credible interval is any interval such that at least 95\% of the posterior samples obtained with MCMC lie in that interval. Naively, one can use the interval from the 2.5\% quantile to the 97.5\% quantile of a parameter as a 95\% credible interval. This can be obtained using the \texttt{quantile} function as shown in section \ref{sec:quantiles}. However, less naively, one may be interested in the smallest interval that includes at least 95\% of the posterior mass. This is commonly known as the highest probability desnity interval (HPDI). To get the HPDI which contains $(1 - \text{a})\%$ of the samples for each population parameter, you can use:
\begin{lstlisting}
hpd(tres, alpha = a)
\end{lstlisting}
To get the same interval for the subject-specific parameters of subject $i$, you can use:
\begin{lstlisting}
hpd(tres, alpha = a, subject = i)
\end{lstlisting}

\subsection{Posterior Plots}

There are a number of plots that you can use to visualize the posterior distribution.
In this section, we'll cover plots related to the parameter estimates: density plots, ridge plots and corner plots.

\subsubsection{Density Plot}

\begin{figure}[h]
    \centering
    \includegraphics[width=0.4\textwidth]{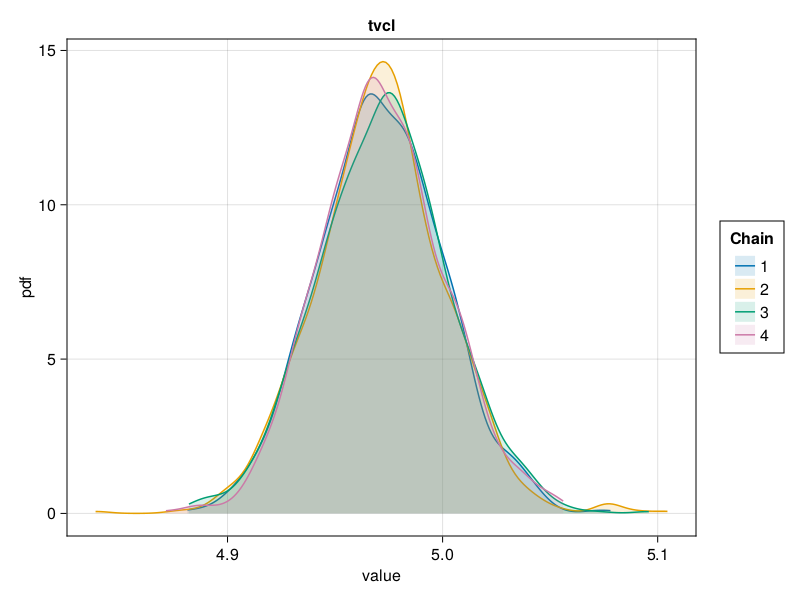}
    \caption{Example of a density plot.}
    \label{fig:density}
\end{figure}

A density plot shows a smoothed version of the histogram of a parameter value, giving an approximate probability density function for the marginal posterior of each parameter.
This helps us visualize the shape of the marginal posterior of each parameter.
If you run multiple Markov chains, the plot will show overlapping densities for each density distinguished by different colors.
You can plot density plots with the function \texttt{density\_plot}.
Listing \ref{lst:density} and Figure \ref{fig:density} show the code and the resulting density plot, respectively.
If you do not specify which parameter you want with the optional keyword argument \texttt{parameters}, the plot will output multiple density plots faceted automatically.
\texttt{parameters} accepts a vector of parameters.
\begin{lstlisting}[caption=Example of a density plot,label=lst:density]
density_plot(pk_1cmp_tfit; parameters = [:tvcl])
\end{lstlisting}

\subsubsection{Ridgeline Plot}

\begin{figure}[h]
    \centering
    \includegraphics[width=0.4\textwidth]{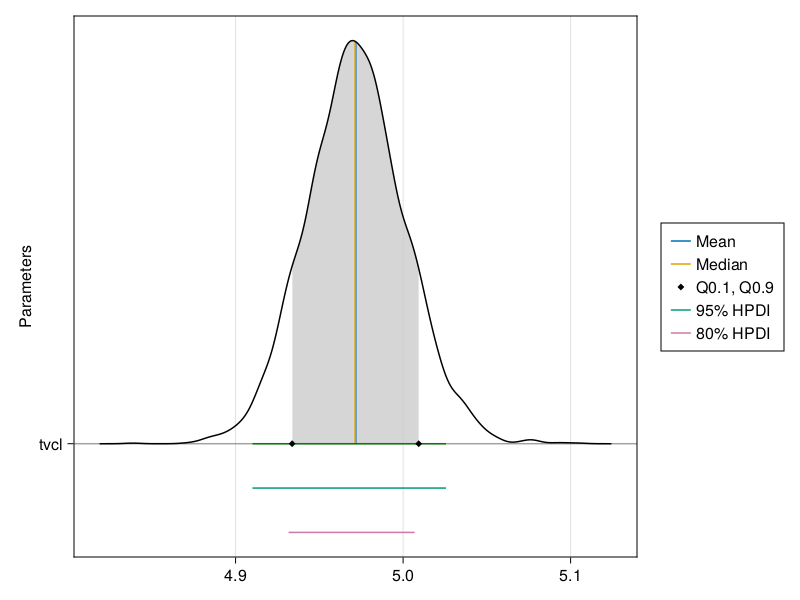}
    \caption{Example of a ridgeline plot.}
    \label{fig:ridge}
\end{figure}

Another common posterior plot is the ridgeline plot, which outputs a single density summarizing all the of the sampled chains along with relevant statistical information about your parameter.
The information that it outputs is the mean, median, 10\% and 90\% quantiles, along with 95\% and 80\%  highest posterior density interval (HPDI).
You can plot ridgeline plots with the function \texttt{ridgeline\_plot}, which has a similar syntax as \texttt{density\_plot}.
Listing \ref{lst:ridge} and Figure \ref{fig:ridge} show the code and the resulting ridgeline plot, respectively.

\begin{lstlisting}[caption=Example of a ridgeline plot,label=lst:ridge]
ridge_plot(pk_1cmp_tfit; parameters = [:tvcl])
\end{lstlisting}

\subsubsection{Corner Plot}

\begin{figure}[h]
    \centering
    \includegraphics[width=0.4\textwidth]{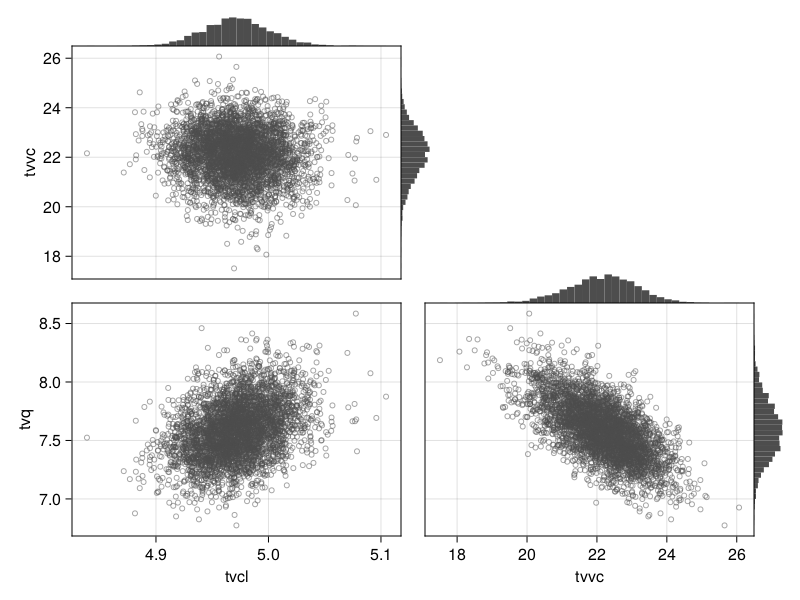}
    \caption{Example of a corner plot.}
    \label{fig:corner}
\end{figure}

The corner plot is a plot that showcases scatter plots between different parameters along with marginal histograms in a well-organized template.
This can be used to investigate a high correlation between parameter values that can be a source of convergence issues for the MCMC sampler.
Listing \ref{lst:corner} shows the code for a corner plot for the parameters \texttt{tvq} and \texttt{tvcl}.
The output is in Figure \ref{fig:corner}.
\begin{lstlisting}[caption=Example of a corner plot,label=lst:corner]
corner_plot(tres, parameters = [:tvq, :tvvc])
\end{lstlisting}

\subsection{Posterior Simulations and Predictions} \label{sec:predictivechecks}

\subsubsection{Existing Subjects}

You can simulate new responses for existing subjects using the parameter values sampled from the posterior stored in the MCMC result. This is not to be confused with prior predictive simulations which use parameter values sampled from the priors. The \texttt{simobs} function can be used to do this:
\begin{lstlisting}
sims = simobs(tres; samples = 100)
\end{lstlisting}
where \texttt{tres} is the output from the \texttt{fit} or \texttt{discard} functions. 
The \texttt{samples} keyword argument is the number of sub-samples taken from the MCMC samples. When not set, all of the MCMC samples will be used.
By default, all the subjects are simulated. If the keyword argument \texttt{subject} is set to any integer index $i$, only the $i^{th}$ subject will be simulated.

In the \texttt{simobs} function, there's also a keyword argument: \texttt{simulate\_error}. If it is set to \texttt{true} (the default value), Pumas will sample from the response's error model in the \texttt{@derived} block (aka simulation), otherwise, it will return the expected value of the error distribution (aka prediction).
The latter is equivalent to using the \texttt{predict} function.

\subsubsection{New Dose or Covariates} \label{sec:new_dose}

It is often useful to do counterfactual simulations by changing some of the variables we have control over and doing what-if analysis. Changing the dose is the most common use case in pharmacometrics but in some cases, covariates may also be changed.

To change the dose and/or covariates and reuse the posterior samples for a particular subject, you will first need to define a new subject. You can either define the new subject manually using the \texttt{Subject} constructor, or you can start from a data frame and use the \texttt{read\_pumas} function to convert it to a Pumas subject. To learn about the manual \texttt{Subject} constructor, please refer to the Pumas documentation (\href{docs.pumas.ai}{https://docs.pumas.ai}). To showcase the second approach, assume the original data frame of the entire population is \texttt{df}. To multiply the dose of subject \texttt{i} by 3 given the data frame \texttt{df} where the dose amount is located in the first row of the \texttt{:amt} field, you can do:
\begin{lstlisting}
subj_df = copy(df[df.id .== i, :])
subj_df[1, :amt] = subj_df[1, :amt] * 3.0
new_subj = read_pumas(subj_df)[1]
\end{lstlisting}
For more on data frame wrangling, please refer to the Pumas documentation (\href{docs.pumas.ai}{https://docs.pumas.ai}) or tutorials (\href{tutorials.pumas.ai}{http://tutorials.pumas.ai}).

After defining the new subject \texttt{new\_subj}, you can call the following method of \texttt{simobs}:
\begin{lstlisting}
simobs(tres, new_subj, subject = i, samples = 100)
\end{lstlisting}
Setting the \texttt{subject} keyword argument to the index \texttt{i} will trigger the use of the MCMC samples for subject \texttt{i}'s parameters while using the dose, covariates and time points from \texttt{new\_subj}. Note that the index \texttt{i} refers to the index of the subject in the training population passed to the \texttt{fit} function which may not match the ID field of the subject.

To simulate for an actually new subject \texttt{new\_subj}, where the subject-specfic parameters are sampled from the prior distribution and the population parameters are sampled from the posterior, you can drop the \texttt{subject} keyword argument:
\begin{lstlisting}
simobs(tres, new_subj, samples = 100)
\end{lstlisting}

\subsection{Visual Predictive Checks and Simulation Plots}
\label{sec:vpc}

A visual predictive check (VPC) of simulations in which the parameter values were sampled from the prior distributions is commonly known as the prior predictive check. Similarly, a VPC of simulations in which the parameter values were sampled from the posterior distribution is commonly known as the posterior predictive check.

After calling a prior or posterior \texttt{simobs} call, the result simulation object \texttt{sims} can be passed to the \texttt{vpc} function to compute all of the quantiles necessary for a VPC plot.
The function also accepts a categorical variable to stratify the VPC results with the keyword argument \texttt{stratify\_by}.
The VPC result can be plotted with the \texttt{vpc\_plot} function.
Listing \ref{lst:vpc} show the code for running a VPC from simulations.

\begin{figure}[h]
  \centering
  \includegraphics[width=0.4\textwidth]{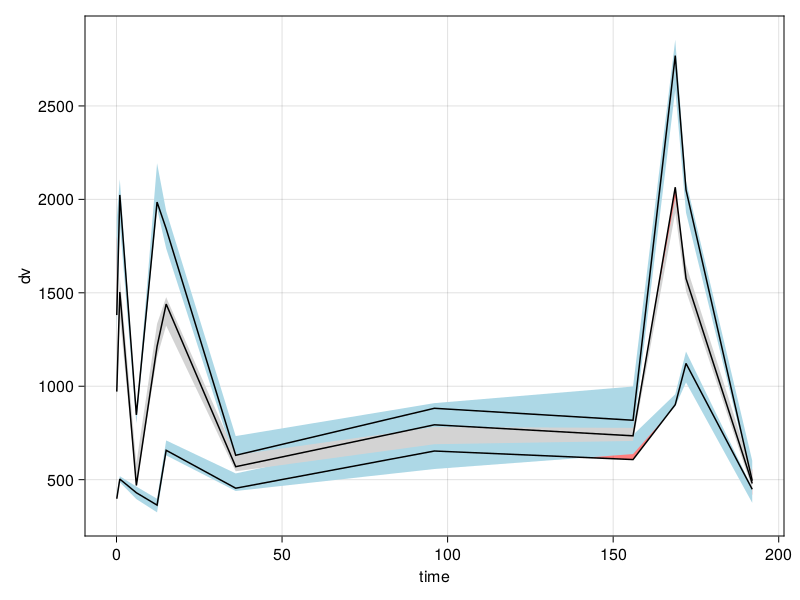}
  \caption{Example of a posterior visual predictive check.}
  \label{fig:vpc1}
\end{figure}

\begin{lstlisting}[caption=Visual predictive check,label=lst:vpc]
vpc_res = vpc(sims)
vpc_plot(vpc_res)
\end{lstlisting}
Figure \ref{fig:vpc1} is an example of an output of such code.

\begin{figure}[h]
  \centering
  \includegraphics[width=0.4\textwidth]{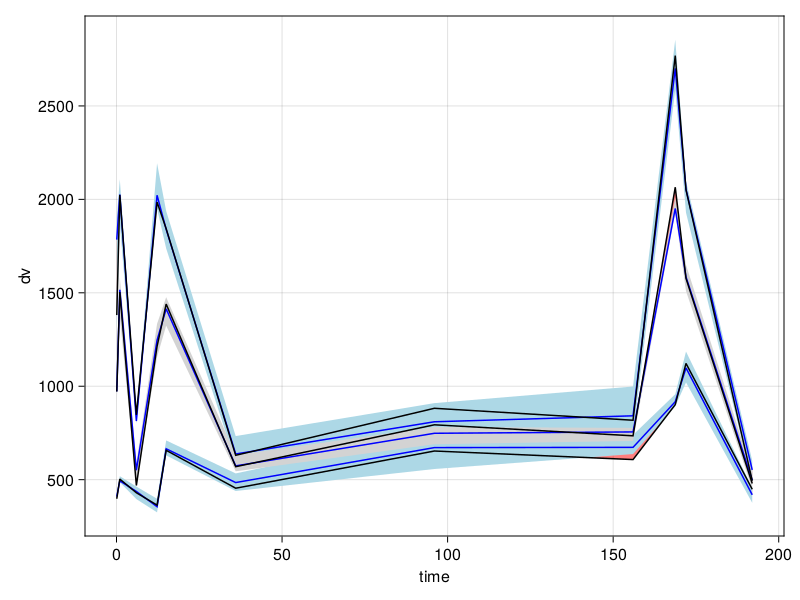}
  \caption{Example of a posterior visual predictive check with simulated quantile medians.}
  \label{fig:vpc2}
\end{figure}

To plot the simulated quantile median lines, you can set the keyword argument \texttt{simquantile\_medians = true}, e.g:
\begin{lstlisting}
vpc_res = vpc(sims)
vpc_plot(vpc_res, simquantile_medians = true)
\end{lstlisting}
The resulting plot will look like Figure \ref{fig:vpc2}.

\begin{figure}[h]
  \centering
  \includegraphics[width=0.4\textwidth]{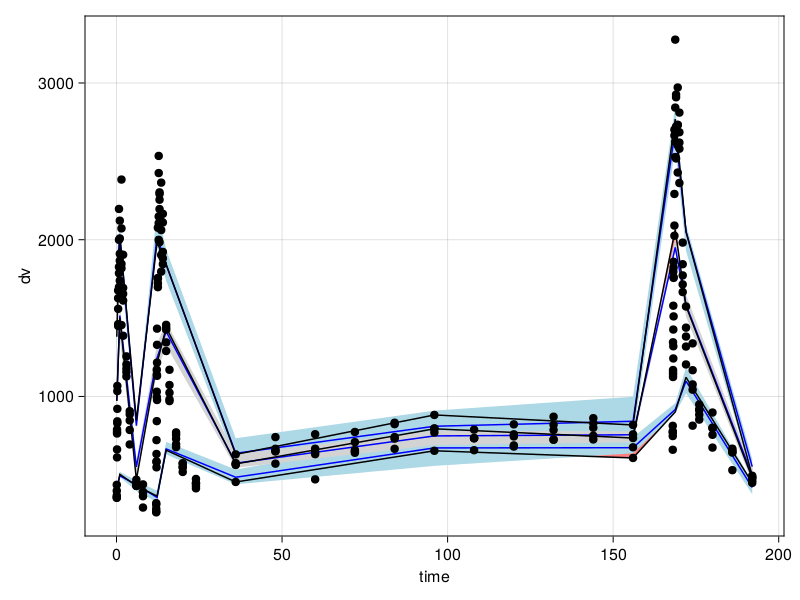}
  \caption{Example of a posterior visual predictive check with simulated quantile medians and observations.}
  \label{fig:vpc3}
\end{figure}

To further display the observations as points on the VPC plot, you can set the keyword argument \texttt{observations = true}, e.g:
\begin{lstlisting}
vpc_res = vpc(sims)
vpc_plot(vpc_res, simquantile_medians = true, observations = true)
\end{lstlisting}
The resulting plot will look like Figure \ref{fig:vpc3}.

For more on the many VPC options available, including changing the covariate or stratification, you can refer to the Pumas documentation (\href{docs.pumas.ai}{https://docs.pumas.ai}).

\begin{figure}[h]
  \centering
  \includegraphics[width=0.4\textwidth]{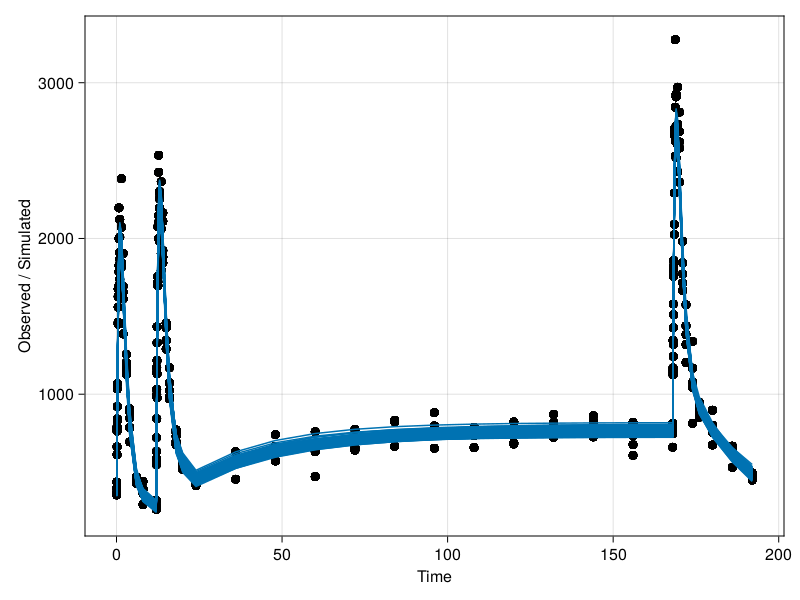}
  \caption{Example of a simple simulation plot.}
  \label{fig:simplot}
\end{figure}

Instead of a full VPC plot, you can also just plot the simulated responses and observations without the VPC colour bands using the \texttt{sim\_plot} function, e.g:
\begin{lstlisting}
sim_plot(sims)
\end{lstlisting}
An example output is shown in Figure \ref{fig:simplot}.

\subsection{Simulation Queries}
\label{sec:sim_queries}

The output of a \texttt{simobs} call stores the simulated observations but also all the intermediate values computed such as: the parameter values used, individual coefficients, dose control parameters, covariates, differential equation solution, etc. There are a number of post-processing operations you can do on the simulation output to compute various queries and summary statistics based on the simulations.

The \texttt{postprocess} function is a powerful tool that allows you to make various queries using the simulation results. There are multiple ways to use the \texttt{postprocess} function. The first way to use the \texttt{postprocess} function is to extract all of the information stored in the simulation result in the form of a vector of named tuples. Each named tuple has all the intermediate values evaluated when simulating 1 run. Let \texttt{sims} be the output of any \texttt{simobs} operation. Listing \ref{lst:postprocess1} shows how to extract all the simulation's intermediate results.
\begin{lstlisting}[caption=Extract intermediate values,label=lst:postprocess1]
generated = postprocess(sims)
\end{lstlisting}

\texttt{generated} is a vector of the outputs from all the simulation runs. Hence, \texttt{generated[i]} stores the outputs from the $i^{th}$ simulation run. Time-dependent variables are stored as a vector with 1 element for each time point where there is an observation. Alternatively, if the keyword argument \texttt{obstimes} was set instead when calling \texttt{simobs}, the time-dependent variables will be evaluated at the time points in \texttt{obstimes} instead.

The second way to use \texttt{postprocess} is by passing in a post-processing function. The post-processing function can be used to: 
\begin{itemize}
    \item Transform the simulated quantities, or
    \item Compare the simulated quantities to the observations.
\end{itemize}

We use the \texttt{do} syntax here which is short for passing in a function as the first argument to \texttt{postprocess}. The `do` syntax to pass in a post-processing function is shown in Listing \ref{lst:postprocess2}.

\begin{lstlisting}[caption=Compare generated quantities and observations,label=lst:postprocess2]
postprocess(sims) do gen, obs
  ...
end
\end{lstlisting}

where \texttt{gen} is the named tuple of all generated quantities from 1 simulation run and \texttt{obs} is the named tuple of observations. For instance to query the ratio of simulated observations \texttt{conc} that are higher than the observed quantity \texttt{conc} at the observations' time points, you can use the code in Listing \ref{lst:postprocess3}. This is sometimes called the Bayesian p-value which is expected to be around 0.5.

\begin{lstlisting}[caption=Bayesian p-value per simulation,label=lst:postprocess3]
postprocess(sims) do gen, obs
  sum(gen.conc .> obs.conc) / length(gen.conc)
end
\end{lstlisting}

\texttt{gen.conc} is the simulated vector of \texttt{conc} whose length is the same as the number of observation time points. \texttt{obs.conc} is the observation vector \texttt{conc}. \texttt{gen.conc .> obs.conc} returns a vector of \texttt{true}/\texttt{false}, with one element for each time point. The sum of this vector gives the number of time points where the simulation was higher than the observation. Dividing by the number of time points gives the ratio. When using \texttt{postprocess} in this way, the output is always a vector of the query results, one number for each simulation. In the query function body, you can choose to use only \texttt{gen} or only \texttt{obs} but the header must always have both \texttt{gen} and \texttt{obs}.

The third way to use the \texttt{postprocess} function is to compute summary statistics of the simulated quantities or of functions of the simulated quantities. Summary statistics can be computed by passing a statistic function as the \texttt{stat} keyword argument. For example in order to estimate the probability that a simulated value is higher than an observation, you can use the code in Listing \ref{lst:postprocess4}.

\begin{lstlisting}[caption=Mean Bayesian p-value,label=lst:postprocess4]
postprocess(sims, stat = mean) do gen, obs
  gen.conc .> obs.conc
end
\end{lstlisting}

This function will do 2 things:
\begin{enumerate}
    \item Concatenate the query results (e.g. \texttt{gen.conc .> obs.conc}) from all the simulation runs into a single vector.
    \item Compute the mean value of the combined vector.
\end{enumerate}

Alternatively, you can use the \texttt{mean} function to do the same thing without using the keyword argument. Listing \ref{lst:postprocess5} will call the \texttt{postprocess} function under the hood.
\begin{lstlisting}[caption=Mean Bayesian p-value using the mean function,label=lst:postprocess5]
mean(sims) do gen, obs
  gen.conc .> obs.conc
end
\end{lstlisting}
The result of this operation will be a scalar equal to the mean value of the \textit{concatenated} vector of queries.

In order to get the probability that the simulated quantity is higher than the observation \textit{for each time point}, you can call the \texttt{mean} function externally as shown in Listing \ref{lst:postprocess6}.
\begin{lstlisting}[caption=Mean Bayesian p-value using the mean function,label=lst:postprocess6]
generated = postprocess(sims) do gen, obs
  gen.conc .> obs.conc
end
mean(generated)
\end{lstlisting}
This returns a vector of probabilities of the same length as the number of time points without concatenating all the queries together.

To compute a summary statistic of all the generated quantities, you can also use the code in Listing \ref{lst:postprocess7}.
\begin{lstlisting}[caption=Mean generated quantities,label=lst:postprocess7]
postprocess(sims, stat = mean)
\end{lstlisting}
without specifying a post-processing function which is also equivalent to the shorter version in Listing \ref{lst:postprocess8}.
\begin{lstlisting}[caption=Mean generated quantities - short version,label=lst:postprocess8]
mean(sims)
\end{lstlisting}

Beside \texttt{mean}, you can also use any of the following summary statistic functions in the same way:
\begin{itemize}
    \item \texttt{std} for element-wise standard deviation
    \item \texttt{var} for element-wise variance
    \item \texttt{cor} for correlation between multiple quantities
    \item \texttt{cov} for covariance between multiple quantities
\end{itemize}

These functions can be passed in as the \texttt{stat} keyword argument to \texttt{postprocess} or they can be used in the short form, e.g.:
\begin{lstlisting}
generated = postprocess(
    sims, stat = std,
) do gen, obs
  ...
end
std(sims) do gen, obs
  ...
end
std(sims)

generated = postprocess(
    sims, stat = var,
) do gen, obs
  ...
end
var(sims) do gen, obs
  ...
end
var(sims)
\end{lstlisting}

The \texttt{cor} and \texttt{cov} statistics are unique in that they require a post-processing function which outputs a vector. For example to estimate the correlation between the \texttt{CL} and \texttt{Vc} parameter in a 1 compartment model, you can use any of the following:
\begin{lstlisting}
postprocess(s, stat = cor) do gen, obs
  [gen.CL[1], gen.Vc[1]]
end
cor(s) do gen, obs
  [gen.CL[1], gen.Vc[1]]
end
\end{lstlisting}
Note that \texttt{gen.CL} is a vector of simulated \texttt{CL} values for all the time points. But since the value is constant across time, we can use the first element \texttt{gen.CL[1]} only. \texttt{cov} can be used instead of \texttt{cor} to compute the covariance matrix. The output of this operation is either a correlation or a covariance matrix.

\subsection{Non-Compartmental Analysis (NCA) Parameters} \label{sec:nca}

You can easily integrate non-compartmental analysis (NCA) parameters such as area-under-curve (\texttt{auc}) and maximum concentration (\texttt{cmax}) in the simulation using the \texttt{@observed} block in the Pumas model, e.g:
\begin{lstlisting}
@derived begin
    cp = @. Central / Vc
    conc {$\sim$} @. LogNormal(log(cp), {$\sigma$})
end
@observed begin
    nca := @nca cp
    auc = NCA.auc(nca)
    cmax = NCA.cmax(nca)
end
\end{lstlisting}
The \texttt{auc} and \texttt{cmax} will now get stored in the output of \texttt{simobs} and can be queried using the various summary statistics queries. For example, the following will be estimating the probability that \texttt{auc} is greater than 20 and \texttt{cmax} is less than 30 given the simulations \texttt{sims} (output from \texttt{simobs}).
\begin{lstlisting}
mean(sims) do gen, obs
  gen.auc > 200 && gen.cmax < 30
end
\end{lstlisting}
For more on NCA integration and parameters, please refer to the Pumas documentation (\href{docs.pumas.ai}{https://docs.pumas.ai}).

\subsection{Crossvalidation and Expected Log Predictive Density} \label{sec:crossvalidation}

Crossvalidation is a technique for evaluating a model's predictive accuracy on unseen data, aka out-of-sample data. This is done by systematically leaving some data out from the training data\footnote{data used during the Bayesian inference to get samples from the posterior} when performing Bayesian inference followed by an evaluation of the average predictive accuracy of the MCMC samples using the data that was left out. Each iteration of the crossvalidation routine leaves a different subset of the data out of training and uses some or all of it for evaluating the prediction accuracy. The metric used for evaluating the prediction accuracy is typically the (conditional or marginal) log likelihood of each of the MCMC samples given the unseen data. The predictive performance metric is then averaged out across the iterations of the crossvalidation routine.

You can do almost all types of crossvalidation for hierarchical models using Pumas. The main function that performs crossvalidation in Pumas is the \texttt{crossvalidate} function. There are 2 inputs to \texttt{crossvalidate}:
\begin{enumerate}
    \item The MCMC result from \texttt{fit} or \texttt{discard}. We will call this \texttt{tres}.
    \item The crossvalidation algorithm. Let's call it \texttt{cv\_method}.
\end{enumerate}
The call syntax is \texttt{cv\_res = crossvalidate(res, cv\_method)}. To estimate the expected log predictive density (ELPD) given \texttt{cv\_res}, you can use the \texttt{elpd} function: \texttt{elpd(cv\_res)}.

There are 2 families of the crossvalidation algorithms available in Pumas:
\begin{enumerate}
    \item Resampling-based CV which performs the MCMC again for each data point or subset left out. This algorithm is constructed using the \texttt{ExactCrossvalidation} struct.
    \item PSIS-based CV \citep{psis_loo} which uses the importance sampling and weight smoothing approach to avoid the need for resampling. This algorithm is constructed using the \texttt{PSISCrossvalidation} struct.
\end{enumerate}
The constructor for a resampling-based CV method is:
\begin{lstlisting}
cv_method = ExactCrossvalidation(; split_method, split_by, ensemblealg = EnsembleThreads())    
\end{lstlisting}
where \texttt{split\_method} and \texttt{split\_by} are keyword arguments that must be set and \texttt{ensemblealg} defaults to the use of multi-threading.

This defines an instance of the \texttt{ExactCrossvalidation} algorithm for crossvalidation. In this algorithm, the fitting is re-run while leaving out a subset of the data each time.

The way by which the data is split between training and validation sets is determined using the keyword arguments \texttt{split\_method} and \texttt{split\_by}. The \texttt{split\_method} argument can be of any of the following types:
\begin{enumerate}
    \item \texttt{LeaveK} for leave-K-out crossvalidation
    \item \texttt{KFold} for K-fold crossvalidation
    \item \texttt{LeaveFutureK} for leaving K future points at a time
\end{enumerate}

The \texttt{split\_by} keyword argument can be of any of the following types:
\begin{enumerate}
    \item \texttt{BySubject} for leaving out subjects
    \item \texttt{ByObservation} for leaving out individual observations per subject
\end{enumerate}

Each of the data-splitting methods will be discussed in the next subsection.

Similar to the resampling-based crossvalidation, the constructor for a PSIS-CV method is:
\begin{lstlisting}
cv_method = PSISCrossvalidation(; split_method, split_by, pareto_shape_threshold = 0.7)
\end{lstlisting}
where \texttt{split\_method} and \texttt{split\_by} are keyword arguments that must be set.

This defines an instance of the \texttt{PSISCrossvalidation} algorithm for crossvalidation. The \texttt{split\_method} and \texttt{split\_by} keyword arguments are similar to the resampling-based CV case. The \texttt{pareto\_shape\_threshold = 0.7} keyword argument will result in the removal of any CV run that leads to a Pareto shape parameter more than 0.7 when computing the expected log predictive density (ELPD) estimate. This can be useful to avoid a few bad runs rendering the whole PSIS-CV method useless. Ideally one would re-run the inference for the subset of CV runs where the shape parameter exceeded the threshold but this is not implemented in Pumas yet as of the time of this writing. Follow the Pumas documentation (\href{docs.pumas.ai}{https://docs.pumas.ai}) for updates on the latest features available.

\subsubsection{Leave-K-Out}

The constructor for the leave-K-out data splitting algorithm is:
\begin{lstlisting}
split_method = LeaveK(; K = 5, shuffle = false, rng = nothing)    
\end{lstlisting}

In this algorithm, the data is split multiple times into 2 disjoint groups, each time starting from the full data set. The 2 groups are typically called training and validation subsets, where the validation subset has K data points. In the next iteration, the whole data set is re-split using another disjoint subset of K data points as the validation subset. This process is done repeatedly until almost each data point has shown up in 1 and only 1 validation subset.

The data is typically a vector of some sort, e.g. observations or subjects, and the splittings are order-dependent. Before performing the splittings, you can randomly shuffle the data vector by setting the \texttt{shuffle} keyword argument to \texttt{true} (default is \texttt{false}) getting rid of the sensitivity to the original order of the data. You can additionally pass an optional pseudo-random number generator \texttt{rng} to control the pseudo-randomness for reproducibility.

Assume some dummy original data \texttt{["A", "B", "C", "D"]} which resembles the subjects or observations. Leave-one-out splitting without shuffling results in the data splittings shown in Table \ref{tab:leave1out}.

\begin{table}[ht]
    \centering
    \begin{tabular}{c|c}
    \hline
    Training subset  & Validation subset \\
    \hline
    \texttt{["A", "B", "C"]}	& \texttt{["D"]} \\
    \hline
    \texttt{["A", "B", "D"]}	& \texttt{["C"]} \\
    \hline
    \texttt{["A", "C", "D"]}	& \texttt{["B"]} \\
    \hline
    \texttt{["B", "C", "D"]}	& \texttt{["A"]} \\
    \hline
    \end{tabular}
    \caption{Leave-one-out splitting.}
    \label{tab:leave1out}
\end{table}

where each data point shows once and only once in a validation subset. Leave-2-out splitting without shuffling results in the data splittings shown in Table \ref{tab:leave2out}.

\begin{table}[ht]
    \centering
    \begin{tabular}{c|c}
    \hline
    Training subset  & Validation subset \\
    \hline
    \texttt{["A", "B"]}	& \texttt{["C", "D"]} \\
    \hline
    \texttt{["C", "D"]} & \texttt{["A", "B"]} \\
    \hline
    \end{tabular}
    \caption{Leave-2-out splittings.}
    \label{tab:leave2out}
\end{table}

\subsubsection{K-Fold}

The constructor for the K-fold data splitting algorithm is:
\begin{lstlisting}
split_method = KFold(; K = 5, shuffle = false, rng = nothing)    
\end{lstlisting}

In this algorithm, the data is split K times into 2 disjoint groups, each time starting from the full data set. The 2 groups are typically called training and validation subsets, where the validation subset has \texttt{floor(N / K)}\footnote{\texttt{floor} is a function that rounds down to an integer.} data points, \texttt{N} being the total number of data points. In the next iteration, the whole data set is re-split using another disjoint validation subset of \texttt{floor(N / K)} different points, disjoint from the previous validation subsets. This process is done iteratively until almost each data point has shown up in 1 and only 1 validation subset. If \texttt{N} is divisible by \texttt{K}, each point will show up in 1 and only 1 validation subset. Otherwise, the remaining points will be part of the training subset for all the splittings and will not show up in any validation subset.

The data is typically a vector of some sort, e.g. observations or subjects, and the splittings are order-dependent. Before performing the splittings, you can randomly shuffle the data vector by setting the \texttt{shuffle} keyword argument to \texttt{true} (default is \texttt{false}) getting rid of the sensitivity to the original order of the data. You can additionally pass an optional pseudo-random number generator \texttt{rng} to control the pseudo-randomness for reproducibility.

Assume some dummy original data \texttt{["A", "B", "C", "D"]} which resembles the subjects or observations. 4-fold splitting without shuffling results in the data splittings shown in Table \ref{tab:4fold},

\begin{table}[ht]
    \centering
    \begin{tabular}{c|c}
    \hline
    Training subset	& Validation subset \\
    \hline
    \texttt{["A", "B", "C"]} & \texttt{["D"]} \\
    \hline
    \texttt{["A", "B", "D"]} & \texttt{["C"]} \\
    \hline
    \texttt{["A", "C", "D"]} & \texttt{["B"]} \\
    \hline
    \texttt{["B", "C", "D"]} & \texttt{["A"]} \\
    \hline
    \end{tabular}
    \caption{4-fold splittings.}
    \label{tab:4fold}
\end{table}

where each data point showed once and only once in a validation subset. 2-fold splitting without shuffling results in the data splittings shown in Table \ref{tab:2fold}.

\begin{table}[ht]
    \centering
    \begin{tabular}{c|c}
    \hline
    Training subset	& Validation subset \\
    \hline
    \texttt{["A", "B"]} & \texttt{["C", "D"]} \\
    \hline
    \texttt{["C", "D"]}& \texttt{["A", "B"]} \\
    \hline
    \end{tabular}
    \caption{2-fold splitting.}
    \label{tab:2fold}
\end{table}

\subsubsection{Leave-Future-K}

The constructor for the leave-future-K data splitting algorithm is:
\begin{lstlisting}
split_method = LeaveFutureK(; K = 1, minimum = 2)
\end{lstlisting}

In this algorithm, the data is assumed to be a time series. The goal is to split the data into "past" and "future". Using this algorithm, the data is split multiple times into 3 disjoint groups where the third group is discarded, each time starting from the full data set. The first 2 groups are typically called the past/training subset and the future/validation subset, where the future validation subset has \texttt{K} future data points. In the next iteration, the whole data set is then re-split using another disjoint subset of \texttt{K} data points as the future validation subset. This process is done iteratively starting from the full data set and moving backward in time until the training subset has less than a pre-set \texttt{minimum} number of points remaining. Using this method, each data point can show up in at most 1 future validation subset. The default values of \texttt{K} and minimum are 1 and 2 respectively.

Assume the original data is \texttt{["A", "B", "C", "D", "E", "F"]}. Leave-1-future-out splitting with \texttt{minimum = 2} results in the data splittings shown in Table \ref{tab:leave1future} where the remaining points are discarded:

\begin{table}[ht]
    \centering
    \begin{tabular}{c|c}
    \hline
    Past subset & Future subset \\
    \hline
    \texttt{["A", "B", "C", "D", "E"]} & \texttt{["F"]} \\
    \hline
    \texttt{["A", "B", "C", "D"]} & \texttt{["E"]} \\
    \hline
    \texttt{["A", "B", "C"]} & \texttt{["D"]} \\
    \hline
    \texttt{["A", "B"]} & \texttt{["C"]} \\
    \hline
    \end{tabular}
    \caption{Leave-1-future-out splittings.}
    \label{tab:leave1future}
\end{table}

Leave-2-future-out splitting with \texttt{minimum = 2} results in the data splittings shown in Table \ref{tab:leave2future}.

\begin{table}[ht]
    \centering
    \begin{tabular}{c|c}
    \hline
    Past subset & Future subset \\
    \hline
    \texttt{["A", "B", "C", "D"]} & \texttt{["E", "F"]} \\
    \hline
    \texttt{["A", "B"]} & \texttt{["C", "D"]} \\
    \hline
    \end{tabular}
    \caption{Leave-2-future-out splittings.}
    \label{tab:leave2future}
\end{table}

\subsubsection{Subject-based Splitting}

The constructor for the subject-based splitting method is:
\begin{lstlisting}
split_by = BySubject(; marginal = LaplaceI())
\end{lstlisting}

Using this method, each subject is treated as a single data point. The predictive log-likelihood computed for each subject can be either the marginal log-likelihood or conditional log-likelihood. This method has one keyword argument, \texttt{marginal}. If \texttt{marginal} is set to \texttt{nothing}, the predictive log-likelihood computed for each subject is the conditional log-likelihood using the typical values for the parameters. Otherwise, the predictive log-likelihood computed for each subject is the marginal log-likelihood using \texttt{marginal} as the marginalization algorithm. The default value of marginal is \texttt{LaplaceI()} which uses the Laplace method to integrate out the subject-specific parameters. Other alternatives include: \texttt{FOCE()} and \texttt{FO()}.

\subsubsection{Observation-based Splitting}

The constructor for the observation-based splitting method is:
\begin{lstlisting}
split_by = ByObservation(; allsubjects = true)
\end{lstlisting}

Using this method, each observation or collection of observations is treated as a single data point. When computing predictive log-likelihoods using this method, the predictive log-likelihood computed is the conditional log-likelihood of one or more observations for one or more subjects.

This method has one keyword argument, \texttt{allsubjects}. If \texttt{allsubjects} is set to \texttt{true} (the default value), the $i^{th}$ observation of each subject are all grouped together into a single data point. This assumes all subjects have the same number of observations. If \texttt{allsubjects} is set to false, then each observation for each subject is its individual data point.

Assume there are 2 subjects and 3 observations per subject. When using \texttt{split\_method = LeaveK(K = 1)} as the splitting method together with \texttt{split\_by = ByObservation(allsubjects = false)}, the training and validation splittings are shown in Table \ref{tab:cv1}.

\begin{table}[ht]
    \centering
    \begin{tabular}{p{0.6\linewidth}|p{0.3\linewidth}}
    \hline
    Training subset & Validation subset \\
    \hline
    \texttt{Subj 1 (obs 1, 2, 3), subj 2 (obs 1, 2)} & \texttt{Subj 2 (obs 3)} \\
    \hline
    \texttt{Subj 1 (obs 1, 2, 3), subj 2 (obs 1, 3)}	& \texttt{Subj 2 (obs 2)} \\
    \hline
    \texttt{Subj 1 (obs 1, 2, 3), subj 2 (obs 2, 3)} & \texttt{Subj 2 (obs 1)} \\
    \hline
    \texttt{Subj 1 (obs 1, 2), subj 2 (obs 1, 2, 3)} & \texttt{Subj 1 (obs 3)} \\
    \hline
    \texttt{Subj 1 (obs 1, 3), subj 2 (obs 1, 2, 3)} & \texttt{Subj 1 (obs 2)} \\
    \hline
    \texttt{Subj 1 (obs 2, 3), subj 2 (obs 1, 2, 3)} & \texttt{Subj 1 (obs 1)} \\ 
    \hline
    \end{tabular}
    \caption{Training and validation splits using \texttt{split\_method = LeaveK(K = 1)} and \texttt{split\_by = ByObservation(allsubjects = false)}.}
    \label{tab:cv1}
\end{table}

On the other hand, if \texttt{allsubjects} is set to \texttt{true}, the training and validation splittings are shown in Table \ref{tab:cv2}.

\begin{table}[ht]
    \centering
    \begin{tabular}{p{0.6\linewidth}|p{0.3\linewidth}}
    \hline
    Training subset & Validation subset \\
    \hline
    \texttt{Subj 1 (obs 1, 2), subj 2 (obs 1, 2)} & \texttt{Subj 1 (obs 3), subj 2 (obs 3)} \\
    \hline
    \texttt{Subj 1 (obs 1, 3), subj 2 (obs 1, 3)} & \texttt{Subj 1 (obs 2), subj 2 (obs 2)} \\
    \hline
    \texttt{Subj 1 (obs 2, 3), subj 2 (obs 2, 3)} & \texttt{Subj 1 (obs 1), subj 2 (obs 1)} \\
    \hline
    \end{tabular}
    \caption{Training and validation splits using \texttt{split\_method = LeaveK(K = 1)} and \texttt{split\_by = ByObservation(allsubjects = true)}.}
    \label{tab:cv2}
\end{table}
    
\subsubsection{Examples}

Assume there are 5 subjects and 10 observations per subject and that \texttt{res} is the result of the \texttt{fit} or \texttt{discard} function. The following are some of the combinations in which the above inputs can be used:

\begin{itemize}
    \item Leave-one-observation-out cross-validation, leaving 1 observation for all the subjects at a time. \texttt{allsubjects = true} means that the same observation index is removed for all the subjects, e.g. the 10th observation for all the subjects is used for validation in the first run, then the 9th observation is used for validation in the second run, etc.
\begin{lstlisting}
split_method = LeaveK(K = 1)
split_by = ByObservation(allsubjects = true)
cv_method = ExactCrossvalidation(; split_method = split_method, split_by = split_by, ensemblealg = EnsembleThreads())
cv_res = crossvalidate(res, cv_method)
\end{lstlisting}
    
    \item Leave-two-future-observations-out cross-validation, leaving 2 future observations per subject such that no less than 4 observations are used in training. \texttt{allsubjects = false} means that only the data of one subject at a time will be split. So in the first run, observations 1 to 8 of subject 1 are used for training, and observations 9 and 10 of subject 1 are used for validation. In the second run, observations 1 to 6 of subject 1 are used for training, and observations 7 and 8 are used for validation. In the third run, observations 1 to 4 of subject 1 get used for training, and observations 5 and 6 are used for validation. For all 3 runs, all the observations of subjects 2 to 5 are used in training. Then in the fourth to sixth runs, subject 2's data gets split. In the seventh to ninth runs, subject 3's data gets split, etc.
\begin{lstlisting}
split_method = LeaveFutureK(K = 2, minimum = 4)
split_by = ByObservation(allsubjects = false)
cv_method = ExactCrossvalidation(; split_method = split_method, split_by = split_by, ensemblealg = EnsembleThreads())
cv_res = crossvalidate(res, cv_method)
\end{lstlisting}

    \item Leave-one-subject-out cross-validation, marginalizing the subject-specific parameters using FOCE when computing predictive likelihoods. Using this method, subjects \texttt{1, 2, 3, 4} are used for training in the first run and subject \texttt{5} is used for validation. In the second run, subjects \texttt{1, 2, 3, 5} are used for training and subject \texttt{4} is used for validation, etc. The predictive likelihood computed is the marginal likelihood of the subject left out, computed using \texttt{FOCE} for marginalizing the subject-specific parameters. \texttt{LaplaceI} or \texttt{FO} could have also been used instead.
\begin{lstlisting}
split_method = LeaveK(K = 1)
split_by = BySubject(marginal = FOCE())
cv_method = ExactCrossvalidation(; split_method = split_method, split_by = split_by, ensemblealg = EnsembleThreads())
cv_res = crossvalidate(res, cv_method)
\end{lstlisting}

\end{itemize}

\subsection{Information Criteria} \label{sec:information}

The ELPD model evaluation metric computed from the crossvalidation output is theoretically similar to the so-called Widely Applicable Information Criteria (WAIC) \citep{psis_loo}. More precisely, -2 times the ELPD is comparable to the WAIC. A higher ELPD Is better and a lower WAIC is better. When the ELPD is estimated using the PSIS leave-one-out (LOO) crossvalidation method, -2 times the ELPD estimate is sometimes known as the LOOIC. Besides the ELPD estimates, Pumas also supports common information criteria \citep{Burnham2002-fr}, such as the Akaike information criteria (AIC), the corrected AIC (AICc), the Bayesian information criteria (BIC), and the WAIC. To estimate these, we first need to compute the pointwise log-likelihoods by some definition of ``pointwise''. To do this, you can call:
\begin{lstlisting}
pl = loglikelihood(tres; split_method, split_by)
\end{lstlisting}
where \texttt{tres} is the output of \texttt{fit} or \texttt{discard} and \texttt{split\_method} and \texttt{split\_by} are the keyword arguments explained in Section \ref{sec:crossvalidation} which define what a point is and which log-likelihood to compute.
To calculate the information criteria using the pointwise log-likelihoods, you can then use any of the following functions:
\begin{lstlisting}
Pumas.aic(pl)
Pumas.aicc(pl)
Pumas.bic(pl)
Pumas.waic(pl)
\end{lstlisting}

\label{sec:crossval}

%% file: sections/3-background-and-intuition.tex

\section{Background and Intuition} \label{sec:background}

In this section, the notation used and the mathematical background of Bayesian inference in pharmacometrics will be presented.
We include a brief introduction of Bayesian statistics and where it's useful in pharmacometrics, followed by an intuitive explanation of Markov Chain Monte Carlo (MCMC) and the No-U-Turn sampler (NUTS) algorithm \citep{hoffman2014no,multlinomial}.
We then discuss the intuition and some of the math behind prior selection, MCMC convergence diagnostics, and cross-validation and model selection.
This section should prepare the readers for using the Bayesian workflow in Pumas or any standard Bayesian analysis tool by giving them \textit{working knowledge} of highly technical concepts, using intuition and light mathematics.

\subsection{Notation}
\label{sec:notation}

For convenience of notation, for the rest of this paper we use: 

\begin{enumerate}
    \item \textbf{$\theta$ to refer to all the population-level parameters including all of $(\theta, \Omega, \sigma)$} \item $\eta$ to refer to the patient-specific parameters for all the subjects, where $\eta_i$ refers to the subject-specific parameters of subject $i$.

    \item $x$ to refer to the covariates for all the subjects, where $x_i$ refers to the subject-specific covariates of subject $i$.
    To be more rigorous, $x_i$ also includes all the time points at which the observations were made for subject $i$.

    \item $y$ to refer to the observed response for all the subjects, where $y_i$ refers to the subject-specific response of subject $i$

    \item $p(A = \alpha \mid B = \beta)$ to denote the probability density/mass of the random variable $A$ taking the value $\alpha$ conditional on the variable $B$ taking the value $\beta$.
    $B = \beta$ can generally be replaced with multiple variables, e.g. $p(A = \alpha \mid B = \beta, C = c)$.
    If $A$ is a continuous random variable, $p(A = \alpha \mid B = \beta)$ refers to the probability \textit{density} of $A = \alpha$ conditioned on $B = \beta$. If $A$ is a discrete random variable, $p(A = \alpha \mid B = \beta)$ refers to the probability \textit{mass} of $A = \alpha$ conditioned on $B = \beta$.
    When $\alpha$ and/or $\beta$ are dropped, they can be replaced by the value $A$ and/or $B$ respectively, e.g. $p(A = A \mid B = B$ to be understood from the context.
    This is a slight abuse of notation using the same symbol $A$/$B$ to refer to both the random variable and the specific value in its support but this is common in probability theory.

    \item $p(A, B \mid C)$ to denote $p(A \mid B, C) \times p(B \mid C)$ which is equal to $p(B \mid A, C) \times p(A \mid C)$ which could be the product of probability densities and/or masses depending on the supports of $A$ and $B$.

    \item $D$ to refer to all the observed data including both $x$ and $y$.

    \item $p(y_i \mid x_i, \eta_i \theta)$ to denote the \textit{conditional} likelihood of $(\eta_i, \theta)$ given subject $i$'s observations $(x_i, y_i)$. Recall that the likelihood is a function of the parameters given the data but it is also the probability of observing the data given the model's parameters.

    \item $p(y \mid x, \eta, \theta)$ to denote the \textit{conditional} likelihood of $(\eta, \theta)$ given all the subjects' observations $(x, y)$. Given the hierarchical nature of pharmacometric models, this is equal to $\prod_i p(y_i \mid x_i, \eta_i, \theta)$.

    \item $p(y_i \mid x_i, \theta)$ to denote the \textit{marginal} likelihood of $\theta$ after marginalizing $\eta_i$ given subject $i$'s observations $(x_i, y_i)$. This is equal to $\int p(y_i \mid x_i, \eta_i \theta) \cdot p(\eta_i \mid \theta) d\eta_i$.

    \item $p(y \mid x, \theta)$ to denote the \textit{marginal} likelihood of $\theta$ given all the subjects' observations $(x, y)$. Given the hierarchical nature of pharmacometric models, this is equal to $\prod_i p(y_i \mid x_i, \theta)$.
\end{enumerate}

Some additional assumptions to keep in mind are that: 

\begin{enumerate}
	\item $y_i$ may not be a scalar, instead, it could and often is a subject-specific time series response or multiple such time series responses.
	\item $\eta_i$ is not generally a scalar, it can be composed of multiple subject-specific parameters with a different prior distribution assigned to each parameter.
	\item $x_i$ is not generally a scalar, it can be multiple time-independent values or a combination of time-independent values and some time series.
	      It also includes all the time points at which its corresponding $y_i$ is observed.
	\item $p(A \mid B)$ will be used in equations to denote the probability density/mass function, but in text it may be used to also refer to the distribution itself as an object/concept, depending on the context.
\end{enumerate}

\begin{figure*}
	\centering
	\includegraphics[width=0.8\textwidth]{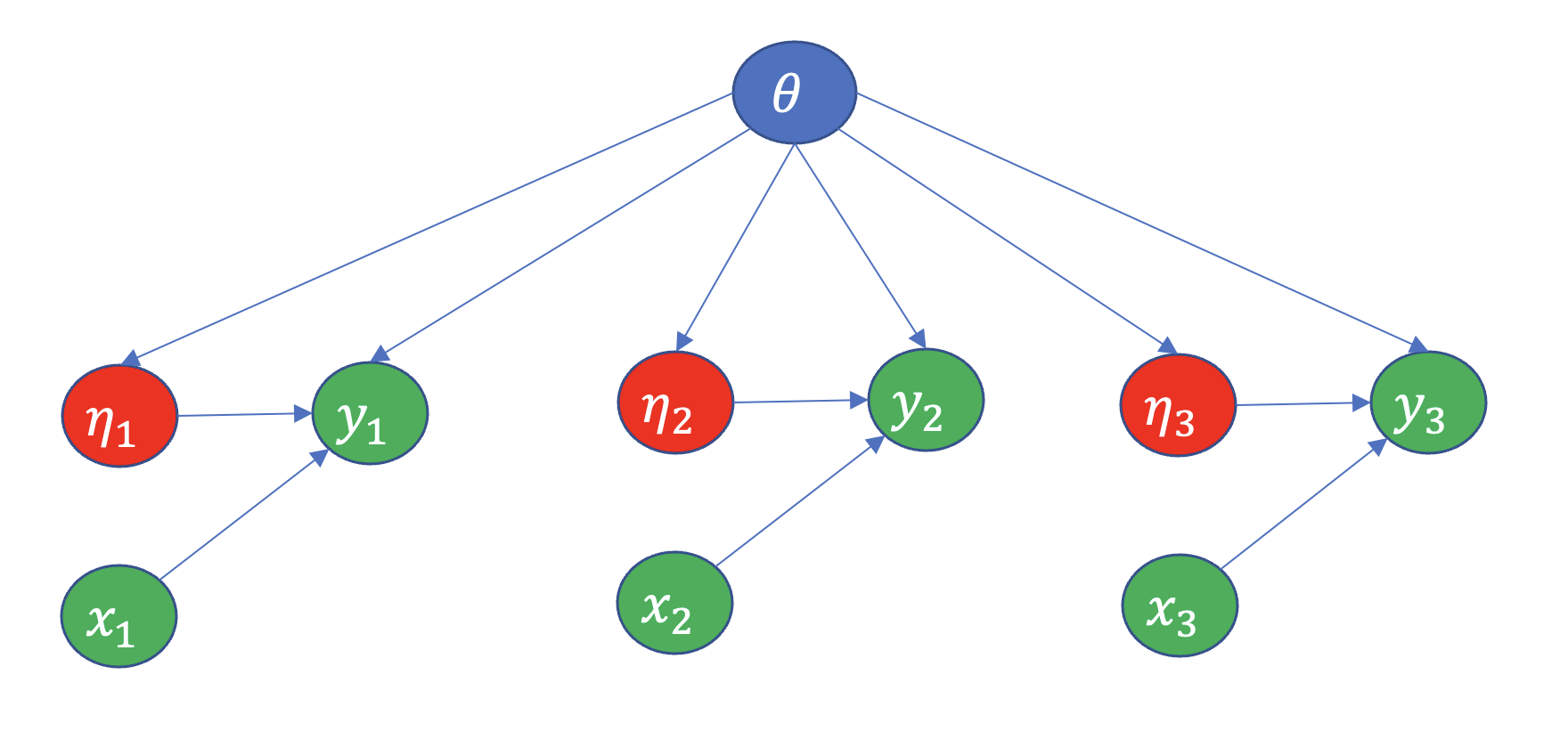}
	\caption{Schematic of the hierarchical structure of models typically used in pharmacometrics when there are only 3 subjects in the population.
		The schematic can be trivially extended to more subjects.
		See the notation section (\ref{sec:notation}) to understand the notations.
	}
	\label{fig:nlme}
\end{figure*}

\begin{figure*}
	\centering
	\includegraphics[width=0.8\textwidth]{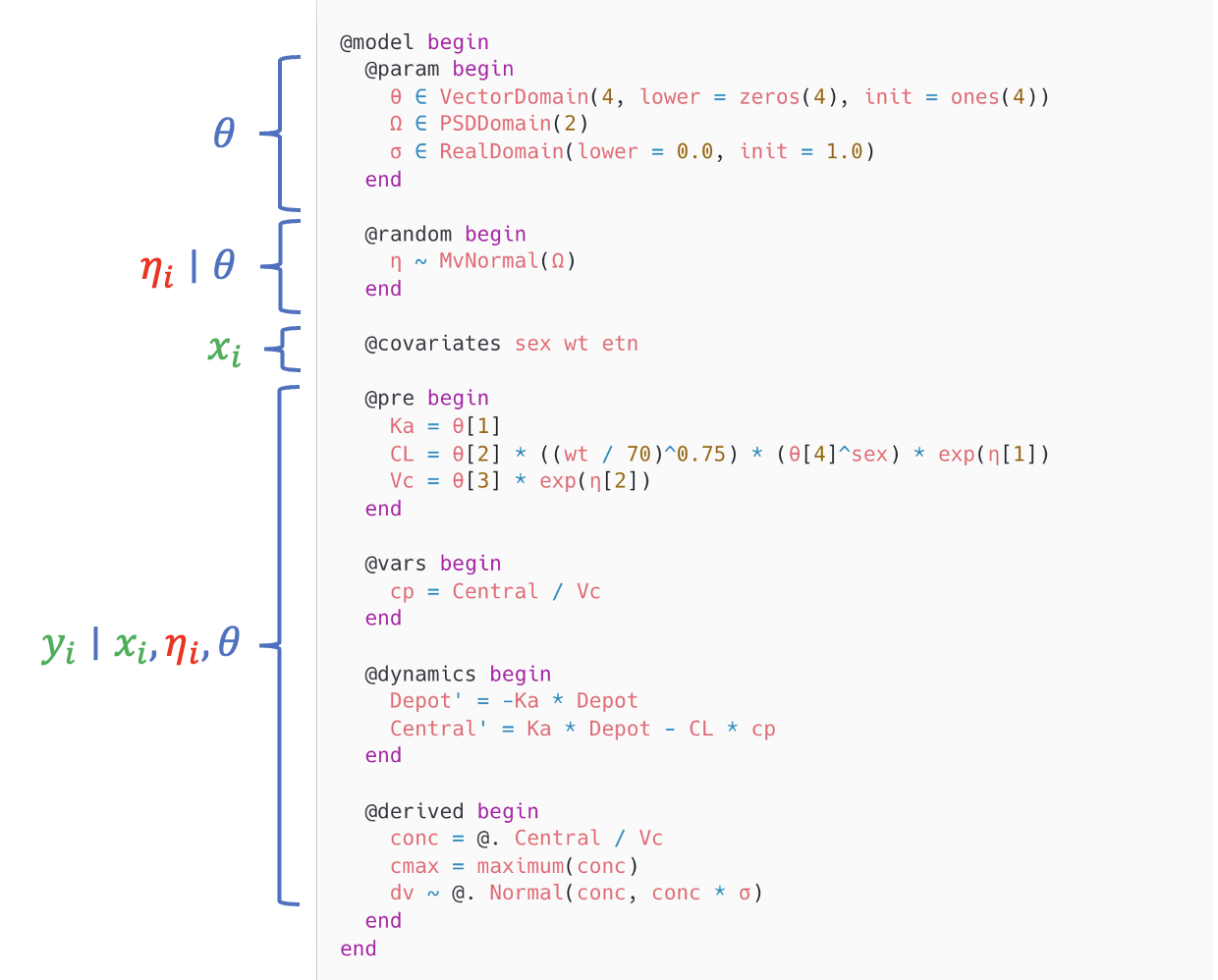}
	\caption{A dummy Pumas model showing where each variable in Figure \ref{fig:nlme} is defined.}
	\label{fig:pumas_model}
\end{figure*}

Figure \ref{fig:nlme} shows the typical model structure in pharmacometrics using the above notation when there are 3 subjects in the population.
Additionally, Figure \ref{fig:pumas_model} shows a dummy Pumas model highlighting where each variable in Figure \ref{fig:nlme} is defined.

\clearpage

\clearpage

\subsection{Bayesian Statistics} \label{sec:bayes_stats}

Bayesian Statistics is the use of \textbf{Bayes theorem} as the procedure to estimate parameters of interest or unobserved data \citep{gelman2013bayesian}.
Bayes' theorem, named after Thomas Bayes\footnote{\textbf{Thomas Bayes} (1701 - 1761) was a statistician, philosopher, and Presbyterian minister who is known for formulating a specific case of the theorem that bears his name: Bayes' theorem.
	Bayes never published what would become his most famous accomplishment; his notes were edited and published posthumously by his friend \textbf{Richard Price}.
	The theorem's official name is \textbf{Bayes-Price-Laplace}, because \textbf{Bayes} was the first to discover, \textbf{Price} got his notes, transcribed into mathematical notation, and read to the Royal Society of London, and \textbf{Laplace} independently rediscovered the theorem without having previous contact in the end of the XVIII century in France while using probability for statistical inference with census data in the Napoleonic era.
},
tells us how to ``invert'' conditional probabilities going from $p(B \mid A, C)$ to $p(A \mid B, C)$ where $C$ is optional:

\begin{align}
	p(A \mid B, C) = \frac{p(A, C) \cdot p(B \mid A, C)}{p(B, C)}
\end{align}

In the context of statistics, Bayes' rule can be used to calculate the probability that each hypothesis is true given the observations. Assume we have 10 hypotheses $H_1, \dots, H_{10}$ where each has a prior probability $p(\text{truth} = H_i)$. We can use Bayes' rule to calculate the posterior probability $p(\text{truth} = H_i \mid \text{data})$ for each hypothesis $H_i$ using:

\begin{multline}
	p(\text{truth} = H_i \mid \text{data}) = \\ \frac{p(\text{data} \mid \text{truth} = H_i) \cdot p(\text{truth} = H_i)}{p(\text{data})}
\end{multline}
where the denominator can be written as
\begin{multline*}
    p(\text{data}) = \\ \sum_{i=1}^{10} p(\text{data} \mid \text{truth} = H_i) \cdot p(\text{truth} = H_i)
\end{multline*}
which is the sum of the likelihoods of all the hypotheses, $p(\text{data} \mid \text{truth} = H_i)$, weighted by their respective prior probabilities $p(\text{truth} = H_i)$. While the denominator has a profound statistical meaning, it can also be viewed pragmatically as a normalization constant chosen such that $\sum_{i=1}^{10} p(\text{truth} = H_i \mid \text{data}) = 1$. Since the denominator is the sum of the numerator terms for all $i$, the sum of the resulting posterior probabilities is guaranteed to be 1.

$p(\text{truth} = H_i)$ describes what is commonly known as the prior probability of a hypothesis.
This can encode the modeller's domain knowledge giving unequal probabilities to different hypotheses upfront prior to observing any data.
Alternatively, assigning equal probability to each hypothesis can also be done.
Given multiple hypotheses and some data, the hypothesis with the highest probability given the observed data $p(\text{truth} = H_i \mid \text{data})$ is the most plausible one.
A hypothesis is typically a combination of a model and parameter values for the model's parameters.

In the pharmacometrics context, let each set of parameter values ($\eta$, $\theta$) given a specific model be a hypothesis. In this case, we have a continuum of hypotheses rather than a discrete set of hypotheses. Assuming a single model which we condition upon by putting it on the right-hand-side of the $\mid$, and using pharmacometrics notation, Bayes' theorem for the hypotheses continuum can be written as: 

\begin{multline}
	p(\eta, \theta \mid x, \text{model}, y) = \\ \frac{p(y \mid x, \text{model}, \eta, \theta) \cdot p(\eta, \theta \mid x, \text{model})}{p(y \mid x, \text{model})}
\end{multline}

where $A$ in the general form is replaced by $(\eta, \theta)$, $B$ is $y$ and $C$ is $(x, \text{model})$\footnote{We can alternatively make the model part of $A$ instead of $C$ when model selection is relevant but don't consider this case for simplicity.
}.
Note that we conditioned on $x$ everywhere in the above equation because we are generally not interested in modelling the probability of $x$ per se, but rather we are interested in the probability of $y$ given $x$ ($y \mid x$).

Also note that $p(\eta, \theta \mid x, \text{model})$, known as the prior probability, can be replaced by $p(\eta, \theta \mid \text{model})$ since the prior typically doesn't depend on the covariates in pharmacometrics. In Pumas syntax, this means that the covariates from the \texttt{@covariates} block are not used in the prior specification in the \texttt{@param} or \texttt{@random} blocks.

When only a single model is considered, so we can drop it from the equations, and the prior is independent of the covariates, Bayes' theorem simplifies to: 

\begin{align}
	p(\eta, \theta \mid x, y) = \frac{p(y \mid x, \eta, \theta) \cdot p(\eta, \theta)}{p(y \mid x)}
\end{align}

To further simplify the notation, we denote $(x, y)$ by $D$, $p(y \mid x)$ by $p(D)$ and $p(y \mid x, \eta, \theta)$ by $p(D \mid \eta, \theta)$\footnote{This is a slight abuse of notation because we chose to put $D$ on the left-hand-side even though it includes $x$ which is on the right-hand-side.}.
This gives us the more standard looking Bayes' theorem: 

\begin{align}
	\label{eq:bayes} p(\eta, \theta \mid D) = \frac{p(D \mid \eta, \theta) \cdot p(\eta, \theta)}{p(D)}
\end{align}

Pharmacometric models typically describe some data generating process from which we can simulate synthetic data $y$ given: 1) a set of covariates $x$, 2) a specific model, and 3) a set of parameters $(\eta, \theta)$.
Such a model describes a probability distribution for the response $y$, $p(D \mid \eta, \theta)$.
This makes computing $p(D \mid \eta, \theta)$ computationally straightforward.
Similarly, the prior probability $p(\eta, \theta)$ is typically defined in terms of standard probability distributions with known and computationally tractable probability density or mass functions.
The main issue when trying to apply Bayes' theorem in practice is the calculation of the denominator term $p(D) = p(y \mid x)$.
When all the parameters are continuous, this can be written as an integral: 

\begin{align}
	\label{eq:evidence} p(D) = \int \int p(D \mid \eta, \theta) \cdot p(\eta, \theta) \, d\eta d\theta
\end{align}

This is known as the marginal probability of the data (also known as the evidence or normalization constant) which is the weighted average of the conditional probabilities $p(D \mid \eta, \theta)$ given all possible values of $(\eta, \theta)$ weighted by their prior probabilities.
$p(D \mid \eta, \theta)$ is typically known as the conditional likelihood and $p(\eta, \theta \mid D)$ is known as the posterior probability of $(\eta, \theta)$ after observing $D$.
To better make sense of $p(D)$, it's helpful to bring back the conditioning on the model and think of $p(D \mid \text{model})$ as the marginal \textit{likelihood}\footnote{Note that in statistics in general, $p(D \mid \theta)$ is the probability of $D$ given $\theta$ and the \textit{likelihood} of $\theta$ given $D$.
} of the model after integrating out all the population and subject-specific parameters.

The computation of the high dimensional integral over $(\eta, \theta)$ is intractable in the general case.
But why do we need the posterior probability?
Often we are more interested in making predictions and using the posterior probability to weigh all the likely hypotheses when making predictions.
Assume $\hat{y}$ is either: 

\begin{enumerate}
	\item The unknown response of a new subject given the subject's known covariates $\hat{x}$, or \item The unknown partial response (e.g. at future time points) of an existing subject with a previously observed response that is part of $D$ and some known covariates $\hat{x}$.
\end{enumerate}

The covariates $\hat{x}$ include the classic pharmacometric covariates, e.g. age and weight, but also include the time points at which the response $\hat{y}$ is defined if it is a time series.
One can write the average prediction for $\hat{y}$ (using the posterior probability as weights) as:
\begin{align}
	\label{eq:expectation1} E[y \mid x = \hat{x}, D] = \int \hat{y} \times p(y = \hat{y} \mid x = \hat{x}, D) \, d{\hat{y}}
\end{align}

where $p(y = \hat{y} \mid x = \hat{x}, D)$ is defined as: 

\begin{multline} \label{eq:expectation2}
    p(y = \hat{y} \mid x = \hat{x}, D) = \\ \int \int p(y = \hat{y} \mid x = \hat{x}, \eta, \theta) \times p(\eta, \theta \mid D) \, d\eta d\theta
\end{multline}
where $p(\eta, \theta \mid D)$\footnote{To use the product rule for probabilities in the standard way, we should have used $p(\eta, \theta \mid \hat{x}, D)$ instead but $\hat{x}$ doesn't contribute to the posterior given that $\hat{y}$ is not observed yet, so the 2 terms are equal.
} is the posterior probability and $D$ refers to all the previously seen data $(x, y)$ excluding $\hat{x}$.
There are 2 problems with the above integration: 

\begin{enumerate}
	\item We cannot evaluate $p(\eta, \theta \mid D)$ using Eq \ref{eq:bayes} because computing $p(D)$ using Eq \ref{eq:evidence} requires a high dimensional integral which is computationally intractable.
	\item Even if we are able to estimate $p(D)$, computing $p(y = \hat{y} \mid x = \hat{x}, D)$ using Eq \ref{eq:expectation2} requires another high dimensional integral over $(\eta, \theta)$.
\end{enumerate}

Both problems are related to the inability to tractably compute high dimensional integrals.

When some or all of the parameters are discrete, the corresponding integrals become summations instead.
The summation is still intractable when the parameters' dimension is high because there is a combinatorial growth of the possible combinations of values of $\eta, \theta$ as their dimension increases.
In pharmacometrics, parameters are usually continuous so we focus on continuous parameters for the rest of this paper.

\subsection{Prior Selection} \label{sec:priors}

In this section, the main focus is on understanding how to choose good priors.
We will attempt to answer the following questions:
\begin{enumerate}
	\item When and how are priors useful?
	\item When and how are priors harmful?
\end{enumerate}
The specifics of which priors are available in Pumas, their parameters, and how to define them can be found in the workflow section (Section \ref{sec:workflow}).

\subsubsection{Overview}

A prior distribution over a parameter in Bayesian statistics represents the state of belief in the values of a parameter \textit{prior} to observing any data.
For instance, a univariate Gaussian prior distribution of $N(0, 2.0)$ with mean 0 and standard deviation 2, when used on a scalar parameter, means that we think this parameter has a probability mass of $\approx 99.7\%$ of being between -6 and 6.
More generally, a prior distribution with probability density function $p(x)$ means that we think the parameter has a probability mass of $\int_{a}^{b} p(x) dx$ (area under the curve) of being between $a$ and $b$.
Once data is observed, the prior distribution is \textit{updated} using the observations and the likelihood values to arrive at the \textit{posterior} distribution.
The posterior distribution represents the state of belief in the values of a parameter \textit{after} the data has been observed.
If even more data is collected, in theory we can use the old posterior as the new prior distribution in the analysis of the new data only\footnote{We can't use the data twice for the same analysis.
	If the old data was already used to define the prior in the new analysis, then only the new data should be used in the new analysis.
}.
In practice, because the posterior distribution typically doesn't have a closed form solution, it can be tricky to use it as a prior in any new analysis. Therefore, a full analysis using all of the old and new data together with the \textit{old priors} may have to be performed.
\footnote{There are algorithms for updating the posterior samples from a previous study given some new data but we don't cover these here in this paper.}

In some ways, the prior distribution is analogical to the domain of a parameter in the non-Bayesian workflow and the posterior distribution is like the maximum likelihood estimate.
Before observing data, we can only speak for the domain of the parameter since we don't know its value.
After observing data, we can fit the model to the data and identify the best fitting parameter values.
In non-Bayesian analyses, we typically choose a domain for each parameter to help the optimization algorithm reach reasonable parameter values without wasting time trying values outside of the reasonable domains.
For instance, we know that any covariance matrix parameter has to be positive definite so the optimization algorithm should never have to explore values for a covariance matrix parameter that violate this constraint.
Prior distributions generalize this idea by not only having an underlying domain implied by the distribution (known as the support of the distribution) but also by allowing the specification of differential preference for certain parameter values over others.
The ability to specify preference in parameter values is a powerful tool that can also be very dangerous if used in a wrong way.
When used right, it can allow domain knowledge and results from similar previous studies to be reused in the current study.
But when used wrong, it can be used to mask bad science and unethical behaviour using sophisticated math that is hard to scrutinize.

\subsubsection{Good, Bad and Harmless Priors}

Loosely speaking, priors can be categorized into 2 categories\footnote{There is another class of priors typically known as \textit{non-informative priors} but these will not be covered here since non-informativeness can only be enforced for a particular model parameterization.
	For instance, a non-informative prior on a standard deviation parameter in a model becomes informative when re-parameterizing the model using a variance parameter instead.
	So we will treat non-informative priors as a special case of weakly informative priors here.
	Finally, the so-called \textit{flat improper priors} were also intentionally left out because they can be trivially transformed to flat proper priors by using a domain truncation around the bounds of numbers representable on a machine using fixed precision floating point numbers.
}:
\begin{enumerate}
	\item Strong priors, i.e. informative priors, and
	\item Weak priors, i.e. weakly informative priors.
\end{enumerate}
These are loose categories because in reality, only the relative strength of a prior compared to the likelihood is significant.
Recall that the joint probability used to drive the MCMC sampling is given by:
\begin{align}
	p(D, \eta, \theta) = p(D \mid \eta, \theta) \times p(\eta, \theta)
\end{align}
which is the product of the prior probability and likelihood value.
If a lot of data exist, the likelihood term will dominate most priors and the posterior distribution will be mostly reflective of the parameter values that fit the data well.
If not enough data exist, the prior and its choice become more significant since it's possible some choices of the prior will lead it to dominate the above term.
When the prior dominates the likelihood, the posterior distribution will be only a slight shift of the prior distribution towards the parameter values that \textit{actually} fit the data well.
In these cases, the danger of abusing the prior in analyses is higher and we will discuss scenarios where that can happen in this section.

Harmless priors are priors that largely mimic the purpose of specifying parameter domains in non-Bayesian workflows.
These priors have very little to no preference between parameter values and they can be easily overpowered by even a few data points.
For instance, a uniform distribution over $[0, 10^{10}]$ for a positive parameter can be considered a harmless prior.
This prior only encodes the domain of the parameter without favouring certain values over others in this particular parameterization.

Good and bad priors share one thing in common: they are both ``\textit{informative}''.
More precisely, good priors are informative and bad priors are mis-informative.
Table \ref{tab:priors1} summarizes the various types of informative and mis-informative priors.
A good informative prior is one that has a good scientific basis and doesn't contradict the data.

\begin{table*}
	\centering
	\begin{tabular}
		{|c|c|p{0.6\linewidth}|}
		\hline
		\textbf{Scientific Basis} & \textbf{Contradicts Data?} & \textbf{Comment}                                                                                                                                                                                                        \\
		\hline
		None                      & Yes                        & This is cheating.
		Using a strong prior that contradicts and over-powers the data with no scientific basis can be used to cook any results we desire.
		For instance, a drug can be ineffective but if the strong prior says it's effective with a very high probability and not enough data exists to counter that strong wrong prior, the conclusion of the ``analysis'' will be that the drug is effective.
		\\
		\hline
		None                      & No                         & This is bad science.
		Using a strong prior that's consistent with the data but that over-powers the likelihood with no scientific basis can lead to over-confident predictions and premature conclusions using fewer data points than necessary.
		This has a similar effect to artificially replicating the data multiple times to artificially reduce the confidence interval in the non-Bayesian workflow.
		\\
		\hline
		Previous studies          & Yes                        & This is a sign that dis-similar previous studies were used to guide the prior choice and that the prior choice should be revised because it contradicts the new data and can possibly bias the results of the analysis. \\
		\hline
		Previous studies          & No                         & When used with care, results from previous studies (e.g. an approximation of its parameters' posterior distribution) can be used to guide the selection of good informative priors for a new similar study.
		These priors should not contradict the new data collected but may help us terminate the new study early using a smaller sample size than otherwise possible.
		Positively concluding a study early, after it's become clear that a drug is effective given all the information available, means that more patients can have access to \textit{truly effective} drugs earlier.
		This is especially important for rare disease drug development where collecting more data in a study often means adding years to the clinical trial duration.
		This is a use case that requires more regulatory and industry agreement on best practices for defining informative prior distributions in such studies with the goal of benefiting the patients.
		\\
		\hline
	\end{tabular}
	\caption{The table shows a description of a number of ways to choose \textbf{informative prior distributions}.
		Only the last case is a good use of informative priors.
		\textbf{This table is only applicable to informative priors} that may dominate the likelihood, since weakly informative priors that are dominated by the likelihood typically don't matter as much.} \label{tab:priors1}
\end{table*}

Given Table \ref{tab:priors1}, there are 2 general ways to define sound priors:
\begin{enumerate}
	\item Define a weakly informative prior\footnote{
        Very weak priors are sometimes called diffuse priors.
    } and ensure that the conclusion of the study does not change in the limit as the prior gets weaker and weaker.
	      This is a case where we are letting the data speak for itself without imposing our own bias on the results.
	      In this case, we are only using Bayesian inference for its probabilistic soundness and ability to quantify the total uncertainty in the parameters and response even when the model is non-identifiable and/or we only have a few data points.
	\item Use similar previous studies to guide the prior choice and test that it doesn't contradict the new data.
	      In this case, more justification of the prior distribution is necessary and a proof that the prior doesn't contradict the data is required.
\end{enumerate}
One way to show that the prior doesn't contradict the data is to start with no observations at all and with a weakened version of the strong prior, e.g. by increasing the standard deviation.
You can then incrementally make the prior stronger again (e.g. by decreasing the standard deviation back) until it reaches the desired level of strength, followed by incrementally adding the (previously removed) observations back to the analysis.
If doing so and re-running the analysis at every increment shows a consistent trend in all of the significant statistics (e.g. the probability that the drug is effective), then the prior's strength is aligned with the story told by the data.
This can be done using a sequence of prior simulations (when all the data is removed) followed by a combination of MCMC runs and posterior simulations (when the data is added back).
For more methods for detecting prior-data conflicts and model mis-informativeness, the readers are referred to \cite{prior_sens}.

Another simple way to detect the potential conflict between data and informative priors is to simulate from the following distribution:
\begin{align}
	\begin{split}
		(\eta, \theta) & \sim p(\eta, \theta) \\ y & \sim p(y \mid \eta, \theta)
	\end{split}
\end{align}
where $p(\eta, \theta)$ is the prior distribution.
You can then do a simulation or VPC plot checking the consistency of the data and prior simulations.
If the prior is weakly informative or nearly non-informative and we have a lot of data such that the likelihood dominates the prior, prior simulations that are inconsistent with the data in a VPC plot can be ignored.
However if the prior is informative, it is based on previous studies and there are not enough data points in the new study to fully dominate the prior, the VPC plot of the prior simulations next to the data can be a useful diagnostic to inspect. Refer to Section \ref{sec:prior_sims_and_preds} for how to do this in Pumas.

When using a weakly informative prior, you may be inclined to use such simulation plots to select a good prior with a good (but not necessarily tight) coverage of the data.
In general, any fine-tuning of the prior based on the data is frowned upon.
This is because we would then be using the same data twice, once to fine-tune the prior and once to update the prior to get the posterior.
This can result in under-estimating the posterior's variance, i.e. over-confident posteriors, which in turn leads to over-confident posterior predictions.
This is analogical to replicating some or all of the data points in your data in a frequentist workflow which under-estimates the confidence intervals and standard errors.

In cases where sampling fails due to numerical errors, an overly weak prior may be a cause.
In this case, one may try changing the prior, e.g. truncating its support to reasonable bounds, to be more consistent with the data.
However, only minimal such changes can be allowed in the final analysis and a good post-sampling sensitivity analysis study would be needed to ensure that the conclusion of the study is not locally sensitive to the prior.
More generally, numerical errors are often a sign that the model is too sensitive to some of the parameter values which may imply a structural problem in the model itself.
In this case, one should also consider better fixes than simply truncating the priors, e.g. by simplifying the model or re-parameterizing it to avoid numerical instability.
The Bayesian workflow paper by \cite{bayesian_workflow} has excellent recommendations and case studies to take heed from when diagnosing failing MCMC runs.

\subsubsection{Correlation vs Covariance}
\label{sec:corvscov}

When defining models that have a covariance matrix parameter (e.g. the covariance parameter of the multivariate normal prior distribution typically used for subject-specific parameters in pharmacometrics), one is always faced with the following 2 equivalent parameterizations:
\begin{enumerate}
	\item Use a covariance matrix parameter $\Omega$, or \item Use a vector of standard deviations $\omega$ and a correlation matrix parameter $C$.
\end{enumerate}
One can easily recover $\Omega$ from $(\omega, C)$ and vice versa.
Let $D_{\omega}$ be the diagonal matrix whose elements are $\omega$, $\omega_i$ be the $i^{th}$ element in $\omega$, and $\Omega[i,i]$ be the $i^{th}$ diagonal element of $\Omega$.
The relationships between $C$, $\omega$ and $\Omega$ is given by:
\begin{align}
	\begin{split}
		\Omega & = D_{\omega} \times C \times D_{\omega} \\ C & = D_{\omega}^{-1} \times \Omega \times D_{\omega}^{-1} \\ \omega_i & = \sqrt{\Omega[i,i]}
	\end{split}
\end{align}
In the non-Bayesian context, the 2 parameterizations are equivalent.
However in Bayesian analysis, one should define a prior distribution on the parameters.
Since prior distributions are supposed to encode the domain knowledge and state of belief about the values of the parameters, using more intuitive/interpretable parameterizations is generally recommended to better make sense of the prior distributions used.
For this reason, some people prefer to use the second parameterization with separate standard deviations vector and a correlation matrix since they are more interpretable.
We saw examples of prior distributions that can be used for standard deviation, correlation and covariance matrix parameters in Section \ref{sec:workflow}.

\subsection{Markov Chain Monte Carlo (MCMC) Intuition} \label{sec:mcmc_intuition}

In this section, the use of MCMC will be motivated showing how MCMC can be used to bypass the need for high dimensional integrals (discussed in Section \ref{sec:bayes_stats}) for all practical purposes.

\subsubsection{Inference}

Markov Chain Monte Carlo (MCMC) \citep{brooksHandbookMarkovChain2011} bypasses the need to solve the numerical integration problem by sampling\footnote{A sample of a random variable or a sample from its distribution is an instantiation of said random variable. For instance, $[0, 1, 0, 0, 1, 1]$ are 6 samples from a Bernoulli distribution whose support is the set $\{0, 1\}$. The support of a distribution is the domain of the random variable. For example, the set $\{0, 1\}$ is the support of the Bernoulli distribution and the set of all real numbers is the support of the normal distribution. A sample from the posterior distribution can be interpreted as the likely parameter values that could have generated the data that we observed. The term \textit{sample} can sometimes be used to refer to multiple such samples, to be understood from the context.} from the posterior probability distribution $p(\eta, \theta \mid D)$ directly using only the tractable numerator in Eq \ref{eq:bayes}.
This numerator is sometimes called the \textit{joint} probability since it is just $p(D, \eta, \theta) = p(D \mid \eta, \theta) \cdot p(\eta, \theta)$. Note the difference between the terms with and without conditioning $\mid$.
Having samples from the posterior allows us to estimate quantities such as: 

\begin{align}
	p(f(\eta, \theta) > 0 \mid D)
\end{align}

for an arbitrary function $f$ using a simple count by checking all the samples from the posterior and counting how many satisfy the particular conditions of interest.
The ratio of samples satisfying the condition $f(\eta, \theta) > 0$ is the unbiased estimate of the above probability.
More concretely, this probability could be $p(\theta_i > 0 \mid D)$ where $\theta_i$ corresponds to the effect size of an experiment treatment arm compared to the placebo arm.

\subsubsection{Prediction}

Besides estimating probabilities of events using the posterior samples, the posterior samples can also be used to make predictions.
The functions for performing posterior predictions in Pumas were presented in Section \ref{sec:predictivechecks}.
Assume we have $N$ samples from the posterior $p(\eta, \theta \mid D)$: 

\begin{align*}
	\{(\eta^{(j)}, \theta^{(j)}): j \in 1 \dots N\}
\end{align*}

Recall the intractable term in Eq \ref{eq:expectation1} was $p(y = \hat{y} \mid x = \hat{x}, D)$ which can be written as:

\begin{multline*}
  \int \int p(y = \hat{y} \mid x = \hat{x}, \eta, \theta) \times p(\eta, \theta \mid D) \, d\eta d\theta \approx \\ \frac{1}{N} \sum_{j = 1}^N p(y = \hat{y} \mid x = \hat{x}, \eta = \eta^{(j)}, \theta = \theta^{(j)})
\end{multline*}
where $p(y = \hat{y} \mid x = \hat{x}, \eta = \eta^{(j)}, \theta = \theta^{(j)})$ is the conditional probability of $\hat{y} \mid \hat{x}$ evaluated using the given parameter values.

Using the samples, the intractable integral was therefore reduced to a tractable average of $N$ terms since we can easily evaluate $p(\hat{y} \mid \hat{x}, \eta^{(j)}, \theta^{(j)})$ given $(\hat{x}, \hat{y}, \eta^{(j)}, \theta^{(j)})$ and the model.
The expectation term in Eq \ref{eq:expectation1} can therefore be approximated by Eq \ref{eq:expectation3},
\begin{figure*}
\begin{align} \label{eq:expectation3}
    E[y \mid x = \hat{x}, D] & = \int \hat{y} \times p(y = \hat{y} \mid x = \hat{x}, D) \, d{\hat{y}} \nonumber \\
    & \approx \int \hat{y} \times \Bigg( \frac{1}{N} \sum_{j = 1}^N p(y = \hat{y} \mid x = \hat{x}, \eta = \eta^{(j)}, \theta = \theta^{(j)}) \Bigg) d\hat{y} \\
    & = \frac{1}{N} \sum_{j = 1}^N \Bigg( \int \hat{y} \times p(y = \hat{y} \mid x = \hat{x}, \eta = \eta^{(j)}, \theta = \theta^{(j)}) d\hat{y} \Bigg) \nonumber \\
    & = \frac{1}{N} \sum_{j = 1}^N E[y \mid x = \hat{x}, \eta = \eta^{(j)}, \theta = \theta^{(j)}] \nonumber
\end{align}
\end{figure*}
where $E[y \mid x = \hat{x}, \eta^{(j)}, \theta^{(j)}]$ is just the mean value of the conditional distribution $p(y = \hat{y} \mid x = \hat{x}, \eta = \eta^{(j)}, \theta = \theta^{(j)})$ which can be evaluated from a single run of the model. In Pumas syntax, this is the output of the \texttt{predict} function (or the output of \texttt{simobs} with \texttt{simulate\_error = false}) with parameters $(\eta^{(j)}, \theta^{(j)})$, and covariates $\hat{x}$.

More generally, one can estimate the expectation of an \textbf{arbitrary function $g(\eta, \theta)$} with respect to the posterior distribution using:
\begin{align*}
	E[g(\eta, \theta) \mid D] & = \int \int g(\eta, \theta) \times p(\eta, \theta \mid D) \, d\eta d\theta \\  & \approx \frac{1}{N} \sum_{j = 1}^N g(\eta^{(j)}, \theta^{(j)})
\end{align*}
For instance, $g$ could be computing some NCA parameters (Section \ref{sec:nca}) based on the model's prediction or computing any other deterministic quantity that is a deterministic function of the parameters.

When defining $g$ as $g(\eta', \theta') = E[y \mid x = \hat{x}, \eta = \eta', \theta = \theta']$, we recover the prediction special case.
And when defining $g$ as $g(\eta, \theta) = \mathbbm{1}_{f(\eta, \theta) > 0}$ for another arbitrary function $f$, where $\mathbbm{1}_{f(\eta, \theta) > 0}$ is the indicator function that is 1 when the condition $f(\eta, \theta) > 0$ is satisfied and 0 otherwise, we recover the special case of estimating the probability $p(f(\eta, \theta) > 0 \mid D)$.

In other words, samples from the posterior are almost everything you may ever need to estimate all the quantities of interest needed to make decisions.
So how do we obtain such samples without computing the intractable integrals?
We use MCMC.

\subsubsection{Simulation}

We saw how given $N$ samples from the posterior, we can compute the average prediction $E[g(\eta, \theta) \mid D] \approx \frac{1}{N} \sum_{j = 1}^N g(\eta^{(j)}, \theta^{(j)})$ for any particular choice of the function $g$.
Alternatively, you can also obtain a distribution of predictions: 
\begin{align}
	\{ g(\eta^{(j)}, \theta^{(j)}) \text{ for } j \in 1 \dots N\}
\end{align}
where $g(\eta', \theta') = E[y \mid x = \hat{x}, \eta = \eta', \theta = \theta']$.
This is the MCMC approximation of the distribution of $g(\eta, \theta)$ where $(\eta, \theta) \sim p(\eta, \theta \mid D)$.
For the above choice of $g$, this distribution of predictions is typically known as the posterior predictive distribution.
When $(\eta, \theta)$ are sampled from the prior instead, the distribution of $g(\eta, \theta)$, for the above choice of $g$, is known as the prior predictive distribution.

Beside sampling predictions or more generally deterministic functions of the parameters $(\eta, \theta)$, one may also sample from the following distribution of $\hat{y}$: 

\begin{align*}
  (\eta, \theta) & \sim p(\eta, \theta \mid D) \\ \hat{y} & \sim p(y = \hat{y} \mid x = \hat{x}, \eta, \theta)
\end{align*}

In Pumas syntax, this is the output of the \texttt{simobs} function using the posterior parameter values and covariates $\hat{x}$.
Alternatively, $(\eta, \theta)$ may be sampled from their prior distributions instead or just fixed to particular \textit{ground truth} values.
These prior/posterior/ground truth simulations can be used to do any of the following: 

\begin{enumerate}
	\item Generate synthetic data to test the MCMC algorithm on synthetic data before using the real data\footnote{This is known as ``simulation-based calibration''.
		      A good reference is \cite{Talts_Betancourt_Simpson_Vehtari_Gelman_2020}.
	      }.
	\item Identify extremely poor choices of priors to minimally guide the selection of priors by inspecting the similarity of the prior simulations and real data, e.g. using a visual predictive check (VPC) plot, also known as prior predictive check.
	      See section \ref{sec:priors} for more details on prior selection.
	\item Quantify the quality of model fit by comparing posterior simulations to the real data using a VPC plot, also known as posterior predictive check, and estimating the so-called Bayesian $p$-value.
\end{enumerate}

The code for doing prior simulations and predictions in Pumas was presented in Section \ref{sec:prior_sims_and_preds}. Similarly, the code for doing posterior simulations and predictions was presented in Section \ref{sec:predictivechecks}. Finally, the codes for performing VPC, various simulation queries, e.g. the Bayesian $p$-value, and NCA were presented in Sections \ref{sec:vpc}, \ref{sec:posterior_queries} and \ref{sec:nca} respectively.

\subsection{No-U-Turn Sampler (NUTS) Algorithm} \label{sec:nuts}

In this section, an intuitive explanation of the No-U-Turn Sampler (NUTS) \citep{hoffman2014no,multlinomial} MCMC algorithm will be given.
The focus of the explanation will be to develop a strong intuition for how the algorithm works and how to tune its hyper-parameters\footnote{The term \textit{hyper-parameters} generally refers to any parameters that are not being inferred by the Bayesian inference algorithm and that need to be pre-specified by the user before the Bayesian analysis. These can generally be model hyper-parameters, e.g. the parameters of the prior distributions on the population 
 parameters $\theta$, or they can be algorithm hyper-parameters such as the settings of the NUTS algorithm.}.
We don't want to clutter the minds of the readers with equations that can be found in numerous other resources and which are not strictly necessary to be able to effectively use the algorithm.
For MCMC beginners or when reading this for the first time, you may skip subsections \ref{sec:uturns}, \ref{sec:velocity} and \ref{sec:hpriors} without a significant loss of context.

\subsubsection{High Level Description}

\begin{figure*}
	\centering
	\begin{tikzpicture}
		\begin{axis}[every axis plot, line width=2pt,
				ylabel=PDF,
				domain=-4:4,samples=200,
				ymax = 0.6, ytick={0, 0.2, 0.4},
				axis x line*=bottom, 
				axis y line*=left, 
				enlargelimits=true,
			] 

			\addplot [blue] {gaussian(0, 1)};
			\node[inner sep=0pt] (hikerlower) at (-2,0.13){\Strichmaxerl[2pt]};
			\node[inner sep=0pt] (hikerupper) at (0,0.5){\Strichmaxerl[2pt]};
			\node[inner sep=0pt] (hikerlower2) at (2,0.13){\Strichmaxerl[2pt]};
			\draw[->, green, line width=2pt] (hikerlower) to [out=90,in=135] node[above left] {\large$P=1$} (hikerupper);
			\draw[->, red, line width=2pt] (hikerupper) to [out=45,in=135] node[right] {\large$P\approx\frac{0.1}{0.4}\approx\frac{1}{4}$} (hikerlower2);
		\end{axis}
	\end{tikzpicture}
	\caption{Random walk visualization for a normally distributed random variable (on the x-axis) where the probability density function (PDF) is shown on the y-axis.
		Firstly a proposal is made, then it is accepted with a specific acceptance probability.
		When the proposal is for the hiker to climb up, the move is accepted with a probability of 1.
		When the proposal is for the hiker to climb down, it is accepted with a probability equal to the ratio of PDFs at the 2 positions.
	} \label{fig:random_walk}
\end{figure*}
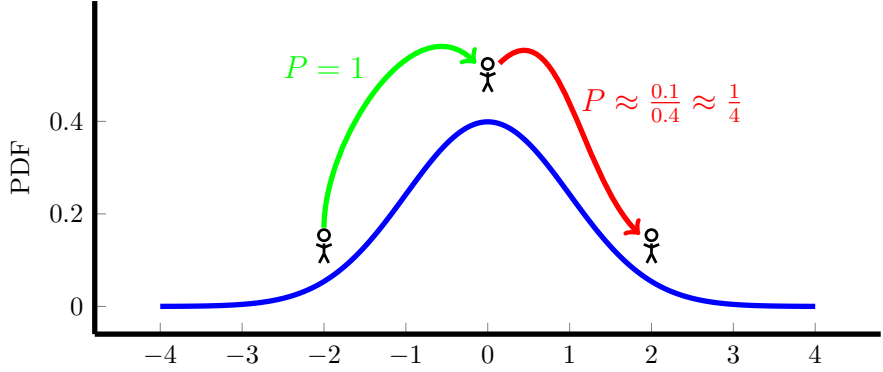

MCMC sampling uses a stochastic process (random walk) in the $(\eta, \theta)$ space to collect samples from the posterior $p(\eta, \theta \mid D)$ in an iterative manner using nothing but ``local information available''.
Figure \ref{fig:random_walk} shows a simple random walk for a 1-dimensional Gaussian random variable.
In the $j^{th}$ iteration of the algorithm, the local information available is basically the numerator in Eq \ref{eq:bayes} evaluated at a particular value $(\eta = \eta^{j-1}, \theta = \theta^{j-1})$ and its gradient\footnote{Not all MCMC algorithms use the gradient but the ones that scale to many parameters do.
} with respect to $(\eta, \theta)$ both of which can be computed easily.
Given that this iterative algorithm only uses information from the previous iteration $j - 1$, it is a so-called Markov chain by definition.
The goal of the MCMC family of algorithms is to make it such that the individual steps $\{(\eta^{(j)}, \theta^{(j)}): j \in 1 \dots N\}$ are valid samples from the posterior.
In this section, we focus on the so-called No-U-Turn sampler (NUTS) algorithm \citep{hoffman2014no,multlinomial} which is a variant of the Hamiltonian Monte Carlo (HMC) \citep{neal2011mcmc} family of MCMC algorithms.
We will not cover these algorithms in details but we will explain the intuition behind them for you to make sense of their hyper-parameters to be able to informatively tinker with them when needed.

Imagine the current position of the sampler $(\eta^{(j - 1)}, \theta^{(j - 1)})$ is a particle in the $(\eta, \theta)$ space.
In the NUTS algorithm, the random walk process goes like this\footnote{The real algorithm includes many performance enhancements which are not discussed here.
}:
\begin{enumerate}
    \item The imaginary particle $(\eta^{(j - 1)}, \theta^{(j - 1)})$ is given a random imaginary speed in a random direction, i.e. it is given a random imaginary velocity. The velocity is sampled from a multivariate Gaussian distribution.

    \item The gradient of the log of the joint probability (log of the numerator in Eq \ref{eq:bayes}) with respect to $(\eta, \theta)$ acts as an imaginary force field locally pushing the imaginary particle towards regions of high (log) prior and/or (log) likelihood and away from regions of low (log) prior and/or (log) likelihood.

    \item The imaginary particle's motion is \textbf{approximately} simulated using time discretization and an approximate ODE solver\footnote{
        The particle dynamics simulation boils down to simulating a set of ODEs with the gradient of $\log p(D, \eta, \theta)$ as the driving force. An approximate ODE solver called the leapfrog method is used to do the simulation. The leapfrog method with a large step size is approximate because its solution violates the law of conservation of energy, even though it is so-called volume preserving. However, this is a desirable property in this case and can help fully explore the posterior even with disconnected regions of high probability mass.
        For the effect of the time step size on the sampling behaviour, see Section \ref{sec:time_step}.
    } for a total of $T$ simulated time steps\footnote{
        In the NUTS algorithm, the $T$ time steps are randomly split between forward time steps and backward time steps, such that the total number of forward and backward time steps taken is $T$. Backward time steps are equivalent to forward time steps after the initial velocity has been multiplied by $-1$.
    } under the influence of the imaginary force field, where each simulated time step is of size $\epsilon$.
    This simulation only requires the initial position, initial velocity and being able to calculate the force applied at any arbitrary point $(\eta, \theta)$ which is equivalent to evaluating the gradient of $\log p(D, \eta, \theta)$ with respect to $(\eta, \theta)$, i.e. $\frac{d\log p(D, \eta, \theta)}{d(\eta, \theta)}$.

    \item A randomly chosen position $(\eta, \theta)$ on the simulated imaginary trajectory of $T$ time steps becomes the \textbf{proposal}.

    \item The proposal is then accepted with a carefully chosen probability to ensure that the random walk gives us correct samples from the posterior. If the proposal is accepted, it becomes the next sample $(\eta^{(j)}, \theta^{(j)})$, otherwise the previous value is sampled once more, i.e. $(\eta^{(j)}, \theta^{(j)}) = (\eta^{(j - 1)}, \theta^{(j - 1)})$.\footnote{In the state-of-the-art variant of NUTS \citep{multlinomial}, the proposal selection and acceptance/rejection steps are combined into a single step which samples from the imaginary trajectory, which includes the previous particle's position $(\eta^{(j - 1)}, \theta^{(j - 1)})$. Sampling is done in a way that ensures the chosen next position $(\eta^{j}, \theta^{j})$ is a correct sample from the posterior. However, we chose to separate the 2 steps conceptually to aid with the explanation.}
\end{enumerate}

The above algorithm is still a random walk but it is biased towards high probability regions because it uses the gradient of the log joint probability to push the particle in the right direction even if it started with a random velocity.
The above sampling algorithm is done in 2 phases: an adaptation phase and a sampling phase.
In the adaptation phase, the sampler is adapting its time step size and initial velocity distribution while performing the above sampling procedure.
In the sampling phase, the algorithm's hyper-parameters seize to adapt and sampling continues using the same procedure above.
It is generally recommended to discard the adaptation steps after sampling as \textit{burn-in}\footnote{Also called \textit{warm-up}.
} as they may not be representative samples from the posterior.
In Pumas, this can be done using the \texttt{discard} function as explained in Section \ref{sec:fit}. The number of adaptation steps can also be specified using the \texttt{nadapts} option as shown in Section \ref{sec:fit}.

\subsubsection{MCMC Visualization}

To interactively visualize how an MCMC sampler works, \href{chi-feng.github.io/mcmc-demo/}{https://chi-feng.github.io/mcmc-demo/} is a great resource to look at where you can change the target distribution and sampling algorithm to develop an intuition for how MCMC samplers work in different cases. You can select the "No-U-Turn Sampler" as the algorithm and then change the target distribution to visualize how NUTS behaves when sampling from different posterior distributions.

\subsubsection{Proposal Acceptance}

While the state-of-the-art variant of NUTS \citep{multlinomial} does not use an explicit acceptance/rejection test (also known as Metropolis-Hastings test) of a proposal, what it does is analogical to a traditional proposal selection followed by an acceptance/rejection test. For pedagogical reasons, we assume these are 2 separate steps. The acceptance probability of a proposal in step 5 of the algorithm depends on:
\begin{enumerate}
	\item The prior probability, and \item The likelihood function (how well the proposal fits the data)
\end{enumerate}
A proposal leading to bad predictions that don't fit the data well compared to the previous sample $(\eta^{(j - 1)}, \theta^{(j - 1)})$, or a proposal that is more improbable according to the prior compared to the previous sample is more likely to be rejected.
On the other hand, a proposal that fits the data better than the previous sample and/or is more probable according to the prior will be more likely to be accepted.

\subsubsection{Effect of Time Step Size} \label{sec:time_step}

Generally speaking, the larger the time step size in the simulation, the more approximate the ODE solver is and the more exploratory/adventurous the proposals will be which lead to a lower ratio of accepted proposals.
On the other hand, smaller step sizes generally lead to less exploratory proposals which are more likely to be accepted increasing the acceptance ratio.
The sampler's exploration ability partly comes from the ability of the approximate ODE solver to over-shoot and jump from one area of high posterior probability\footnote{High posterior probability regions have high joint probability values (the numerator in Eq \ref{eq:bayes}).
	The joint probability is the product of the prior probability and likelihood.
	So parameters values with a high prior probability and/or high likelihood will have high joint and posterior probabilities.
} to another when making proposals, thus exploring multiple modes even if there is a 0 probability region between the 2 modes.
A zero probability $p(\eta, \theta \mid D)$ implies zero probability $p(D, \eta, \theta)$ (Eq \ref{eq:bayes}) which will result in an infinite force pushing the particle away from that region.
Therefore exact simulation will never be able to make the jump across such a region, hence the need for approximate solvers and over-shooting.

\subsubsection{Time Step Size Adaptation and Target Acceptance Ratio} \label{sec:exploration}

In the NUTS algorithm, you don't set the step size yourself.
The NUTS algorithm adapts its step size to encourage a certain fraction of the proposals to get accepted on average.
This target acceptance ratio is a hyper-parameter of NUTS.
In Pumas, you can set the target acceptance ratio using the \texttt{target\_accept} option as shown in Section \ref{sec:fit}.
A value of 0.99 means that we want to accept 99\% of the proposals the sampler makes.
This will generally lead to a small distance between the proposal and the current sample since this increases the chance of accepting such a proposal.
On the other hand, a target acceptance fraction of 0.1 means that we want to only accept 10\% of the proposals made on average.
The NUTS algorithm will therefore attempt larger step sizes to ensure it rejects 90\% of the proposals.
In general, a target acceptance ratio of 0.6-0.8 is recommended to use.
The default value used in Pumas is 0.8.

In sampling, there is usually a trade-off between exploration and exploitation.
If the sampler is too adventurous, trying aggressive proposals that are far from the previous sample in each step, the sampler would be more likely to explore the full posterior and not get stuck sampling near a local mode of the posterior.
However on the flip side, too much exploration will often lead to many proposal rejections due to the low joint probability $p(D, \eta, \theta)$ of the data and the adventurous proposals.
This can decrease the ratio of the effective sample size (ESS)\footnote{The ESS is an approximation of the ``number of independent samples'' generated by a Markov chain, when estimating the posterior mean. A low ESS per sample ratio is caused by high auto-correlation in the MCMC samples and is often a bad indicator.} to the total number of samples (also known as relative ESS) since a large number of samples will be mere copies of each other due to rejections.

On the other hand if we do less exploration, there are at least 2 possible scenarios:
\begin{enumerate}
	\item The first scenario is if we initialize the sampler from a mode of the posterior.
	      Making proposals only near the previous sample will ensure that we accept most of the samples since proposals near a mode of the posterior are likely to be good parameter values.
	      This local sampling behavior around known good parameter values is what we call here exploitation.
	      While the samples generated via high exploitation around a mode may not be representative of the whole posterior distribution, they might still give a satisfactory approximation of the posterior predictive distributions, which is to be judged with a VPC plot.
	\item The second scenario is if we initialize the sampler from bad parameter values.
	      Bad parameter values and low exploration often lead to optimization-like behavior where the sampler spends a considerable number of iterations moving towards a mode in a noisy fashion.
	      This optimization-like, mode-seeking behavior causes a high auto-correlation in the samples since the sampler is mostly moving in the same direction (towards the mode).
	      A high auto-correlation means a low ESS because the samples would be less independent from each other. \footnote{In Pumas, the ESS values of the population parameters are displayed in the default display of the MCMC result as shown in Figure \ref{fig:fit_result}.}
	      Also until the sampler reaches parameter values that actually fit the data well, it's unlikely these samples will lead to a good posterior predictive distribution.
	      This is a fairly common failure mode of MCMC algorithms when the adaptation algorithm fails to find a good step size that properly explores the posterior distribution due to bad initial parameters and the model being too complicated and difficult to optimize, let alone sample from its posterior.
	      In this case, all the samples may look auto-correlated and the step sizes between samples will likely be very small (low exploration).
            In Pumas, the step size is displayed as part of the live progress information during sampling as shown in Figure \ref{fig:live_progress}. 
	      It's often helpful to detect such a failure mode early in the sampling and kill the sampling early.
\end{enumerate}

\subsubsection{Optional: Number of Time Steps and U-Turns} \label{sec:uturns}

Consider a single direction in the $(\eta, \theta)$ space, e.g. the axis of a particular parameter.
For relatively flat regions of the posterior where a lot of the values along this direction are almost equally likely, i.e. they all the fit the data to the same level and are almost equally probable according to the prior, proposals far away from the current sample may still be accepted most of the time.
This is especially likely in the parts of the posterior where the model is (almost) non-identifiable causing high parameter correlations, and the prior is indiscriminate (e.g. due to being a weak prior).
On the other hand, regions of the posterior that are heavily concentrated around a mode with a high curvature often require a smaller step size to achieve reasonable acceptance ratios, since proposals that are even slightly far from the current sample may be extremely improbable according to the prior or may lead to very bad predictions.
This is especially likely in regions of the posterior where the model is highly sensitive to the parameter values or if the prior is too strongly concentrated around specific parameter values.

To account for such variation in curvature along the \textit{same direction}\footnote{Different curvatures along different directions is accounted for using the distribution of the initial velocity that the particle is given at the beginning of each NUTS iteration.
} in different regions of the posterior, the NUTS algorithm uses a multi-step proposal mechanism with a fixed time step size (determined during the adaptation phase and then fixed) and a dynamic number of time steps (dynamic in both the adaptation and sampling phases).
More specifically, the sampler simulates a trajectory of $T$ time steps before choosing a proposal randomly from this trajectory, where $T$ is different for each proposal made.

\begin{table*}
	\centering
	\begin{tabular}{|c|c|c|}
		\hline
		Increment $j$ & Simulation Direction & Interval of the Time Steps Simulated after $j$ Increments \\
		\hline
		0             & -                    & $[0, 0]$                                                  \\
		1             & Forward              & $[0, 0 + 1] = [0, 1]$                                     \\
		2             & Forward              & $[0, 0 + 1 + 2] = [0, 3]$                                 \\
		3             & Reverse              & $[0 - 4, 0 + 1 + 2] = [-4, 3]$                            \\
		4             & Reverse              & $[0 - 4 - 8, 0 + 1 + 2] = [-12, 3]$                       \\
		5             & Forward              & $[0 - 4 - 8, 0 + 1 + 2 + 16] = [-12, 19]$                 \\
		6             & Reverse              & $[0 - 4 - 8 - 32, 0 + 1 + 2 + 16] = [-44, 19]$            \\
		\hline
	\end{tabular}
	\caption{The table shows the incremental simulations of the NUTS algorithm for $j \in [1, 6]$.
	Notice how an increasing power of 2 is added to the positive direction or subtracted from the negative direction in each increment.
	The total number of time steps made after increment $j$ (excluding the initial time point $t = 0$) is $1 + 2 + 4 + 8 + 16 + \dots = 2^0 + 2^1 + 2^2 + \dots + 2^{j - 1}= \sum_{i = 0}^{j - 1} 2^{i} = 2^{j} - 1$.
				\textbf{Check}: $2^1 - 1 = 2^0 = 1$, $2^2 - 1 = 2^0 + 2^1 = 3$, $2^3 - 1 = 2^0 + 2^1 + 2^2 = 7$, etc.
				Note that the intervals above are of the number of time steps.
				Each time step has a simulated time step size of $\epsilon$.
		} \label{tab:nuts}
\end{table*}

The number of time steps $T$ simulated by NUTS is determined by an incremental simulation of: $T = 1 + 2 + 4 + 8 + 16 + \dots$ time steps where the number of time steps in each incremental simulation is an increasing power of 2.
Each incremental simulation can be either:
\begin{enumerate}
	\item Forward in time starting from the future-most state, or \item Backward in time starting from the past-most state.
\end{enumerate}
The direction of each incremental simulation is sampled randomly with 0.5 probability assigned to each direction.
Table \ref{tab:nuts} shows an example of the incremental simulations for the particular choice of simulation directions: [Forward, Forward, Reverse, Reverse, Forward, Reverse].

So when do we stop simulating?
The NUTS algorithm typically stops simulating when one of the following 4 conditions is met:
\begin{enumerate}
    \item It finds a so-called U-turn, that is when the sampler begins to move back towards one end of the trajectory from the other end.
    \item It reaches a pre-set maximum number of simulation steps.
    \item The log prior probability and/or log likelihood drops rapidly in one of the steps, dropping by more than a pre-set threshold.
    \item A numerical error occurs.
\end{enumerate}
The third and fourth termination criteria are often called ``divergence''.

After the simulation terminates, a point randomly chosen on the trajectory simulated becomes the next proposal and the search is terminated.
\textit{Terminating by finding a U-turn is typically considered a sign of successful exploration.}
The number of evaluations of $\log p(D, \eta, \theta)$ in each NUTS iteration is determined by the length of the simulated trajectory which is $\sum_{i = 0}^{j - 1} 2^{i} = 2^{j} - 1$, if $j$ incremental simulations were required to find a U-turn\footnote{In the efficient implementation of NUTS, once a U-turn is found, the final incremental simulation is interrupted so the number of model evaluations is actually somewhere between $2^{j-1}$ and $2^{j} - 1$}.

In the efficient implementations of the NUTS algorithm, a binary tree data structure of depth $j$ is used to help with the efficient selection of a proposal from all of the states $(\eta, \theta)$ visited during the simulation until a U-turn was found\footnote{The number of states visited excluding the initial state is at most $2^j - 1$.
	Adding the initial state, we have $2^j$ possible states any of which could be the proposal.
	These can in theory be stored as the leaf nodes of a binary tree of depth $j$ which has $2^j$ leaf nodes.
	However in practice, only a subset of such states are stored and the tree idea is used to ensure the proposal can be efficiently chosen at random from all $2^j$ possible states while satisfying the theoretical requirements of a proposal in MCMC, which is often called the detailed balance condition.
}, without storing all of the states.
This is an efficiency enhancement trick but the term \textit{tree depth} stuck and became synonymous to the number of incremental simulations ran so far, $j$.
In the case where the sampler is unable to find a U-turn even after a pre-specified maximum $j$ is reached, the sampler terminates the simulation anyways and makes a proposal.
The term maximum tree depth is commonly used to refer to the maximum number of incremental simulations $j$ allowed before having to make a proposal even if no U-turn was found.

\subsubsection{Optional: Distribution of the Initial Velocity} \label{sec:velocity}

Recall that in each NUTS iteration, we are sampling a random initial velocity for the $(\eta, \theta)$ particle before simulating the dynamics to arrive at a proposal.
Hypothetically, assume that we already have samples from the posterior $p(\eta, \theta \mid D)$.
If you were to go back and re-do the sampling using NUTS, how would you sample the initial velocity of the imaginary $(\eta, \theta)$ particle to make sampling as efficient as possible?
In general, it would make sense to move faster along directions in the posterior that have a higher variance and slower along directions that have a lower variance.
For instance, we can compute the variance along each parameter's axis and sample higher speeds for the parameters that change more, and lower speeds for the parameters that change less.
In practice, you can think of different parameters having different scales where 1 parameter may be in the 10s while another one may be in the 1000s.
In that case, it makes sense to use different speeds along different directions to more efficiently sample from the posterior distribution.
More generally, one may even compute the sample covariance matrix from the (hypothetical) samples available, compute the principal components and sample higher speeds along directions with more variance than the other directions.

If we encode how slow we want the particle to go along each direction $d_i$ by a number $s_i$, setting the standard deviation of the speed along this direction to $1/s_i$ can be used to achieve the desired slowness.
Assume each $d_i$ is an axis along a specific parameter $i$ (which could be part of $\eta$ or $\theta$).
The distribution of the velocity $v_i$ along $d_i$ can be the following univariate Gaussian:
\begin{align}
	v_i \sim N(0, (1/s_i)^2)
\end{align}
with mean 0 and standard deviation $1/s_i$.
This distribution will have us sampling speeds along the direction $d_i$ that are on average inversely proportional to $s_i$.

Writing it for all the parameters together, we can write:
\begin{align}
	v \sim N(0, M^{-1})
\end{align}
where $M$ is a diagonal matrix of elements $s_i^2$ on the diagonal and $M^{-1}$ is the covariance matrix of the velocity vector $v$.
Using a diagonal $M$ is equivalent to adapting the speeds' standard deviations along the parameters' axes.
While using a dense matrix $M$ is equivalent to the more general case of adapting the speeds' standard deviations along more optimized directions $d_i$ (e.g. from principal components of the covariance matrix).

It turns out that when simulating the ``imaginary dynamics'' in HMC/NUTS after sampling the initial velocity, the analogical \textit{kinetic energy} is given by:
\begin{align}
	K(v) = v^T M v / 2
\end{align}
hence the natural inclination to call the above matrix $M$ a ``mass matrix'' in the HMC/NUTS literature.
Recall that in physics, the kinetic energy of a particle with a scalar speed $v$ and mass $m$ is $\frac{m v^2}{2}$.

To summarize, directions with a higher ``mass'' will be explored more slowly than directions with a lower mass.
The ideal mass matrix $M$ is one that approximates the \textit{global} precision matrix of the posterior distribution, i.e. the inverse of the covariance matrix.
Equivalently, the ideal $M^{-1}$ is one that approximates the global covariance matrix of the posterior.

So far we assumed that we have samples from the posterior and are able to adapt the mass matrix manually.
In practice, the NUTS algorithm adapts the mass matrix for you during the adaptation phase, and you only need to select the structure of the matrix, e.g. diagonal or dense.
For large problems, a diagonal matrix is typically used in practice since the computational cost of using a dense matrix is $O(D^3)$, where $D$ is the total number of parameters in $(\eta, \theta)$ combined.
On the other hand, the computational cost of using a diagonal matrix is only $O(D)$.
When we have many subjects in the hierarchical model, $D$ can be quite large.

Before we conclude this section, it is important to note that the HMC/NUTS algorithm is typically explained in the literature using a so-called momentum vector $p = M v$ while we chose to use the more intuitive velocity vector $v$ in this paper to explain the intuition behind the algorithm.
The two parameterizations are equivalent but the momentum one is the one typically used in implementations and HMC/NUTS research.
When $v \sim N(0, M^{-1})$, the corresponding distribution of the momentum vector $p$ is $N(0, M)$.

\subsubsection{Optional: Hierarchical Priors and NUTS} \label{sec:hpriors}

\begin{figure}
	\centering
	\includegraphics[width=0.4\textwidth]{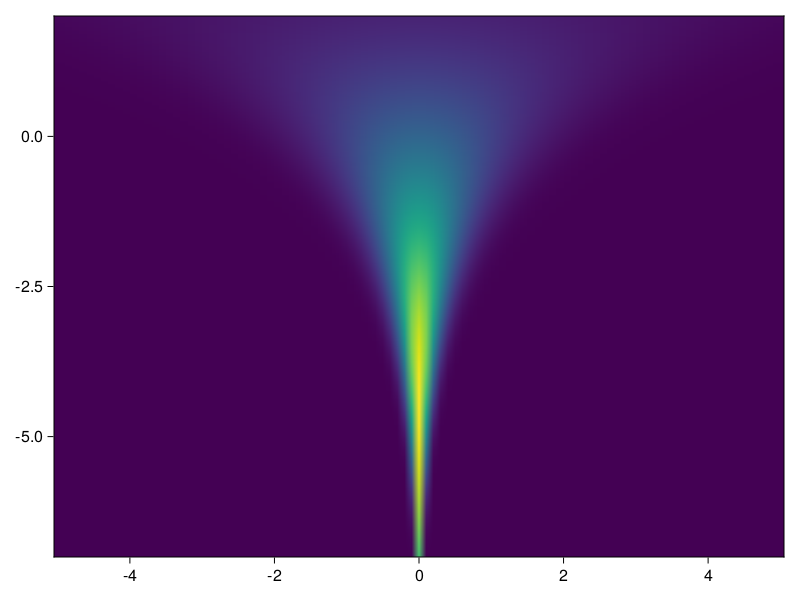}
	\caption{The PDF heatmap of the prior distribution of $\text{log}\omega$ on the y-axis and $\eta$ on the x-axis.}
	\label{fig:funnel}
\end{figure}

Consider the following toy model which has no observations:
\begin{align}
	\begin{split}
		\text{log}\omega &\sim \text{Normal}(0, 1.5) \\ \eta &\sim \text{Normal}\left( 0, (e^{\text{log}\omega}) \right)
	\end{split}
\end{align}
This is a model with 2 parameters $(\text{log}\omega, \eta)$, a prior distribution on each of them and some exponential dependence between them in that the standard deviation of the prior on $\eta$ depends exponentially on the value of $\text{log}\omega$.
Figure \ref{fig:funnel} shows the PDF heatmap of the joint prior distribution of the parameters $\text{log}\omega$ (y-axis) and $\eta$ (x-axis).
Recall that the NUTS algorithm uses a multi-step trajectory with a fixed time step size in the imaginary dynamics simulation to account for variation in curvature along the same direction.
Consider the direction along the x-axis in the figure.
The curvature along the x-axis changes depending on where along the y-axis the imaginary $(\text{log}\omega, \eta)$ particle is.
Lower values of $\text{log}\omega$ lead to exponentially higher curvatures (reflected through the tight color band in the heatmap) along the $\eta$ direction.
So if we try to use NUTS to sample from this prior, two bad things can happen:
\begin{enumerate}
	\item The sampler may adapt its step size to very small values to be able to sample from the lower regions of the prior and it will use a large number of time steps $T$ to effectively explore the upper regions of the prior.
	      In such cases, more often than not, the maximum number of allowed time steps $2^j - 1$ may not be enough to find the U-turn.
	      This will hurt the performance since we will be doing many model evaluations per proposal and we may need multiple steps to fully traverse the upper regions of the prior.
	\item The sampler may adapt its step size to values not small enough to sample from the lower regions in the prior.
	      In this case, the sampler may skip sampling from the significant lower part in the prior leading to potential bias in the results.
\end{enumerate}
In other words, the above prior may lead to slow and biased NUTS sampling, a clearly terrible outcome.

Note that of course we are not trying to sample from the prior using NUTS, because we can just sample directly from the standard distributions in the model using more direct and efficient methods.
However, studying the prior's PDF and how it interacts with NUTS can help us understand how NUTS will interact with the posterior when there are a few data points available.
Also note that the above model is using an explicit log scale parameterization for the standard deviation for pedagogical purposes.
In reality, models may be written directly in terms of the standard deviation $\omega$ (instead of its log) or more generally the covariance matrix $\Omega$ for multivariate $\eta$.
However, the above example is still relevant in those cases because implementations of the NUTS algorithm do the log-scale transformation behind the scenes and so NUTS actually samples unconstrained parameters all the time even if the original model's parameters were constrained to a different support.
So the same insight we build for the model above is applicable to general hierarchical models when using NUTS.

The standard deviation parameters (or more generally the covariance matrix) of the prior on the subject-specific parameters $\eta_i$ (commonly known as the between-subject variability) in pharmacometrics are typically assigned weak priors to avoid bias.
This means that they tend to have a wide variability in their value in the posterior distribution unless enough data is collected to precisely identify the value of the parameters.
This combination of:
\begin{enumerate}
	\item Weak priors on the covariance matrix parameters to avoid bias, \item Not having enough data to precisely identify the parameter values, \item The dependence between parameters' priors (in the standard deviation) in hierarchical models, \item The log scale domain transformation of standard deviation and covariance parameters used by NUTS, and \item The fixed step size used by the NUTS sampler,
\end{enumerate}
is an unfortunate but very common combination of factors that can lead to wrong and very slow inference.

So what's the solution?
One solution is to reparameterize the model as such:
\begin{align}
	\begin{split}
		\text{log}\omega &\sim \text{Normal}(0, 1.5) \\ \eta\text{std} &\sim \text{Normal}\left( 0, 1 \right) \\ \eta &= e^{\text{log}\omega} \times \eta\text{std}
	\end{split}
\end{align}
This reparameterization de-couples the priors of the parameters and resolves the issue in the PDF heatmap.
Note that this model transformation does not change the data generating process.
That is if you sample values for $\eta\text{std}$ and $\text{log}\omega$, the values of $\eta$ simulated will be identical to the values simulated from the original model's prior.
However, the latter parameterization is more friendly to the NUTS algorithm.
In the context of pharmacometrics, the following covariance-based model:
\begin{align}
	\begin{split}
		\theta & \sim p(\theta) \\ \Omega & \sim p(\Omega) \\ \eta_i & \sim N(\theta, \Omega)
	\end{split}
\end{align}
can be transformed to:
\begin{align}
	\begin{split}
		\theta & \sim p(\theta) \\ \Omega & \sim p(\Omega) \\ \eta\text{std}_i & \sim N(0, I) \\ \eta_i & = \text{chol}(\Omega) \times \eta\text{std}_i + \theta
	\end{split}
\end{align}
where $\text{chol}(\Omega)$ is the lower triangular Cholesky factor (a generalized square root for matrices) of the covariance matrix $\Omega$.
Similarly, using the standard deviation and correlation matrix parameterization instead, the original model becomes:
\begin{align}
	\begin{split}
		\theta & \sim p(\theta) \\ \omega & \sim p(\omega) \\ C & \sim p(C) \\ \eta_i & \sim N(\theta, D_{\omega} \times C \times D_{\omega})
	\end{split}
\end{align}
where $D_{\omega}$ is the diagonal matrix whose elements are the standard deviations vector $\omega$.
The above correlation-based model can be transformed to de-couple the priors as such:
\begin{align}
	\begin{split}
		\theta & \sim p(\theta) \\ \omega & \sim p(\omega) \\ C & \sim p(C) \\ \eta\text{std}_i & \sim N(0, I) \\ \eta_i & = D_{\omega} \times \text{chol}(C) \times \eta\text{std}_i + \theta
	\end{split}
\end{align}
When using Pumas to define these models, a transformation equivalent to the above transformation is done automatically behind the scenes even if you write the model in the coupled way.
\footnote{This equivalence is in exact arithmetic but when running computation on the computer, floating point arithmetic is done.
	This means that the results may not be identical depending on how sensitive the model is to round-off errors in the floating point arithmetic.
}

Before we conclude this section, we note that models where the priors on the population and subject-specific parameters are coupled are sometimes called centered parameterization (CP) models in the Bayesian literature.
While de-coupling the priors using the transformations discussed above is often called the non-centered parameterization (NCP).
These terms can be confusing though so we mostly avoid their use in this paper.

\subsection{Basic Summary Statistics} \label{sec:summary_stats3}

There are a few summary statistics that one can view to assess the convergence of the MCMC chains. These include:
\begin{itemize}
    \item \textbf{Effective Sample Size (ESS)}: an approximation of the ``number of independent samples'' generated by a Markov chain, when estimating the posterior mean. A low ESS per sample ratio is caused by high auto-correlation in the MCMC samples and is often a bad indicator.
    \item \textbf{$\widehat{R}$ (Rhat)}: potential scale reduction factor, a metric to measure if the Markov chains have mixed, and, potentially, converged. Chain mixing refers to the case when different chains include samples from the same regions in the posterior as opposed to each chain including samples from a separate region of the posterior.
    \item \textbf{Monte Carlo Standard Error (MCSE)}: the standard deviation divided by the ESS, which is a measure of estimation noise in the posterior mean.
\end{itemize}

The formula for the effective sample size (ESS) when estimating the posterior mean is:

\begin{align*}
\widehat{\eta}_{\text{eff}} = \frac{mn}{1 + \sum^T_{t=1}\widehat{\rho}_t}
\end{align*}

where $m$ is number of Markov chains, $n$ is total samples per Markov chain, and $\widehat{\rho}_t$ is an auto-correlation estimate.

This formula is an approximation of the ``number of independent sample'' generated by a Markov chain when estimating the mean values of the parameters. Since we don't have a way to recover the true auto-correlation $\rho$, instead we rely on an estimate $\widehat{\rho}$. The higher the auto-correlation in the chains, the lower the ESS will be for the same number of MCMC samples. High auto-correlation can result from too many rejections or optimization-like behaviour where sampler is moving towards a mode. Both of these are signs of lack of convergence. In general, a high auto-correlation alone is not a sign of lack of convergence though so care must be taken when interpreting the ESS to root-cause why it might be low.

The formula for the $\widehat{R}$ is:

\begin{align*}
\widehat{R} = \sqrt{\frac{\widehat{\operatorname{var}}^+ \left( \psi \mid y \right)}{W}}
\end{align*}

where $\widehat{\operatorname{var}}^+ \left( \psi \mid y \right)$ is the Markov chains' sample variance for a certain parameter $\psi$.
We calculate it by using a weighted sum of the within-chain variance $W$ and between-chain variance $B$:

\begin{align*}
\widehat{\operatorname{var}}^+ \left( \psi \mid y \right) = \frac{n - 1}{n} W + \frac{1}{n} B
\end{align*}

Intuitively, the value is $\widehat{R} = 1.0$ if all chains are totally convergent.
As a heuristic, if $\widehat{R} > 1.1$, you need to worry because probably the chains have not converged adequately.

\subsection{Convergence} \label{sec:convergence}


\subsubsection{Signs of Lack of Convergence}

MCMC has an interesting property that it will asymptotically converge to the target distribution.
That means, if time is not a limited resource, it is guaranteed that, irrelevant of the target distribution posterior geometry, MCMC will give you the right answer.
However, for all real-world scenarios, time is a limited resource.
Different MCMC algorithms, like NUTS, can reduce the sampling (and adaptation) time necessary for convergence to the target distribution.

In the ideal scenario, the NUTS sampler will converge to the true posterior and doesn't miss on any mode.
But, can we prove convergence?
Unfortunately, this is not easy to prove in general.
All the convergence diagnostics are only tests for symptoms of lack of convergence.
In other words if all the diagnostics look normal, then we can't prove that the sampler didn't converge.

There are some signs of lack of convergence:
\begin{itemize}
    \item Any of the moments (e.g. the mean or standard deviation) is changing with time.
    This is diagnosed using stationarity tests by comparing different parts of a single chain to each other.
    \item Any of the moments is sensitive to the initial parameter values.
    This is diagnosed using multiple chains by comparing their summary statistics to each other.
\end{itemize}
While high auto-correlation is not strictly a sign of lack of convergence, samplers with high auto-correlation will require many more samples to get to the same efficiency as another sampler with low auto-correlation.
So a low auto-correlation is usually more desirable.

\subsubsection{When Does Convergence Matter?}

Broadly speaking, there are 2 main classes of models we can use:
\begin{enumerate}
	\item Causal models, sometimes known as mechanistic models.
	\item Black-box models, sometimes known as regression models or machine learning models.
\end{enumerate}

Simply put, causal/mechanistic models are models that make sense in the domain of interest.
This means that:
\begin{enumerate}
	\item All of the variables in the model have a meaning.
	\item Each relationship in the model is well thought out, minimal and based on a claimed causal relationship between a subset of the variables.
\end{enumerate}

The goal of mechanistic models is to understand the system of interest.
Each model is typically designed to answer questions about some of the parameters in the model.
For example in PK models, we typically have absorption and clearance individual parameters.
So after fitting the model to the data, one can answer questions about the probability of a specific individual parameter being greater than or less than a certain threshold.
Another common example is dose response models, where we typically have a coefficient that represents the effect of the dose on the disease.
If the probability that this parameter is more than 0 (according to the posterior distribution) is higher than a certain threshold, we can claim that this drug is effective.
Correct causal/mechanistic models are supposed be good in both interpolation and extrapolation.\footnote{Note that causality is never implied by the model alone, instead it is based on the scientist's intuition and understanding of the model and its variables. In other words, models are nothing more than tools that can be used to express some claimed causal relationships between quantities that the scientist has in mind.}

On the other end of the spectrum, we have black-box models.
These are models commonly characterized by:
\begin{enumerate}
	\item Many intermediate variables that have no meaning.
	\item Dense relationships between all the variables without having a precise reason for each relationship upfront.
\end{enumerate}
These models are often called machine learning models.
Think of a linear regression model with polynomial bases up to the 5th order.
Simple linear regression models with linear terms only can arguably be in the first class of causal models if the covariates are claimed to cause the response in a linear fashion.
But once you get to the 3rd or 4th order polynomial bases, the higher order polynomial terms and their coefficients start losing meaning and the model becomes more black-box.
In Bayesian black-box models, prior distributions are typically arbitrary (e.g. a standard Gaussian) and used only for regularization.
The hyper-parameters of the prior distributions can even be optimized for maximum average posterior predictive accuracy\footnote{Using a validation data set that wasn't used in the training/inference of the model's parameters}.

There are many techniques to systematically build complicated black-box models.
Some examples include:
\begin{itemize}
	\item Using polynomial series terms as bases, e.g. Taylor polynomial series or the Chebyshev polynomial series \item Using Fourier series terms as bases \item Using deep neural networks adding neurons and layers as needed \item Using a Gaussian process for nonlinear regression
\end{itemize}
These are models that, given enough data for $(x, y)$ and given enough terms in the model, can fit any arbitrary function $y = f(x)$ without having any causal reasoning or meaning built into the model.
They are purely prediction machines that can be used to do interpolation and sometimes very limited extrapolation.
The ability of a model class to fit any function $f$ with a model large enough is sometimes called the universal approximation property which a number of machine learning model classes have \citep{hornikMultilayerFeedforwardNetworks1989}.

In practice, some models may combine components from causal/mechanistic models and black-box models.
For example, a causal model can be used to define which variables depend on which other variables but then the functional form of the dependence can be approximated using a black-box model.
Combining mechanistic and black-box models is sometimes known as scientific machine learning \citep{Rackauckas_Ma_Martensen_Warner_Zubov_Supekar_Skinner_Ramadhan_Edelman_2021}.

The reason why we are talking about different types of models here is because the types of diagnostics to use should be consistent with the goal of the analysis you are running.
If the goal is to make good predictions, regardless of the model's interpretability, then we can treat the model as a black-box and mostly rely on predictive diagnostics.
In this case, good predictions are sufficient even if the model doesn't make sense or if the inference process was imperfect.
To give an example, in Bayesian neural networks, extremely crude approximations are often done when inferring the posterior so long as the posterior predictions are consistent with the data \citep{bnn_review,subspace_bnn}.
On the other hand, if the purpose of the analysis is to understand the system and to learn about the values of the parameters in your model because they are significant in and of themselves, then causal/mechanistic models should have been used and extra care must be taken to ensure that we correctly sample from the posterior distribution and that priors were not too strong.

\subsection{Crossvalidation and Model Selection} \label{sec:model_selection}

In the Bayesian workflow, it is common to evaluate and compare models using their predictive power for out-of-sample data, i.e. data not used for the fitting or inference of the model parameters. One popular model evaluation metric for out-of-sample prediction accuracy is the so-called expected log predictive density (ELPD). Other common model selection criteria include various information criteria \citep{Burnham2002-fr} such as the Widely Applicable Information Criteria (WAIC). For a discussion of the ELPD as well as other model evaluation criteria, refer to \cite{Vehtari2012,PiiVeh2017,scoring_raftery}. 

Intuitively, the ELPD is some average measure of predictive accuracy across all posterior samples, averaged over a number of prediction tasks. Let $\mathcal{M}$ be the pharmacometrics model with parameters $(\eta, \theta)$ that describe the data generating process of the observed data $y \mid x$. The ELPD is defined as:
\begin{align*}
    \text{ELPD} = \int \log p(\hat{y} | \hat{x}, D, \mathcal{M}) \cdot p_t(\hat{y} | \hat{x}) d\hat{y}
\end{align*}
where $\hat{y}$ is unobserved data, e.g. future data points, $p_t(\hat{y} \mid \hat{x})$ is the true data generating distribution of $\hat{y}$ (unknown in practice) and $p(\hat{y} | \hat{x}, D, \mathcal{M})$ is the posterior predictive density defined as:
\begin{align*}
    p(\hat{y} | \hat{x}, D, \mathcal{M}) = \int p(\hat{y} | \hat{x}, \eta, \theta, \mathcal{M}) \cdot p(\eta, \theta | D, \mathcal{M}) d\theta
\end{align*}
where $p(\eta, \theta | D, \mathcal{M})$ describes the posterior distribution of $(\eta, \theta)$ given the previously observed data $D$ and the model $\mathcal{M}$. Since the true data generating distribution is unknown, it is common to approximate the ELPD by an empirical distribution over the observed data. One such estimator is the log pointwise predictive density (lppd).

Let $(x_i, y_i)$ be the $i^{th}$ observation by some arbitrary splitting of the data $D$ (not necessarily by subjects) into $S$ pieces and let $(\eta^{(j)}, \theta^{(j)})$ be the $j^{th}$ sample draw from the posterior $p(\eta, \theta | D, \mathcal{M})$, for $j \in 1,\dots,N$. The lppd can be calculated using Equation \ref{eq:lppd}.
\begin{figure*}
\begin{align} \label{eq:lppd}
    \text{lppd} & = \frac{1}{S} \sum_{i=1}^S \log p(y = y_i | x = x_i, D, \mathcal{M}) \nonumber \\
    & = \frac{1}{S} \sum_{i=1}^S \log \int p(y = y_i | x = x_i, \eta, \theta, \mathcal{M}) p(\eta, \theta | D, \mathcal{M}) d\theta \\
    & \approx \frac{1}{S} \sum_{i=1}^S \log  \Big( \frac{1}{N} \sum_{j=1}^N p(y = y_i | x = x_i, \eta = \eta^{(j)}, \theta = \theta^{(j)}, \mathcal{M}) \Big) \nonumber
\end{align}
\end{figure*}

A shortcoming of the lppd is that it is not representative of predictive accuracy on unseen data, since $(x_i, y_i)$ is used both for inference on the posterior and to evaluate the model out-of-sample.

\subsubsection{Leave-K-Out Crossvalidation}

Crossvalidation overcomes this problem by ensuring that $(x_i, y_i)$ is not used for inference on the posterior when evaluating the out-of-sample performance for $y_i \mid x_i$. The simplest way to divide the data into in-sample and out-of-sample subsets is the leave-one-out (loo) crossvalidation where in each outer iteration, one data point is considered out-of-sample and the remaining are in-sample. The leave-one-out, log predictive density (loo-lpd) is defined in Equation \ref{eq:loolpd},
\begin{figure*}
\begin{align} \label{eq:loolpd}
    \text{loo-lpd} & = \frac{1}{S} \sum_{i=1}^S \log p(y = y_i | x = x_i, D = D_{-i}, \mathcal{M}) \nonumber \\
    & = \frac{1}{S} \sum_{i=1}^S \log \int p(y = y_i | x = x_i, \eta, \theta, \mathcal{M}) \cdot p(\eta, \theta | D = D_{-i}, \mathcal{M}) d\theta \\
    & \approx \frac{1}{S} \sum_{i=1}^S \log  \Big( \frac{1}{N} \sum_{j=1}^N p(y = y_i | x = x_i, \eta = \eta^{(j)}_{-i}, \theta = \theta^{(j)}_{-i}, \mathcal{M}) \Big) \nonumber
\end{align}
\end{figure*}
where $D_{-i}$ is all data excluding $(x_i, y_i)$ and $(\eta^{(j)}_{-i}, \theta^{(j)}_{-i})$ is the $j^{th}$ sample draw from the posterior $p(\eta, \theta | D = D_{-i}, \mathcal{M})$. This can be generalised to leave $K$-out cross validation where $(x_i, y_i)$ is interpreted as $K$ observations, e.g $K$ subjects or $K$ drug concentration observations.

\subsubsection{Leave-Future-K-Out Crossvalidation}

When working with time-series data, it can often be more useful to evaluate models based on their ability to predict future values using nothing but past values for training. This gives rise to another variant of crossvalidation called leave-future-one-out (lfoo) crossvalidation and the lfoo-lpd which is defined in Equation \ref{eq:lfoolpd},
\begin{figure*}
\begin{align} \label{eq:lfoolpd}
    \text{lfoo-lpd} & = \frac{1}{S - t} \sum_{i=t+1}^S \log p(y = y_i | x = x_i, D = D_{1:i-1}, \mathcal{M}) \nonumber \\
    & = \frac{1}{S - t} \sum_{i=t+1}^S \log \int p(y = y_i | x = x_i, \eta, \theta, \mathcal{M}) \cdot p(\eta, \theta | D = D_{1:i-1}, \mathcal{M}) d\theta \\
    & \approx \frac{1}{S - t} \sum_{i=t+1}^S \log \frac{1}{N} \sum_{j=1}^N p(y = y_i | x = x_i, \eta = \eta^{(j)}_{-(i:S)}, \theta = \theta^{(j)}_{-(i:S)}, \mathcal{M}) \nonumber
\end{align}
\end{figure*}
where $t$ is the minimum number of data points used for training/inference, $D_{1:i-1}$ is the past data and $(\eta^{(j)}_{-(i:S)}, \theta^{(j)}_{-(i:S)})$ is the $j^{th}$ sample draw from the posterior $p(\eta, \theta | D = D_{1:i-1}, \mathcal{M})$ which is obtained by excluding the future data $D_{i:S}$ from the inference.

\subsubsection{Crossvalidation for Hierarchical Models}

When performing crossvalidation in a hierarchical model, there are multiple ways to measure the predictive power of the model. For instance in hierarchical pharmacometric modeling, the goal is to learn a population model to make predictions on new patients while simultaneously learning subject-specific models to make future predictions for specific subjects having seen their past response to drugs. These models are useful for dose selection and dose adaptation for new or existing patients with the objective of maximizing the therapeutic effect while avoiding toxicity.

Depending on the prediction task of interest, one may choose to treat each time observation as a data point or each entire patient/subject as a data point. If the goal is to evaluate the model's ability to predict responses for new patients, leave-one-subject-out crossvalidation should be used. Alternatively, if the goal is to evaluate the model's ability to predict future drug concentrations or any other observable time-dependent quantity the model predicts, then leaving future observations out for each subject makes more sense. This will be called leave-one-observation-out or leave-one-future-observation-out crossvalidation.

The choice of what constitutes a point to leave out when doing crossvalidation affects the way the predictive likelihoods are computed:
\begin{align*}
    p(y = y_i | x = x_i, \eta = \theta^{(j)}, \theta = \theta^{(j)}, \mathcal{M})
\end{align*}
When leaving subjects out, we are often interested in the marginal likelihood of this subject given a posterior sample draw of the population parameters $\theta$, marginalizing out the subject-specific parameters $\eta$. Alternatively, the conditional likelihood can also be used for some default or typical values of the subject-specific parameters, e.g. the mode of the prior distribution. To marginalize subject-specific parameters, approximate integration methods such as \texttt{LaplaceI} and \texttt{FOCE} can be used to obtain the marginal likelihood. On the other hand when leaving observations out in a single subject, the quantity of interest is often the conditional likelihood given each sample from the joint posterior of population and subject-specific parameters of individual subjects given previous data of the subject.

\subsubsection{Pareto Smoothed Importance Sampling Crossvalidation}

Evaluating the loo-lpd or lfoo-lpd is expensive since one needs to draw samples from $N$ or $N - t$ different posteriors, e.g. from $p(\eta, \theta | D = D_{-i}, \mathcal{M})$ for loo-lpd. Typically this will be done by MCMC, e.g. the NUTS algorithm, which in spite of recent progress, remains computationally expensive when the number of parameters is large and the curvature of the posterior is uneven along one or more dimensions.

One approach to overcome this difficulty, is the Pareto-smoothed importance sampling method for leave-one-out, crossvalidation (PSIS-LOO-CV) \citep{psis_loo}. In PSIS-LOO-CV, MCMC is run only once on the full data. The same samples are then re-used in each outer iteration of CV but using different weights. The weights are determined using importance sampling (IS) by comparing the likelihood with one data point left out to the likelihood of the full dataset. The raw importance weights are then smoothed by fitting them to a generalized Pareto distribution. The smoothed weights can then be used to estimate the ELPD contribution of each data point.

Beside the ability to approximate the ELPD, PSIS-LOO-CV also provides a useful diagnostic which is the shape parameter of the Pareto distribution fitted to the raw weights when leaving out each data point. Data points that when removed lead to a large shape parameter are more influential than data points which have a low shape parameter. For highly influential points where the Pareto shape parameter is higher than 0.7, the ELPD contribution for this point can be considered unreliable. In those cases, resampling from the posterior after removing the influential point is recommended.

%% file: sections/4-example_models.tex

\section{Example Models} \label{sec:example_models}

Listings \ref{lst:example_pk}, \ref{lst:example_poppk}, \ref{lst:example_hcv} and \ref{lst:example_tte} are examples of some common models in pharmacometrics.

\begin{figure*}[t]
\begin{lstlisting}[caption=Single Subject PK Model,label=lst:example_pk]
@model begin
    @param begin
        tvcl {$\sim$} LogNormal(log(10), 0.25) # CL
        tvq  {$\sim$} LogNormal(log(15), 0.5)   # Q
        tvvc {$\sim$} LogNormal(log(35), 0.25) # V1
        tvvp {$\sim$} LogNormal(log(105), 0.5) # V2
        tvka {$\sim$} LogNormal(log(2.5), 1)   # ka
        {$\sigma$}    {$\sim$} truncated(Cauchy(), 0, Inf) # sigma
    end
    @pre begin
        CL = tvcl
        Vc = tvvc
        Q  = tvq
        Vp = tvvp
        Ka = tvka
    end
    @dynamics begin
        Depot'      = -Ka * Depot
        Central'    =  Ka * Depot  - (CL+Q)/Vc * Central + Q/Vp * Peripheral
        Peripheral' =                     Q/Vc * Central - Q/Vp * Peripheral
    end
    @derived begin
        cp  := @. Central/Vc
        conc {$\sim$} @. LogNormal(log(cp), {$\sigma$})
    end
end
\end{lstlisting}
\end{figure*}

\begin{figure*}[t]
\begin{lstlisting}[caption=Population PK Model,label=lst:example_poppk]
@model begin
  @param begin
    tvcl {$\sim$} LogNormal(log(10), 0.25) # CL
    tvq  {$\sim$} LogNormal(log(15), 0.5)   # Q
    tvvc {$\sim$} LogNormal(log(35), 0.25) # V1
    tvvp {$\sim$} LogNormal(log(105), 0.5) # V2
    tvka {$\sim$} LogNormal(log(2.5), 1)   # ka
    {$\sigma$}    {$\sim$} truncated(Cauchy(0, 5), 0, Inf) # sigma
    C    {$\sim$} LKJCholesky(5, 1.0)
    {$\omega$}    {$\sim$} Constrained(
      MvNormal(zeros(5), Diagonal(0.4^2 * ones(5))),
      lower=zeros(5),
      upper=fill(Inf, 5),
      init=ones(5),
    )
  end
  @random begin
    {$\eta$}std {$\sim$} MvLogNormal(I(5))
  end
  @pre begin
    {$\eta$}  = {$\omega$}   .* (getchol(C).L * {$\eta$}std)
    CL = tvcl * {$\eta$}[1]
    Q  = tvq  * {$\eta$}[2]
    Vc = tvvc * {$\eta$}[3]
    Vp = tvvp * {$\eta$}[4]
    Ka = tvka * {$\eta$}[5]
  end
  @dynamics begin
    Depot'      = -Ka * Depot
    Central'    =  Ka * Depot - (CL+Q)/Vc * Central + Q/Vp * Peripheral
    Peripheral' =                    Q/Vc * Central - Q/Vp * Peripheral
  end
  @derived begin
    cp  := @. Central/Vc
    conc {$\sim$} @. LogNormal(log(cp), {$\sigma$})
  end
end
\end{lstlisting}
\end{figure*}

\begin{figure*}[t]
\begin{lstlisting}[caption=Hepatitis C Virus Model,label=lst:example_hcv]
@model begin
    @param begin
        {$\theta$}Ka   {$\sim$} LogNormal(log(0.8), 1)
        {$\theta$}Ke   {$\sim$} LogNormal(log(0.15), 1)
        {$\theta$}Vd   {$\sim$} LogNormal(log(100), 1)
        {$\theta$}n    {$\sim$} LogNormal(log(2.0), 1)
        {$\theta$}{$\delta$}    {$\sim$} LogNormal(log(0.20), 1)
        {$\theta$}c    {$\sim$} LogNormal(log(7.0), 1)
        {$\theta$}EC50 {$\sim$} LogNormal(log(0.12), 1)

        {$\omega^2$}Ka   {$\sim$} truncated(Normal(0.25, 1), 0.0, Inf)
        {$\omega^2$}Ke   {$\sim$} truncated(Normal(0.25, 1), 0.0, Inf)
        {$\omega^2$}Vd   {$\sim$} truncated(Normal(0.25, 1), 0.0, Inf)
        {$\omega^2$}n    {$\sim$} truncated(Normal(0.25, 1), 0.0, Inf)
        {$\omega^2$}{$\delta$}    {$\sim$} truncated(Normal(0.25, 1), 0.0, Inf)
        {$\omega^2$}c    {$\sim$} truncated(Normal(0.25, 1), 0.0, Inf)
        {$\omega^2$}EC50 {$\sim$} truncated(Normal(0.25, 1), 0.0, Inf)

        {$\sigma^2$}PK {$\sim$} truncated(Normal(0.04, 1), 0.0, Inf)
        {$\sigma^2$}PD {$\sim$} truncated(Normal(0.04, 1), 0.0, Inf)
    end
    @random begin
        {$\eta$}Ka   {$\sim$} LogNormal(0.0, sqrt({$\omega^2$}Ka))
        {$\eta$}Ke   {$\sim$} LogNormal(0.0, sqrt({$\omega^2$}Ke))
        {$\eta$}Vd   {$\sim$} LogNormal(0.0, sqrt({$\omega^2$}Vd))
        {$\eta$}n    {$\sim$} LogNormal(0.0, sqrt({$\omega^2$}n))
        {$\eta$}{$\delta$}    {$\sim$} LogNormal(0.0, sqrt({$\omega^2$}{$\delta$}))
        {$\eta$}c    {$\sim$} LogNormal(0.0, sqrt({$\omega^2$}c))
        {$\eta$}EC50 {$\sim$} LogNormal(0.0, sqrt({$\omega^2$}EC50))
    end
    @pre begin
        # constants
        p = 100.0
        d = 0.001
        e = 1e-7
        s = 20000.0

        Ka   = {$\theta$}Ka * {$\eta$}Ka
        Ke   = {$\theta$}Ke * {$\eta$}Ke
        Vd   = {$\theta$}Vd * {$\eta$}Vd
        n    = {$\theta$}n * {$\eta$}n
        {$\delta$}    = {$\theta$}{$\delta$} * {$\eta$}{$\delta$}
        c    = {$\theta$}c * {$\eta$}c
        EC50 = {$\theta$}EC50 * {$\eta$}EC50
    end
    @init begin
        T = (c * {$\delta$}) / (p * e)
        I = max(0, s * e * p - d * c * {$\delta$}) / (p * {$\delta$} * e)
        W = max(0, s * e * p - d * c * {$\delta$}) / (c * {$\delta$} * e)
    end
    @vars begin
        # short-hand notation for A / Vd
        conc = A / Vd
    end
    @dynamics begin
        X' = -Ka * X
        A' = Ka * X - Ke * A
        T' = s - T * (e * W + d)
        I' = e * W * T - {$\delta$} * I
        W' = p / ((conc / EC50 + 1e-100)^n + 1) * I - c * W
    end
    @derived begin
        log10W := @. log10(W)
        yPK    {$\sim$} @. truncated(Normal(conc, sqrt({$\sigma^2$}PK)), 0, Inf)
        yPD    {$\sim$} @. truncated(Normal(log10W, sqrt({$\sigma^2$}PD)), 0, Inf)
    end
end
\end{lstlisting}
\end{figure*}

\begin{figure*}[t]
\begin{lstlisting}[caption=Time-To-Event Model,label=lst:example_tte]
@model begin
    @param begin
        {$\lambda_1$} {$\sim$} LogNormal(0.0, 2.0) # basal hazard
        {$\beta$}  {$\sim$} LogNormal(0.0, 2.0) # fixed effect DOSE
        {$\omega$}  {$\sim$} LogNormal(0.0, 2.0) # inter-subject variability
    end
    @random begin
        {$\eta$} {$\sim$} LogNormal(0.0, {$\omega$})
    end
    @covariates DOSE
    @pre begin
        _{$\lambda_1$} =  {$\lambda_1$} * {$\eta$}      # basal hazard with inter-subject variability
        _{$\lambda_0$} = _{$\lambda_1$} * {$\beta$}^DOSE # total hazard
    end
    @vars begin
        {$\lambda$} = _{$\lambda_0$}
    end
    @dynamics begin
        {$\Lambda$}' = {$\lambda$}
    end
    @derived begin
        dv {$\sim$} @. TimeToEvent({$\lambda$}, {$\Lambda$})
    end
end
\end{lstlisting}
\end{figure*}

\clearpage

%% file: sections/5-conclusion_and_acknowledgements.tex

\section{Conclusion} \label{sec:conclusion}

In this work, we presented a comprehensive Bayesian analysis workflow using Pumas. All the syntax and relevant theory were presented following an intuition-first approach as much as possible, with numerous cross-links. If you are an existing Pumas user and you have further questions, you can reach out to us via the Pumas Discourse platform (\href{discourse.pumas.ai}{https://discourse.pumas.ai}). You can also find more focused tutorials and complete scripts on the Pumas tutorials website (\href{tutorials.pumas.ai}{https://tutorials.pumas.ai}) for your continued learning.

If after reading this paper, you would like to read and learn more about Bayesian statistics, the following are some excellent resources that can be used to further your learning:
\begin{enumerate}
    \item Bayesian Data Analysis \citep{Gelman2013BDA}
    \item Statistical Rethinking: A Bayesian Course with Examples in R and Stan \citep{mcelreath2020statistical},
    \item Regression and Other Stories \citep{Gelman2020ROS},
    \item Data Analysis Using Regression and Multilevel/Hierarchical Models \citep{Gelman2006DA},
    \item Probabilistic Machine Learning: An Introduction \citep{pml1Book}
    \item Probability Theory: The Logic of Science \citep{jaynes2003probability}
\end{enumerate}

\section{Acknowledgements}

We would like to acknowledge all of the reviewers of the early drafts of this work for their valuable feedback. In particular, we would like to thank \textbf{Haden Bunn} (Pumas-AI Inc.), \textbf{Joga Gobburu} (Pumas-AI Inc. and the University of Maryland Baltimore), \textbf{Yoni Nazarathy} (Pumas-AI Inc. and the University of Queensland), \textbf{Vaibhav Dixit} (Pumas-AI Inc.), \textbf{Russell Tsuchida} (CSIRO Data61), \textbf{Anastasios Panagiotelis} (University of Sydney) and \textbf{Mutaz Jaber} (Gilead Sciences Inc.).